# A Review on Mechanics and Mechanical Properties of 2D Materials – Graphene and Beyond

Deji Akinwande[1], Christopher J. Brennan[1], J. Scott Bunch[2], Philip Egberts[3], Jonathan R. Felts[4], Huajian Gao[5], Rui Huang[6*], Joon-Seok Kim[1], Teng Li[7], Yao Li[8], Kenneth M. Liechti[6*], Nanshu Lu[6], Harold S. Park[2], Evan J. Reed[9], Peng Wang[6], Boris I. Yakobson[10], Teng Zhang[11], Yong-Wei Zhang[12], Yao Zhou[9], Yong Zhu[13]

[1]Department of Electrical and Computer Engineering, University of Texas at Austin, Austin, Texas 78712, USA

[2]Department of Mechanical Engineering, Boston University, Boston, MA 02215, USA

[3]Department of Mechanical and Manufacturing Engineering, University of Calgary, 40 Research Place NW, Calgary, AB T2L 1Y6 Canada

[4]Department of Mechanical Engineering, Texas A&M University, College Station, Texas, 77843 USA

[5]School of Engineering, Brown University, Providence, RI 02912, USA

[6]Department of Aerospace Engineering & Engineering Mechanics, University of Texas at Austin, Austin, Texas 78712, USA

[7]Department of Mechanical Engineering, University of Maryland, College Park, MD 20742, USA

[8]Department of Applied Physics, Stanford University, Stanford, CA 94305, USA

[9]Department of Material Science and Engineering, Stanford University, Stanford, CA 94305, USA

[10]Department of Materials Science & NanoEngineering, Department of Chemistry, and the Richard E. Smalley Institute, Rice University, Houston, Texas 77005, USA

[11]Department of Mechanical and Aerospace Engineering, Syracuse University, Syracuse, NY 13244, USA

[12]Institute of High Performance Computing, A*STAR, 138632, Singapore

[13]Department of Mechanical and Aerospace Engineering, North Carolina State University, NC 27695, USA

---

* Corresponding authors. Email: ruihuang@mail.utexas.edu; kml@mail.utexas.edu

**Abstract**

Since the first successful synthesis of graphene just over a decade ago, a variety of two-dimensional (2D) materials (e.g., transition metal-dichalcogenides, hexagonal boron-nitride, etc.) have been discovered. Among the many unique and attractive properties of 2D materials, mechanical properties play important roles in manufacturing, integration and performance for their potential applications. Mechanics is indispensable in the study of mechanical properties, both experimentally and theoretically. The coupling between the mechanical and other physical properties (thermal, electronic, optical) is also of great interest in exploring novel applications, where mechanics has to be combined with condensed matter physics to establish a scalable theoretical framework. Moreover, mechanical interactions between 2D materials and various substrate materials are essential for integrated device applications of 2D materials, for which the mechanics of interfaces (adhesion and friction) has to be developed for the 2D materials. Here we review recent theoretical and experimental works related to mechanics and mechanical properties of 2D materials. While graphene is the most studied 2D material to date, we expect continual growth of interest in the mechanics of other 2D materials beyond graphene.

## 1. Introduction

The isolation of monolayer graphene flakes by mechanical exfoliation of bulk graphite opened the field of two-dimensional (2D) materials [1]. Since then, many other 2D materials have been discovered, such as transition metal-dichalcogenides (TMDs, e.g., $MoS_2$), hexagonal boron-nitride (h-BN), and black phosphorous or phosphorene. The family of 2D materials offers a full spectrum of physical properties, from conducting graphene to semiconducting $MoS_2$ and to insulating h-BN. Moreover, the 2D crystal structures render a unique combination of mechanical properties, with high in-plane stiffness and strength but extremely low flexural rigidity. Together, the 2D materials are promising for a wide range of applications [2, 3].

Here we review recent theoretical and experimental studies related to mechanics and mechanical properties of 2D materials. We emphasize how mechanics is indispensable in the study of mechanical properties including interfacial properties and the coupling between the mechanical and other physical properties. The review is divided into five self-contained sections. As the most basic mechanical properties, elastic properties of 2D materials are discussed in Section 2. Experimental methods to measure the in-plane elastic properties of graphene have been developed [4] and extended to other 2D materials [5, 6]. A recent experiment reported surprising results for the in-plane stiffness of graphene [7], opening a question on the effects of defects and statistical rippling. Direct measurement of the elastic bending modulus is more challenging for 2D materials [8]. Here again, a recent experiment [9] reported orders of magnitude higher values than theoretical predictions, raising a question on the fundamental mechanics of bending an ultrathin membrane with the effect of thermal fluctuations [10]. Theoretically, density functional theory based first-principles calculations have been used to predict linear and nonlinear elastic properties of graphene and other 2D materials. On the other hand, the accuracy of empirical potentials for molecular dynamics (MD) simulations remains to be improved. The effects of thermal rippling on the elastic properties of graphene are discussed based on MD simulations and statistical mechanics of elastic membranes [11].

Section 3 focuses on inelastic properties of 2D materials, starting with a description of defects such as vacancies, dislocations, and grain boundaries [12]. The strength and toughness are then discussed. The strength of a pristine 2D material is usually high [4, 6], but it could vary substantially due to the presence of topological defects and out-of-plane deformation [13, 14]. The fracture toughness of graphene obtained from experiments is relatively low [15], for which potential toughening mechanisms have been explored [16]. Fundamental questions have also been raised on the appropriate definition of fracture toughness for 2D materials based on fracture mechanics.

Section 4 deals with the coupling between mechanical deformation and other physical properties of 2D materials. Recent theoretical and experimental work has shown unprecedented effects of strain on many physical properties of graphene and other 2D materials [17], making "strain engineering" a viable approach for a wide range of potential applications involving 2D materials. As a sampling of the vast literature on this subject, we focus here on pseudomagnetic fields (PMFs) in deformed graphene [18, 19], phase transitions of TMDs under different mechanical constraints [20, 21], phonon and electronic structures of TMDs under hydrostatic pressure and strain [22, 23], and piezo- and flexoelectricity of 2D materials that couple strains and strain gradients (curvature) to polarization [24, 25].

Section 5 is devoted to interfacial properties of 2D materials. Adhesion and friction experiments have been developed to measure the mechanical interactions between graphene and other materials as its substrate or probing tips. In addition to the measurement of adhesion energy [26], more detailed measurements and analysis revealed the strength and range of the interactions in form of traction-separation relations [27], which provided further insight into the underlying mechanisms of the mechanical interactions. While van der Waals interactions have been commonly assumed to be the primary mechanism, experimental evidence suggests that other mechanisms may also have to be considered, such as the effects of water capillary, reactive defects, and surface roughness. Theoretically, it is also possible to unify the adhesion and friction properties of the interface within the framework of mixed-mode, nonlinear fracture mechanics.

In Section 6, a brief account of potential applications related to the mechanics and mechanical properties of 2D materials is presented, including synthesis and transfer for large-scale manufacturing, graphene origami and kirigami, flexible electronics and biomedical applications.

In the final section of this review, we provide an outlook for further studies related to mechanics and mechanical properties of 2D materials including and beyond graphene.

## 2. Elastic properties

Like thin membranes, 2D materials may be deformed by in-plane stretching or by bending out-of-plane. As a result, the elastic properties of 2D materials include both in-plane and bending moduli. With combined stretching and bending, a set of coupling moduli may also be defined theoretically [28], as noted for graphene being rolled into carbon nanotubes [29]. This section reviews recent experiments for measuring the elastic properties of 2D materials as well as theoretical predictions from first principles to continuum mechanics modeling.

### 2.1 Experiments

A direct measurement of mechanical properties of monolayer graphene was first reported by Lee et al. [4], by nanoindentation of suspended monolayer graphene membranes using an atomic force microscope (AFM). The indentation force-displacement behavior (Fig. 1A) was interpreted as a result of the nonlinear elastic properties of graphene, with a 2D Young's modulus of 340 N/m and a third-order elastic stiffness of –690 N/m in the nonlinear regime. A more detailed analysis of the nanoindentation experiment was performed by Wei and Kysar [30] using the finite element method (FEM) with an anisotropic, nonlinearly elastic constitutive model for graphene as predicted by density functional theory (DFT) calculations. Similar AFM indentation experiments were later conducted to study the mechanical properties of polycrystalline graphene films with different grain sizes as grown by chemical vapor deposition (CVD) [13]. It was found that the elastic stiffness of CVD-graphene is identical to that of pristine graphene if post processing steps avoided damage or rippling. A separate study showed that presence of out-of-plane ripples would effectively lower the in-plane stiffness of graphene [31]. Evidently, the AFM indentation method is applicable for other 2D materials beyond graphene, such as $MoS_2$ [5, 6] and h-BN [32].

A rather surprising result was reported recently by Lopez-Polin et al. [7], who conducted similar AFM-based nanoindentation experiments but obtained much higher 2D Young's modulus (up to 700 N/m) of graphene after introducing a controlled density of defects by irradiation. The authors attributed the increasing elastic modulus to the effects of thermal fluctuations and associated strain dependence [33]. However, a detailed analysis by Los et al. [34] based on statistical mechanics found that, while the 2D Young's modulus of graphene could increase significantly with a tensile strain due to suppression of rippling, the maximum value should not exceed the fundamental value (~340 N/m) at the limit of a perfectly flat graphene membrane. Moreover, the same analysis predicted a power-law decrease in the elastic modulus with increasing membrane size at a finite temperature, which has not been observed in any experiment. Another explanation of the surprisingly high elastic modulus was offered by Song and Xu [35], who considered the geometrical effect due to areal expansion of the graphene membranes with defects. While not fully resolved, the counter-intuitive results suggest some uncertainties in the indentation experiments and in particular, the interpretation of the data based on a relatively simple mechanics model.

The elastic properties of 2D materials can also be measured by pressurized blister tests or bulge tests, a common method for thin film materials [36]. Utilizing the remarkable gas impermeability of graphene, Koenig et al. [26] conducted a series of blister tests and obtained elastic moduli of single and multilayered graphene membranes. A similar setup was recently employed for single and multilayered $MoS_2$ as a way to apply large biaxial strains for band gap engineering [37]. An interesting variation of the blister test was developed by Nicholl et al. [38], where suspended graphene membranes were electrostatically pressurized by applying a voltage between the graphene and a gating chip. They found that the in-plane stiffness of graphene is 20-100 N/m at room temperature, much smaller than the expected value (~340 N/m). Moreover, the in-plane stiffness increased moderately when the temperature decreases to 10 K, but it increased significantly (approaching 300 N/m) when the graphene membranes were cut into narrow ribbons. The temperature and geometry (size) dependence of the elastic stiffness again point to the effects of the out-of-plane rippling or crumpling [38].

In both the AFM-based nanoindentation and pressurized blister experiments, the effect of bending modulus of the 2D materials is often considered to be negligible in the mechanics model used to extract the in-plane elastic properties and residual tension, which is theoretically justifiable as the membrane-like behavior with extremely low bending modulus. However, the effect of bending modulus may become substantial for multilayered specimens, which was demonstrated by the indentation experiments on multilayers of $MoS_2$ [39] and mica [40]. As shown in Fig. 1B, with increasing number of atomic layers of mica, the indentation force-displacement data exhibited a transition from the nonlinear membrane-like behavior (bilayers) to a linear plate-like behavior (12 layers). A similar transition was predicted for pressurized graphene blisters of different sizes [41]. A recent review by Castellanos-Gomez et al. [42] highlighted the membrane-to-plate like transition in both static and dynamic responses of 2D materials.

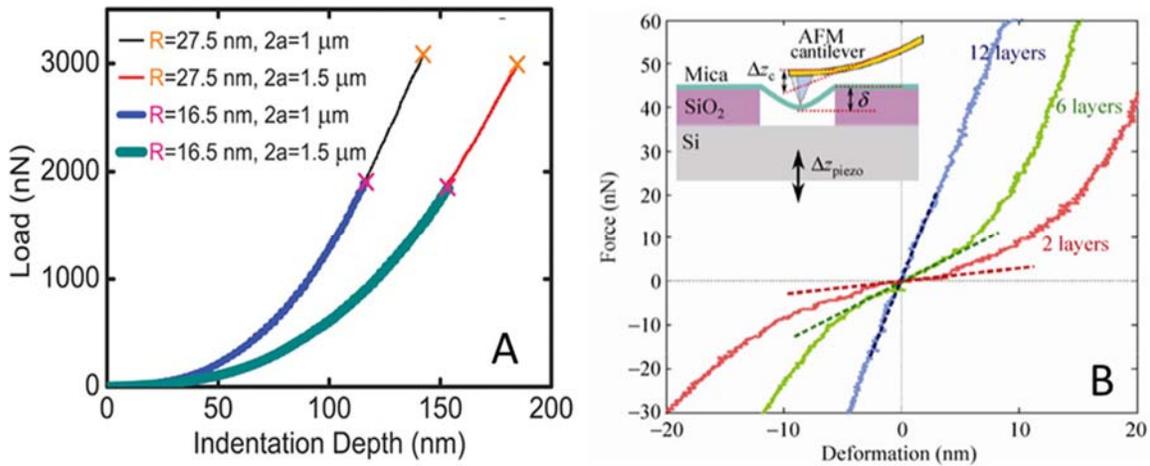

**Fig. 1**: (A) Force-displacment data from AFM nanoindentaiton of suspended monolayer graphene, with different tip radii and specimen diameters; fracture loads are indicated by × marks. (B) Force-displacement data measured by nanoindentaiton of suspended mica nanosheets of 2, 6 and 12 atomic layers. (Inset) schematic diagram of the indentation experiment. Figures adapted from (A) [4] and (B) [40].

Direct measurement of the bending modulus has been challenging for monolayer graphene as well as other 2D materials. The value often quoted for the bending modulus of monolayer graphene (~1.2 eV) was estimated from the phonon spectrum of bulk graphite [43]. Lindahl et al. [8] determined the bending stiffness of double-layer graphene based on measurements of the critical voltage for snap-through of pre-buckled graphene membranes. The obtained value (~35.5 eV) lies in between two theoretical limits, ~3 and 160 eV for two independent monolayers and two fully coupled monolayers [44, 45], respectively. The same method was applied to monolayer graphene membranes, yielding a rough estimate of ~7.1 eV for the bending modulus with higher uncertainties due to rather limited data points [8]. More recently, Blees et al. [9] measured the spring constant of a graphene cantilever structure by using the photon pressure from an infrared laser, based on which the bending modulus of monolayer graphene was inferred (using an elementary mechanics model). They also measured thermal fluctuations of the graphene cantilevers to determine the spring constants based on the equipartition theorem of classical statistical mechanics. Both methods yielded a surprisingly high bending modulus for monolayer graphene, on the order of $10^3$ to $10^4$ eV. They attributed the obtained high bending modulus to the effects of ripples (both static and thermal) in the graphene membranes, which would also suggest a strong size dependence for the bending modulus. The large disparity in these values (from ~1 to $10^4$ eV) for the bending modulus of monolayer graphene calls for further studies, both experimentally and theoretically.

**2.2 Theoretical predictions**

The elastic properties of pristine graphene have been well predicted by first principles based calculations [46-48]. Subject to small in-plane deformation, graphene is isotropic and linearly elastic with a 2D Young's modulus ($Y_{2D}$) and Poisson's ratio (ν) as listed in Table I. The predicted 2D Young's modulus, which is in good agreement with the measured values [4], may be converted to the conventional Young's modulus ($Y = Y_{2D}/h \sim 1.03$ TPa) by assuming a thickness ($h$) for the graphene monolayer, typically the interlayer spacing (0.335 nm) in bulk graphite. The high in-plane stiffness of graphene is a direct result of its hexagonal lattice and the carbon-carbon bonds. When highly deformed, the hexagonal symmetry of the graphene lattice may be broken, leading to anisotropic and nonlinearly elastic properties. To describe the nonlinear elastic behavior of graphene, several continuum mechanics models based on hyperelasticity have been proposed [48-50]. Once calibrated by the first principles calculations, the continuum mechanics models can be used to predict nonlinear elastic behaviors of pristine graphene monolayers up to the limit of intrinsic elastic instability at much larger scales under various loading conditions (such as nanoindentation) [30, 50].

Molecular dynamics (MD) simulations are often used to study the elastic and inelastic behavior of graphene. However, the accuracy of MD simulations depends on parametrization of the empirical potentials that describe the atomic interactions. The reactive empirical bond-order (REBO) potentials [51-53] have been commonly used in MD simulations of graphene and carbon nanotubes (CNTs). Unfortunately, these potentials were not parameterized to yield accurate elastic properties of graphene [54, 55]. As listed in Table I, the 2D Young's modulus and Poisson's ratio obtained by the REBO potentials are considerably different from the values predicted by first-principles calculations. Interestingly, the biaxial modulus, $Y_{2D}/(1-v)$, is well predicted by the REBO potentials. Re-parameterization of the REBO potential improved the predictions of the in-plane phonon-dispersion data for graphite [56], which might also offer a better prediction of the elastic properties of graphene. A few other potentials have also been used for graphene, and their predictions of the linear elastic properties are listed in Table I for comparison. Beyond linear elasticity, the empirical potentials can be used to simulate the nonlinear elastic behavior of graphene [57], in qualitatively good agreement with the first-principles calculations (Fig. 2A), but the quantitative agreement is more challenging as the empirical potentials are usually parameterized with the properties near the undeformed equilibrium state.

Table I. Linearly elastic properties of monolayer graphene predicted by first principles and empirical potential based calculations.

| Method | 2D Young's modulus $Y_{2D}$ (N/m) | Poisson's ratio | Biaxial modulus (N/m) | Bending modulus $D_m$ (eV) | Gaussian modulus $D_G$ (eV) |
|---|---|---|---|---|---|
| DFT [46] | 345 | 0.149 | 406 | 1.49 | - |
| DFT [48] | 348 | 0.169 | 419 | - | - |
| DF-TB [44] | - | - | - | 1.61 | -0.7 |
| DFT [58] | - | - | - | 1.44 | -1.52 |
| REBO-1 [51] | 236 | 0.412 | 401 | 0.83 | - |
| REBO-2 [52] | 243 | 0.397 | 403 | 1.41 | - |
| AIREBO [53] | 279 | 0.357 | 434 | 1.56 | - |
| REBO-LB [56] | 349 | 0.132 | 402 | - | - |
| LCBOPII [59] | 343 | 0.156 | 406 | ~1.1 | - |

Besides the in-plane elastic properties, graphene is highly flexible due to its monatomic thinness. The flexural deformation of graphene is commonly observed in the form of rippling, wrinkling, and folding. A general continuum mechanics formulation was proposed to describe the coupled in-plane and flexural deformation of monolayer graphene [28]. Under relatively small in-plane deformation and moderately large out-of-plane deflection, the general formulation reduces to a form similar to the nonlinear von-Karman plate theory with two elastic bending moduli [41], one for the mean curvature and the other for the Gaussian curvature. Unlike classical plate theory, the bending moduli of monolayer graphene are not directly related to the in-plane Young's modulus and Poisson's ratio. Instead, they are independent properties resulting from multibody interactions among carbon atoms in a monolayer [29, 44, 46, 54, 55, 58]. It was noted that the physical origin of the bending moduli of monolayer graphene is fundamentally different from that in classical plate theory [29]. As listed in Table I, the bending modulus of graphene associated with mean curvature ($D_m$) can be well predicted by the second-generation REBO potential [52], which includes atomic interactions up to the fourth nearest neighbors via bond angle and dihedral angle effects. On the other hand, the bending modulus associated with Gaussian curvature ($D_G$) has received less attention, for which the two reported values differ by more than a factor of two [44, 58]. The challenge to accurately predict the Gaussian curvature modulus may be due to the geometrical coupling between Gaussian curvature and in-plane stretch, which in some case had to be accommodated by different bond structures or defects.

Beyond graphene, the elastic properties of other 2D materials have also been theoretically predicted by both first principles [5, 46, 60-65] and empirical potential based calculations [66-70]. In general, all 2D materials with monatomic or ultrathin crystal membrane structures share similar elastic properties as graphene, characterized by high in-plane stiffness and low flexural rigidity. Table II lists typical values of the linearly elastic properties of several 2D materials. Note that phosphorene is highly anisotropic due to its uniquely puckered atomic structure, and as a result, the elastic properties vary over a wide range depending on the loading direction. Similar to graphene, essentially all 2D materials become nonlinearly elastic (Fig. 2) when subjected to relatively large deformation (typically with an in-plane strain over 5%). Their nonlinear elastic properties can be well predicted by first-principles calculations and then incorporated in the same continuum mechanics formulation as for graphene to predict elastic behaviors at much larger scales [5]. For MD simulations, parameterization of empirical potentials for various 2D materials remains critical to accurately predict the elastic properties [66-70].

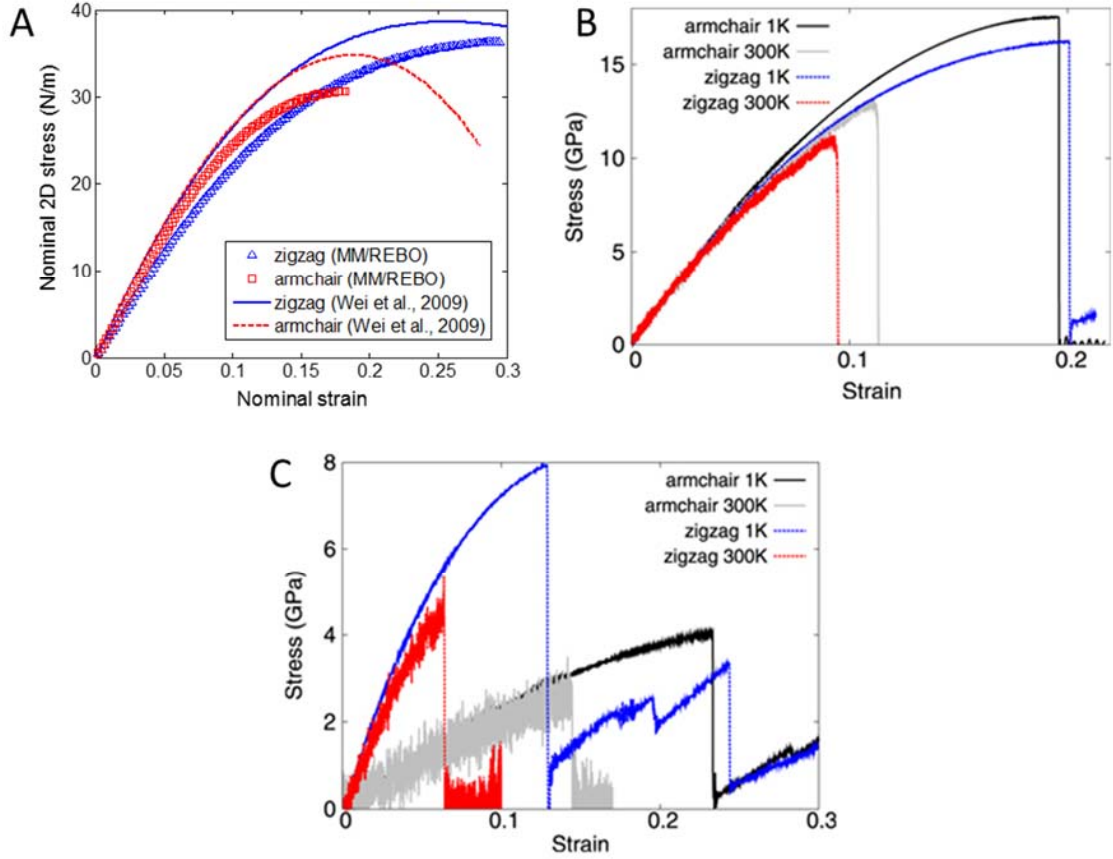

**Fig. 2**: Calculated uniaxial stress-strain diagrams for (A) graphene at 0 K; (B) $MoS_2$ at 1 and 300 K, and (C) phosphorene at 1 and 300 K. Figures adapted from (A) [57] and (B-C) [66].

Table II. Linearly elastic properties of 2D materials predicted by first principles or empirical potential based calculations.

| 2D Materials | $Y_{2D}$ (N/m) | Poisson's ratio | $D_m$ (eV) |
|---|---|---|---|
| graphene [46] | 345 | 0.149 | 1.49 |
| h-BN [46] | 271 | 0.211 | 1.34 |
| $MoS_2$ [5, 71] | 118 ~ 141 | ~0.3 | ~11.7 |
| phosphorene [65, 72] | 23.0 ~ 92.3[a] | 0.064 ~ 0.703[a] | - |
| silicene [62, 63] | ~60 | ~0.4 | - |

[a] Highly anisotropic

Since 2014, researchers have found that some 2D materials may exhibit a negative Poisson's ratio (NPR), both intrinsically and also from extrinsic effects. Materials that exhibit a NPR are known as auxetic materials [73]. Multiple 2D materials have been reported to exhibit an intrinsic

NPR. The first such report was for single-layer black phosphorus or phosphorene [74], which exhibits an intrinsic NPR in the out-of-plane direction due to its anisotropic, puckered crystal structure. Interestingly, this crystal structure is the nanoscale analog of the re-entrant hinged structure proposed by Lakes [75] for NPR in bulk materials. Other puckered 2D materials, such as orthorhombic arsenic, would also be expected to exhibit NPR, which was recently confirmed by Han et al [76]. In addition to puckered 2D materials, graphene also exhibits intrinsic NPR, though for tensile strains exceeding about 6% [77]. As graphene has a planar crystal structure, the mechanism enabling the intrinsic NPR is dependent on a competition between bond angle rotation and bond stretching, with the bond stretching becoming dominant for tensile strains exceeding about 6%, thus enabling the intrinsic NPR. 2D materials can also be tailored extrinsically to exhibit NPR. This has been accomplished through cutting and patterning of graphene [78], buckling of 2D boron (borophene) sheets [79], rippling through introduction of vacancies in graphene [80], and via edge stress-induced warping in graphene nanoribbons [81]. A common feature underlying these extrinsic mechanisms is that they lead to a reference configuration that involves out-of-plane deformation. Upon stretching the 2D materials with the initial out-of-plane deformation, an in-plane expansion occurs in flattening the materials under tension, leading to the NPR phenomenon.

## 2.3 Thermal rippling and thermoelastic properties

While the basic elastic properties of graphene and other 2D materials have been reasonably understood based on first principles, the effects of finite temperature on the elastic properties have not been fully established. Recent experiments have raised fundamental questions on both the in-plane and bending elasticity of the ultrathin membrane materials [7, 9, 33]. It is well known that a graphene monolayer is not perfectly flat at a finite temperature (T > 0 K). Experimental observations have found that suspended graphene membranes often display spontaneous ripples [82-84]. Theoretically, thermal rippling is inevitable [85], which may have profound effects on thermomechanical properties of graphene [10, 11], including thermal expansion and temperature-dependent elastic modulus. MD simulations by Zhao and Aluru [86] predicted that Young's modulus of graphene does not vary significantly with temperature up to about 1200 K, beyond which graphene becomes more compliant. On the other hand, by atomistic Monte Carlo (MC) simulations, Zakharchenko et al. [87] predicted a non-monotonic behavior of the shear modulus of graphene with a maximum at about 900 K, while Chen and Chrzan [88] predicted a monotonic decrease of the elastic modulus of graphene with temperature up to 4000 K. More recently, Los et al. [34] predicted a power law scaling of the in-plane elastic modulus, which decreases with increasing membrane size at a finite temperature. Another manifestation of thermal rippling is the reduction of the projected area, which has been suggested as the cause of the negative in-plane thermal expansion of graphene [87, 88]. Based on DFT calculations and a quasiharmonic approximation, Mounet and Marzari [89] predicted negative in-plane thermal expansion for graphite and graphene, which was attributed to the lowest transversal acoustic (ZA) phonon modes (also called bending modes). Negative thermal expansion of graphene was also predicted by a nonequilibrium Green's function approach [90] and ab initio molecular dynamics (AIMD) simulations [91].

A unified approach of nonlinear thermoelasticity was proposed to study both the temperature dependence of elastic properties and the thermal expansion of graphene [11]. Following classical theories of statistical mechanics and thermodynamics, the Helmholtz free energy of a graphene membrane can be obtained as a function of temperature, macroscopically averaged in-plane strain, and size of the membrane, which would include entropic contributions due to in-plane and out-of-plane thermal fluctuations. As a result, the stress-strain relation derived from the Helmholtz free energy function would be temperature and size dependent in general. With zero stress, the in-plane strain as a function of temperature gives thermal expansion or contraction, depending on the sign of thermal strain. With zero or any prescribed strain, thermal stress can be obtained as a function of temperature as well. It was found by MD simulations that the in-plane thermal fluctuations alone lead to a positive thermal expansion coefficient ($\sim 5.5 \times 10^{-6}$ $K^{-1}$), independent of the membrane size or temperature (up to 1000 K) [11]. With out-of-plane thermal rippling, however, a transition from negative to positive thermal expansion was predicted at a critical temperature ($\sim 400 - 600$ K), which may depend on the membrane size due to the size dependence of thermal rippling. Moreover, the rippling amplitude depends sensitively on the mechanical constraints in terms of either strain or stress, which may be imposed by boundary conditions or interfaces with a substrate. Therefore, the coefficient of thermal expansion (with the effects of thermal rippling) may not be considered an intrinsic material property of graphene, but the positive CTE due to in-plane fluctuations alone is intrinsic. Furthermore, the effective in-plane elastic properties of graphene at a finite temperature are closely related to thermal rippling. In particular, the presence of thermal rippling effectively lowers the in-plane stiffness of graphene. Since the amplitude of thermal rippling decreases nonlinearly with increasing tensile strain (Fig. 3A), graphene becomes nonlinearly elastic even at infinitesimal strain, with a tangent modulus increasing with increasing strain until thermal rippling is significantly suppressed by tension (Fig. 3B). Similarly, since the amplitude of thermal rippling increases with increasing membrane size, the in-plane elastic modulus is predicted to decrease with increasing membrane size at a finite temperature [11, 34]. Looking beyond graphene, the effects of thermal rippling are expected to be equally important for other 2D materials due to the low flexural rigidity, although experimental evidence of such an effect has been limited.

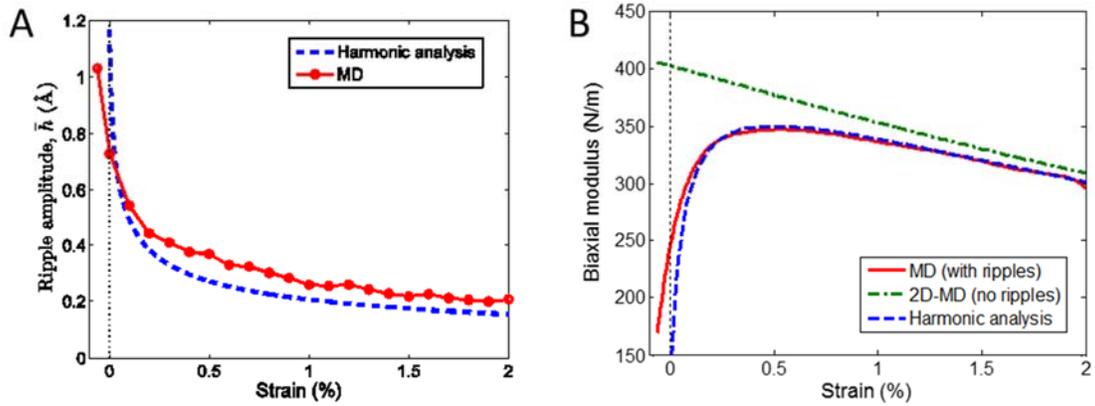

**Fig. 3**: (A) Decreasing ripple amplitude under increasing tensile strain; (B) Tangent modulus of graphene at a finite temperature (T = 300 K). Figures adapted from [11].

## 3. Inelastic Properties: Defects, Strength and Toughness

Mechanical properties of 2D materials beyond elasticity depend sensitively on the presence of defects (e.g., vacancies, dislocations, grain boundaries, and crack-like flaws) and their evolution during deformation. Strength and toughness are two distinct mechanical properties describing the onset of failure in terms of stress and energy, respectively. While the toughness is defined as the energy per unit volume of the material with a unit of $J/m^3$ in elementary mechanics, it is defined more rigorously based on fracture mechanics as the energy per unit area of crack growth in the material with a unit of $J/m^2$. For 2D materials, the fracture toughness may be defined as the energy per unit length of crack growth with a unit of J/m or eV/nm, in the same spirit of the 2D Young's modulus (unit: N/m). This section reviews recent studies on the types of defects in 2D materials, their strength and toughness, and potential toughening mechanisms.

### 3.1 Types of defects

Defects in 2D materials can have profound influence on their physicochemical, electronic, and mechanical properties [12, 92-94]. In thermodynamic equilibrium, the second law of thermodynamics predicts the necessary presence of defects, but usually would suggest extremely low concentration at any realistic temperatures (below the melting point or sublimation). In non-equilibrium states, defects are introduced unintentionally or intentionally into 2D materials, often as undesirable departure from perfection or possibly to engineer the material properties, especially mechanical [12]. In general, there are two types of intrinsic defects in 2D materials: point defects (vacancies, self-interstitials, dislocations and topological defects) and line defects (grain boundaries). All these defects can be present in different 2D materials. However, due to the difference in lattice structures and bonding energies, the configurations of these defects may take different forms. The family of 2D materials is expanding rapidly [2, 3]. Here, we primarily focus on three members of the 2D materials family: graphene, $MoS_2$ and phosphorene, with emphasis on three typical types of defects: mono-vacancies, dislocations and grain boundaries.

### Vacancies

Typical configurations for mono-vacancies (MVs) in graphene, $MoS_2$ and phosphorene are shown in Fig. 4. It is seen from Fig. 4A that there is only one chemical type of MV in graphene. For $MoS_2$, there are two types of MVs: Mo vacancy ($V_M$) and S vacancy ($V_S$), as shown in Fig. 4B. For phosphorene, as shown in Fig. 4C, there are two types of MV with the same degenerate energy level [95]. An important quantity in discussing vacancies is their formation energy ($E_f$), which essentially dictates their thermodynamic equilibrium concentration. The value of $E_f$ for MV in phosphorene is 1.65 eV, significantly smaller than that of graphene of 7.57 eV [95]. Since the defect population depends exponentially on $E_f$, the equilibrium concentration of MVs in phosphorene should be exceedingly larger than that of graphene under the same conditions. The much lower value of $E_f$ in phosphorene is associated with the inherently softer P-P bonds compared with the much stronger C-C bond, and also the curvature effect related to the structural buckling. For $MoS_2$, the value of $E_f$ for $V_S$ in monolayer $MoS_2$ (from 1.22 to 2.25 eV) is smaller than or comparable to that in phosphorene [96, 97]. This may explain the high $V_S$-concentration often observed experimentally. Here, the variability of the chemical potential (elemental rich and poor conditions) defines the upper and lower bounds of $E_f$.

The mobility of MV is determined by their diffusion energy barrier, $E_b$. The sequence of $E_b$ for MV diffusion in the above 2D materials follows: phosphorene (0.40 eV) < graphene (1.39 eV) < $MoS_2$ ($V_S$, 2.27 eV) [95]. Therefore, phosphorene has the lowest value of $E_b$, which is only one third of that of graphene. According to the Arrhenius formula, the hopping rate ($v$) can be calculated by $v = v_s \exp(-E_b/kT)$, where $v_s$ is the characteristic frequency, normally around $10^{13}$ Hz, k is the Boltzmann constant and $T$ is the temperature. At room temperature, the rate-limiting process of the migration of MV in phosphorene is estimated to be amazingly rapid, around 16 orders of magnitude faster than that in graphene, signifying a significantly greater vacancy activity [95].

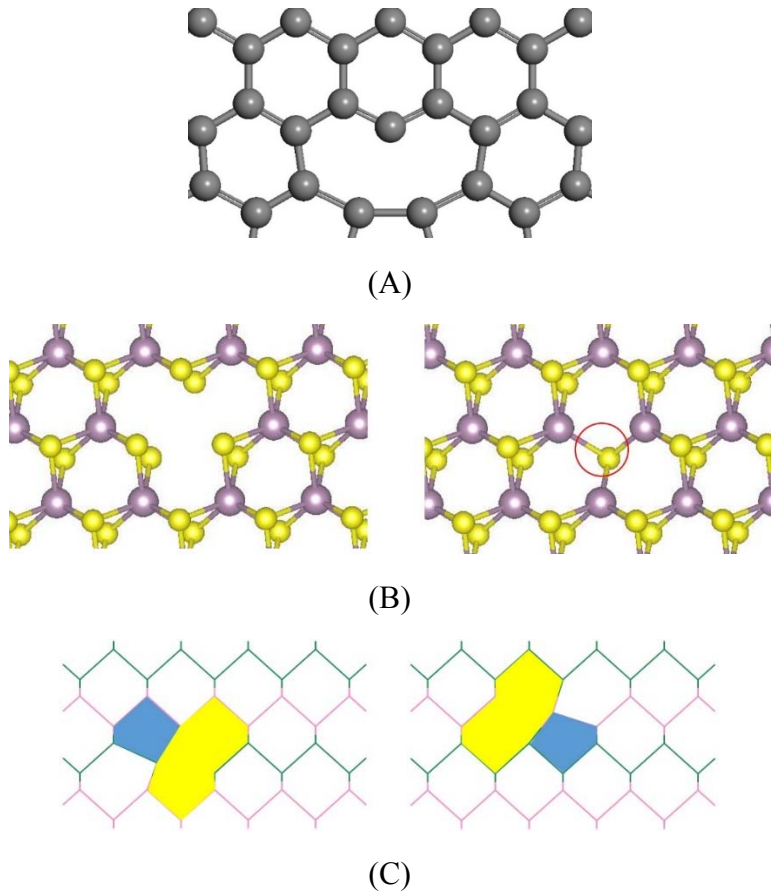

(A)

(B)

(C)

**Fig. 4**: Vacancy configurations in 2D materials: (A) graphene, (B) $MoS_2$ (Mo purple and S yellow), and (C) phosphorene.

**Dislocations**

Another type of defect in 2D materials is dislocation. Configurations for dislocations in graphene, $MoS_2$, and phosphorene are shown in Fig. 5. In general, there are two types of dislocations: edge and screw. In the latter case, the Burgers vector points in the out-of-plane dimension, turning the 2D layer into a 3D form [98, 99]. The edge dislocation in 2D materials can be represented by a pentagon-heptagon pair (5|7) with two basic disclinations, 5−pentagon and 7−heptagon [96, 97]. Since strain energy of a dislocation is proportional to the square of its Burgers vector $|\mathbf{b}|^2$, the 5|7 with smallest $\mathbf{b} = (1, 0)$ is in general the most energetically favorable configuration in 2D materials. For phosphorene, due to its strong structural anisotropy and uniquely buckled atomic structure, its 5|7 dislocation is heavily distorted compared with graphene, as shown in Fig. 5B.

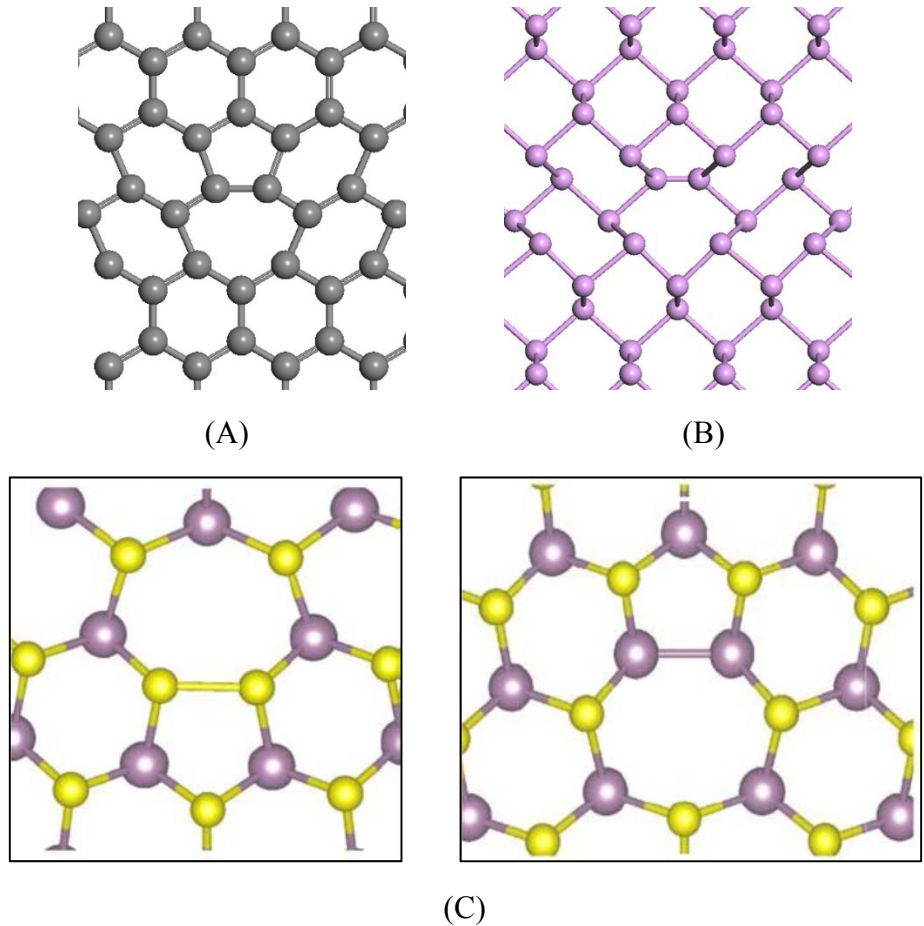

**Fig. 5**: Dislocation configurations in 2D materials: (A) graphene, (B) phosphorene, and (C) MoS$_2$.

Tri-atomic layer MoS$_2$ brings additional complexity to the dislocation structures [96, 97]. For monolayer MoS$_2$, the smallest dislocations can be classified into Mo- or S-rich ones with their frontal view resembling 5|7 or 7|5. Owing to the tri-atomic-layer structure of MoS$_2$, the dislocation cores are essentially 3D concave polyhedra, as shown in Fig. 5C. Both the strong stress field around dislocation cores and the flexible coordination number of S atoms render the interaction between dislocations and point defects (interstitials, vacancies, or substitutions) highly effective, endowing rich chemical variability to the core structures. For example, it was shown that a 5|7 dislocation is able to react with 2 S vacancies to form a 4|6 defect, which is energetically more favourable at a certain chemical potential of S.

In general, dislocations in 2D materials are immobile at room temperature due to high gliding energy barriers. As a result, plastic deformation of 2D materials is generally insignificant at room temperature [100].

**Grain boundaries**

Typical grain boundary (GB) configurations in graphene, $MoS_2$, and phosphorene are shown in Fig. 6. It is seen that grain boundaries can be considered as an array of dislocations arranged in a linear manner. In general, grain boundaries can be classified as low-angle and high-angle grain boundaries. For the former, dislocation cores are kept at a certain distance; while for the latter, dislocation cores may become crowded and even overlap. It was found that graphene sheets with large-angle tilt boundaries that have a high density of defects are as strong as the pristine material and are stronger than those with low-angle boundaries having fewer defects [101]. This trend can be easily understood by considering the critical bond at the strained seven-membered carbon rings that lead to failure; the large-angle boundaries appear stronger because they are able to better accommodate these strained rings, through simple cancellation of tensile and compressive strain in the grain-boundary dislocation sequence [102, 103]. It was shown that GB strength can either increase or decrease with the tilt, indicating that it is not just the density of defects that affects the mechanical properties, but the detailed arrangements of defects are also important [104, 105]. For polycrystalline graphene containing a network of many GBs, it was shown that grain boundary junctions are the weakest link that fails first. As a result, the strength of polycrystalline graphene can follow a pseudo Hall-Petch [104] or an inverse pseudo Hall-Petch [105] relation. That is, the smaller the average grain size, the higher or lower the strength can be, depending on how specifically the grain boundaries and the domains they are separating, are organized.

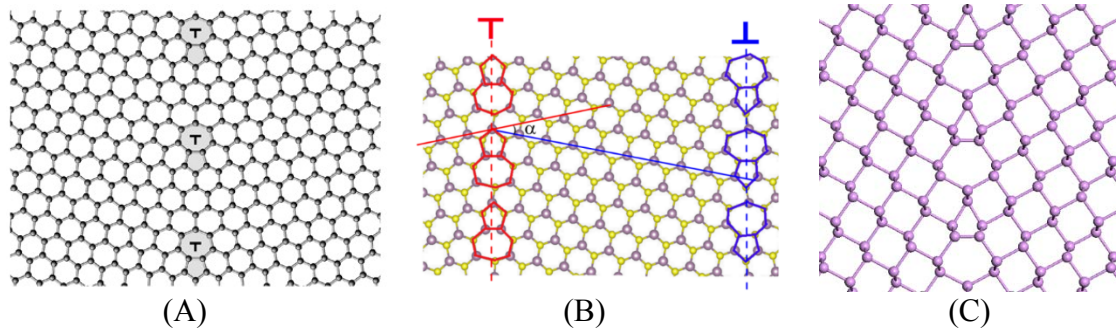

**Fig. 6**: Grain boundaries in 2D materials: (A) graphene, (B) MoS$_2$, the left red is Mo-rich, the right blue is S-rich [96]; and (C) phosphorene, where one still can recognize the strongly-puckered 5|7's [106].

Typical GBs with the (5|7) dislocation atomic structures have been observed in MoS$_2$. The plane view of a GB model is shown in Fig. 6B, in which two grains with a misorientation angle of ~20.6° form two GBs with opposite directions because of the trigonal symmetry: one is Mo-rich labeled as "⊤" and the other is S-rich labeled as "⊥". Using this GB model, the calculated formation energy is about 2.50 eV/unit, which is nearly the same as its crystalline counterpart (~2.53 eV/unit). The very small difference in these two energies may explain the high density of GBs in MoS$_2$ observed experimentally. The calculated GB energy is ~ 0.05 eV/Å. This low GB energy again signifies the easy formation of GBs in MoS$_2$ [100].

Phosphorene, a monolayer of black phosphorus, possesses a puckered honeycomb lattice. Currently, grain boundary structures are still unavailable in experimental imaging, and may be absent from the available samples. A typical grain boundary structure obtained from first-principles calculations is shown in Fig. 6C. Based on first-principles calculations, the atomic structure and thermodynamic stability of phosphorene GBs were studied [106, 107]. It was found that the GBs in phosphorene are energetically stable with formation energies much lower than those in graphene [107].

## 3.2 Strength and toughness

Fracture is one of the most prominent concerns for large-scale applications of 2D materials. A substantial amount of effort has been dedicated to understanding the fundamental fracture mechanisms and exploring ways to enhance the fracture toughness of 2D materials. In this section, we present a brief overview of some recent studies on strength and toughness of 2D materials. A more complete review of this subject on graphene can be found in [16].

For 2D materials, the theoretical strength is commonly defined as the maximum stress that the material can sustain in the absence of any defects. As a representative 2D material, graphene has been found to be the strongest material in existence, with a strength as high as 130 GPa (assuming a thickness of ~0.335 nm) [4, 47], making it an ideal candidate for many potential applications, such as wear-resistance coatings [108], ballistic barriers for body armor [109, 110] and reinforcements for novel composites [111, 112]. The strength of pristine graphene and other 2D materials with prefect lattices can be reasonably measured or predicted by AFM nano-indentation experiments [4, 6, 32], DFT calculations [5, 47, 48, 60-65, 113-117] and MD simulations [57, 66-68, 86, 118, 119], as summarized in Table III. It was shown from DFT calculations that most pristine 2D materials fail via an elastic instability of atomic bonds under in-plane stretch (uniaxial or biaxial) or a phonon instability associated with the out-of-plane relaxations of atoms in MoS$_2$ [61] and borophene [79]. Since fracture of 2D materials involves highly nonlinear deformation prior to the rupture of atomic bonds, with strain reaching more than 20% (see Table III), nonlinear elasticity should be taken into account when interpreting the experimental measurements of strength in these materials, such as those based on the force-displacement curves from nano-indentation tests [4, 6, 32].

Table III. Theoretical strength and critical strain of pristine 2D materials

| 2D Materials | $\sigma_a^c$ (N/m) | $\sigma_z^c$ (N/m) | $\epsilon_a^c$ | $\epsilon_z^c$ |
|---|---|---|---|---|
| graphene [47] | 32.93 | 36.23 | 0.19 | 0.27 |
| h-BN [113] | 29.04 | 33.66 | 0.18 | 0.29 |
| MoS$_2$ [61] | 15.37 | 15.13 | 0.28 | 0.36 |
| phosphorene [64] | 20.26 | 12.98 | 0.08 | 0.15 |
| silicene [63] | 5.90 | 6.00 | 0.15 | 0.21 |
| borophene [79] | 9.99 | 4.44 | 0.27 | 0.30 |
| 2D silica [115] | 35.30 | 38.30 | 0.34 | 0.40 |

Note: The subscripts "a" and "z" denote the uniaxial strength or strain in the armchair and zigzag directions, respectively. Borophene (monolayer boron) has a triangular lattice structure where armchair and zigzag directions are not well defined. Here they are used to represent two typical orthogonal directions.

Most large-scale 2D materials contain various topological defects [120-122] that make the prediction of their failure strength extremely challenging [13, 14, 31, 101, 102, 104, 105, 123-133]. For example, the strength of a bi-crystal graphene is highly dependent on the grain boundary [101, 102, 123], which was recently explained with a disclination dipole theory [102]. For polycrystalline 2D materials, the co-existence of grain boundaries, triple junctions, and vacancies can significantly complicate their strength prediction. From an experimental point of view, the commonly used AFM nano-indentation technique cannot always identify the weakest grain boundary or triple junction that governs the failure strength of a sample being tested. Although atomistic simulations have been performed to help understand the dependence of strength on grain size in polycrystalline graphene, there is not even a qualitative agreement among the reported studies [104, 105, 129, 130]. For a polycrystalline graphene with randomly distributed grains, it has been found that the failure strength exhibits significant statistical fluctuations following a Weibull distribution, according to MD simulations and theoretical modeling based on the 'weakest-link' statistics [130]. This means that, for an accurate prediction of strength, it might be necessary to map out all the topological defects in the material, which is not always possible or desirable. The findings for graphene are expected to be applicable also for other 2D materials, as demonstrated by a recent study on fracture in polycrystalline boron-nitride (BN) sheet [133]. Net charge can exist in some defects (such as five-seven rings) in BN sheet [134], whose effect on the material fracture has not yet been fully explored in the literature. Defect mobility may also be activated in 2D materials made of relatively weak atomic bonds, such as $MoS_2$ and monolayer silica [135], especially at high temperature or exposure to irradiation, leading to plastic deformation that can strongly affect the fracture properties.

The strength of 2D materials can also be influenced by out-of-plane deformation induced by either external loading [136] or intrinsic topological defects [137-142]. It has been reported that graphene under shear loading develops significant out-of-plane wrinkles [136], leading to a failure strength around 60 GPa while the shear strength of graphene in the absence of out-of-plane deformation is around 97 GPa [136]. A sinusoidal graphene sample with periodically distributed disclination quadrupoles exhibited a strength near 30 GPa [141], and pronounced out-of-plane deformation due to grain boundaries can dramatically alter the mechanical response of graphene [142]. The competition between the out-of-plane and in-plane deformation in an elastic membrane is usually characterized by its flexibility, which is known to be determined by the bending stiffness, in-plane stretching modulus and sample size. Since the effective bending stiffness and in-plane stretching modulus exhibit substantial differences for various 2D materials [29, 46, 71, 115, 143], the out-of-plane effect on fracture is expected to vary substantially across different 2D materials and deserves further study.

Toughness is a key property that characterizes the strength of a material in the presence of an existing crack-like flaw. In large scale applications of 2D materials, crack-like flaws (cracks, notches, corners and holes) are unavoidable and some of them are even intentionally introduced to achieve specific functions, such as water desalination [144, 145], gas separation [146, 147] and DNA sequencing [148, 149]. Using a custom designed tension test platform, Zhang et al. [15] measured the fracture toughness of graphene as low as 15.9 J/m$^2$ (assuming a thickness of ~0.335 nm), close to that of ideally brittle materials like glass and silicon. The reported toughness values of pristine single-crystal graphene based on atomistic simulations are in general agreement with experimental measurements [150-152]. However, the toughness of polycrystalline graphene exhibits large scattering [130, 153, 154] and seems to depend on the grain size as well as the detailed distribution of topological defects [130, 154]. There are only a few studies on the fracture toughness of 2D materials beyond graphene [155, 156], due to a general lack of accurate atomistic potentials for MD simulations of 2D materials [67, 68]. Usually, atomistic simulation of fracture requires sufficiently large samples that exceed the typical capacity of DFT calculations.

Since 2D materials are nanoscale thin membranes, tearing is a very important fracture mode as well. It has been shown that the tearing process may significantly influence the cleavage of multilayer graphene or graphene from a solid substrate [157]. Moura and Marder [158] derived an analytical formula to link the fracture toughness and tearing force for a general thin membrane and suggested a new way to measure the toughness of graphene through tearing. However, so far there has been no systematic study to compare the fracture toughness under in-plane tensile loading and tearing. Similar to the shear strength of graphene [136], out-of-plane wrinkles can play important roles during crack propagation under shear loading. It was recently demonstrated through MD simulations that out-of-plane wrinkles can lead to additional crack nucleation at crack surfaces in MoS$_2$ under mixed mode loading [156]. Interesting questions for crack propagation in 2D materials include: Is there a well-defined toughness for 2D materials under shear? What is the relationship among fracture toughness under different loading conditions, including in-plane tension, shear and out-of-plane tearing?

The intrinsically nanoscale nature of 2D materials also calls for a careful examination of the fundamental concept of fracture toughness. It has been shown that the calculated fracture toughness of graphene is higher than its surface energy [150] due to the discrete lattice trapping effect [159, 160]. A non-zero surface stress is expected to exist along the free edge of open crack surfaces [161]. Furthermore, atomic structure reconstruction from hexagon to pentagon at the crack tip has been observed in MD [152] and DFTB [162] simulations of quasi-static crack growth in graphene, which can modify the stress field near a crack tip. The additional stress contributed by the nanoscale surface effect and atomic reconstruction may alter the energy release rate that drives crack propagation and thus influence the fracture toughness of the material. 2D materials can have a stable crack as small as a few lattice spacings because of the covalent nature of their atomic bonds. As the crack size is reduced to nanoscale, the applicability of Griffith criterion requires careful re-examination. Yin et al. [163] found that a local bond strength failure criterion seems to perform better than the Griffith criterion for a graphene sheet containing a crack shorter than 10 nm. Brochard et al. [164] found a transition from energy-governed (Griffith criterion) to stress-governed failure in graphene, and proposed a composite failure criterion including both stress and energy.

### 3.3 Toughening mechanisms

The low toughness of graphene has raised concerns about the applications of graphene in large-scale devices where crack-like flaws are inevitable. Similarly low toughness is expected for other 2D materials. It is thus of great interest to explore ways of toughening 2D materials from both fundamental science and engineering points of view. Defect engineering has proven to be a versatile approach to enhancing the toughness of classic materials including metals [165, 166] and ceramics [167, 168], and may also have great potential for 2D materials. It has been recently demonstrated that a graphene containing periodically distributed disclination quadrupole with equally spaced positive and negative disclinations, leading to a sinusoidal rippling profile with wavelength and amplitude of 4 nm and 0.75 nm, respectively, can be twice as tough as the pristine graphene [141]. In comparison, grain boundaries in graphene can lead to an enhancement of toughness by up to 50% [154]. Shekhawat and Ritchie [130] showed that polycrystalline graphene is tougher than pristine graphene across a wide range of grain sizes. Although the topological defects induced toughness enhancement has been reported in graphene, a detailed systematic investigation of the toughening mechanism(s) is still lacking [169]. In addition, a trade-off of topological toughening is that it tends to lower the stiffness and failure strength of the material [141]. Therefore, there is an urgent need to develop a full theoretical understanding of toughening mechanisms to guide the design of tough 2D materials with balanced strength and stiffness. Possible toughening mechanisms include interactions of defects with a moving crack, out-of-plane deformation and atomic scale bridging [169], as shown in Fig. 7A-C. Very recently, Meng et al. [170] investigated crack-dislocation interactions in graphene with theoretical modeling and MD simulations, and proposed a formula for the corresponding toughness enhancement. Shekhawat and Ritchie [130] also attributed the toughness enhancement in their atomic simulations of polycrystalline graphene to crack tip interaction with dislocations at grain boundaries or triple junctions. Mitchell et al. [171] studied crack initiation and propagation in a thin elastic membrane conforming to a curved rigid substrate, with results showing that the initial curvature and stress can dramatically alter crack behavior and even guide the crack path, although the effective toughness of a curved thin elastic membrane was not thoroughly discussed. In short, the pronounced effects of dislocations and curvature on the crack propagation in a thin membrane can be reasonably predicted by continuum mechanics models, which may pave a way for developing a theoretical framework to quantify toughness contributions from different mechanisms including defects, curvature, nano-cracks and atomic bridging. The motion, interaction and nucleation of topological defects may also play important roles in the fracture of 2D materials.

Another promising method to toughen 2D materials is by introducing patterned cuts, the so-called kirigami design [172-174]. Graphene kirigami recently created by Blees et al. [9] was found capable of undergoing very large stretching strains without failure (Fig. 7D). This finding has stimulated interest in designing graphene kirigami as stretchable electrodes, springs and hinges. MD simulations have been performed to study the deformation and fracture of graphene [175, 176] and $MoS_2$ kirigami structures [177]. In classical fracture mechanics, it has been well documented that micro-cracks can have significant shielding effect on a main crack and thus increase the fracture toughness [178]. The micro-crack induced toughening has been observed in various materials ranging from ceramic [179-181] to bio-composites [182], and may apply also to 2D materials. During the loading process, nanoslits introduced in the kirigami design act as microcracks and they also cause out-of-plane buckling in the 2D material, especially in the vicinity of a crack tip [9, 175-177] (see Fig. 7E), making the local crack tip stress field more complicated even under simple uniaxial tension in the far field. It remains an interesting question as to how much toughness enhancement can be achieved through kirigami design.

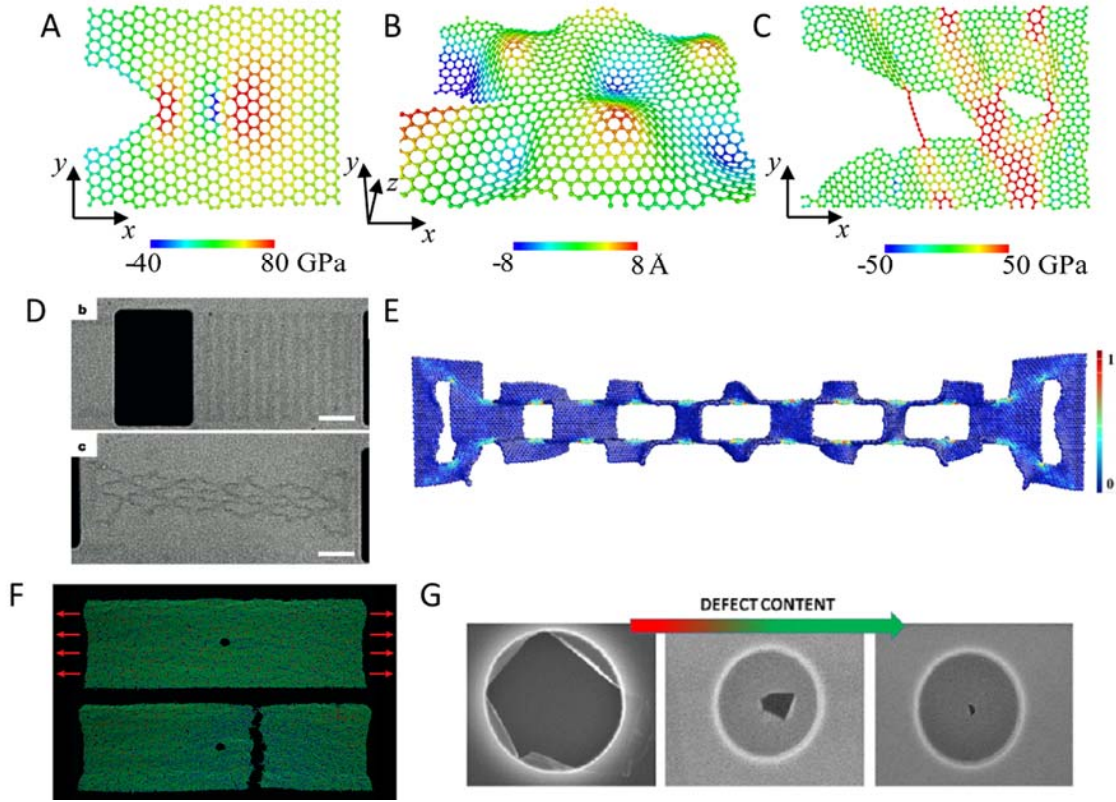

**Fig. 7**: Toughening mechanisms in 2D materials. (A-C) Observed mechanisms of toughness enhancement due to topological defects in graphene [169]. (D) Highly stretchable graphene kirigami [9]. (E) MD simulation snapshots for stress distribution in a stretched graphene kirigami [175]. (F) Flaw insensitive fracture in a nanocrystalline graphene sample [153]. (G) Transition from catastrophic to localized failure with increasing defect content [183]. Figures adapted from: (A-C) [169], (D) [9], (E) [175], (F) [153], and (G) [183].

A concept closely related to toughness enhancement is flaw insensitivity of a material. Zhang et al. [153] investigated fracture of a nanocrystalline graphene with average grain size of 2 nm containing a circular or elliptical hole as a pre-existing geometrical flaw through MD simulations and theoretical modeling [184, 185]; it was found that the strength of the nanocrystalline graphene becomes insensitive to the pre-existing flaw if the strip width falls under 17.6 nm (Fig. 7F). In a recent experimental work on crack propagation in a graphene membrane with different concentrations of defects, Lopez-Polin et al. [183] showed that the defects can change the catastrophic failure of a pristine graphene (i.e., the final crack spanning the whole membrane) to a localized failure mode under a nano-indenter, as shown in Fig. 7G. These studies provided a new perspective for designing robust graphene-based devices where a certain concentration of defects may be considered desirable, such as the nanoporous graphene for water desalination and gas separation.

## 4. Electromechanical Coupling

### 4.1 Electromechanics of graphene and strain engineering

Many of the properties of graphene are strongly tied to its lattice structure. The large elastic deformability of graphene (e.g., 20%) allows for substantial change of the graphene lattice structure, therefore opening up fertile opportunities to tailor the electronic properties (e.g., charge carrier dynamics) of graphene through mechanical strain. For example, despite its many exceptional physical properties, graphene suffers from one key drawback in potential usage in electronics – it is gapless in the band structure. Strain engineering has been shown to be a viable solution to open a bandgap in graphene. The end goal of using this approach is the development of an all-graphene electronic circuit, where the flow of electrons through the circuit can be controlled by strain [186]. Despite early erroneous predictions of strain-induced bandgaps at small levels of tensile strain [187, 188], recent consensus has emerged that graphene does not exhibit a bandgap until more than 20% tensile strain is applied [189], which unfortunately is close to its mechanical failure limit. For this reason, other methods have been proposed by using different types of inhomogeneous strain, including shear [190] and periodic ripples [191], to open a bandgap in graphene. The effects of the intrinsic wrinkles in graphene on its electronic properties have been investigated [192, 193], finding that the wrinkles generally reduce graphene's electrical conductivity. While many works have been performed on tuning graphene's band structure with strain, we do not intend to cover all aspects of the strain effects on the electronic properties of graphene; comprehensive reviews of graphene's electronic properties can be found elsewhere [17, 194-196]. Instead, we focus here on a more recent form of electromechanical coupling – that of so-called pseudomagnetic fields (PMFs) that arise in strained graphene.

Magnetic fields are intrinsic to life on earth, and are important from a condensed matter point of view because of the way they impact the motion of electrons. For example, Earth's magnetic field, which originates in its core and extends out into space, has a magnitude on the order of $10^{-5}$ Teslas (T) on the earth's surface. For a point of reference, the strongest man-made magnetic field recorded on earth is about 100 T [197]. It has recently been experimentally reported that graphene is able to exhibit strong PMFs with intensities up to hundreds of Teslas. These PMFs exhibit both similarities and differences in comparison with real magnetic fields. Electrons in graphene that exhibits a PMF behave identically to those subject to a real magnetic field. The major difference is that while real magnetic fields are space-filling, PMFs only influence electrons in the plane of graphene, and do not project a field beyond the planar graphene structure. The initial measurements of PMFs in graphene were made on very small graphene bubbles with diameters under 10 nm that were formed on a metallic, i.e., Platinum (111) [18] or Ruthenium (0001) [198] substrate (Fig. 8A). The PMFs were determined experimentally by analyzing the local density of states (LDOS), in which the emergence of Landau levels similar to that seen in the presence of a real magnetic field can be observed. Other experiments using a scanning tunneling microscope tip to deform graphene drumheads via localized strains did not measure PMFs explicitly, but instead inferred that the localized deformation led to electronic signatures similar to those expected from quantum dots, and also PMFs [19]. It is suggested that bending graphene over the surface features on an underlying substrate (e.g., steps or sharp features) also generate PMFs in the locally distorted portion of the graphene [199-201].

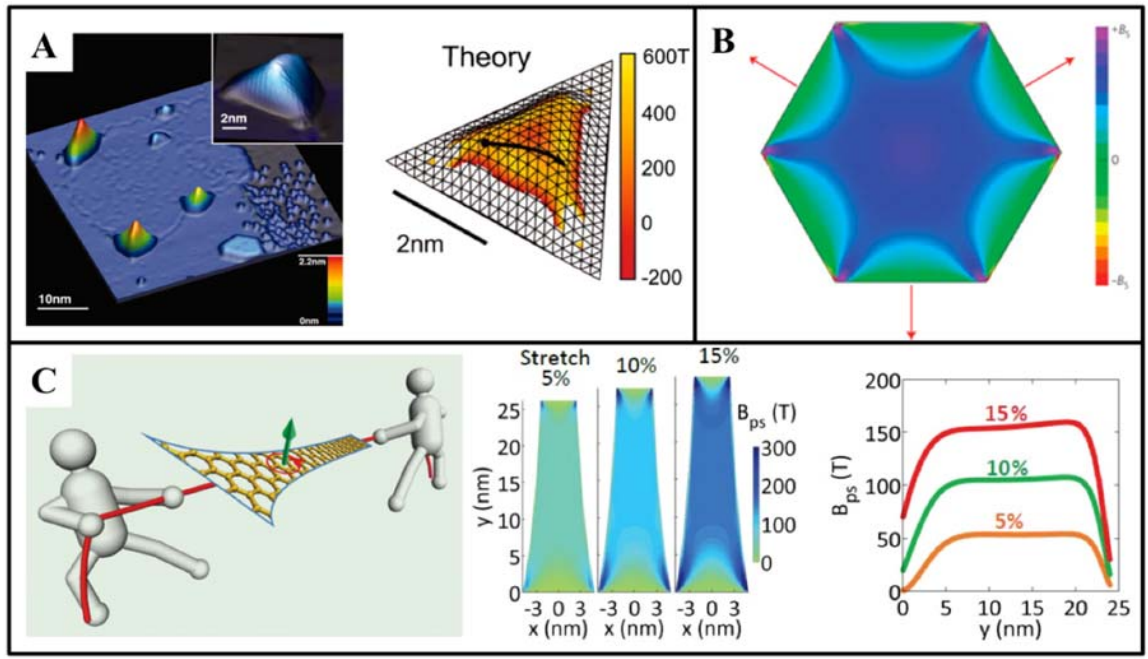

**Fig. 8**: (A) (left) STM images of a graphene monolayer patch on Pt(111) with four nanobubbles at the graphene-Pt border and one in the patch interior. (right) Topography of theoretically simulated graphene nanobubble with calculated pseudomagnetic field distribution. (B) Distribution of the pseudomagnetic field in a graphene under equi-triaxial tension. (C) Subject to uniaxial tension, a suitably patterned graphene

ribbon can have a rather uniform and strong pseudomagnetic field over a large area. Figures adapted from: (A) [18], (B) [202], and (C) [203].

From a theoretical point of view, the low energy electronic states of graphene are accurately modeled by two decoupled 2D Dirac equations. If disorder is applied to graphene, for example in the form of a mechanical strain, the resulting perturbations to the Dirac equations enter in the form of an effective gauge field [204, 205], which bears striking similarity to how a vector potential would be added to describe the effect induced on charge carriers by a real magnetic field [202]. This gauge field $\mathbf{A}_{PMF}$ can be written as [206]

$$\mathbf{A}_{PMF} = \begin{pmatrix} A_x \\ A_y \end{pmatrix} = \frac{\beta}{a} \begin{pmatrix} \varepsilon_{xx} - \varepsilon_{yy} \\ -2\varepsilon_{xy} \end{pmatrix} \quad (1)$$

where $a = ev_F/t$, $t = 2.8$ eV is the hopping energy, $v_F = 10^6$ m/s is the Fermi velocity, $e = 1.6 \times 10^{-19}$ C is the electron charge, $\beta \approx 3$ is a dimensionless parameter that represents the coupling between Dirac electrons and lattice deformation, and $\varepsilon_{ij}$ represent the components of the strain in the plane of graphene. The PMF can then be calculated by $\mathbf{B}_{PMF} = \nabla \times \mathbf{A}_{PMF}$, whose magnitude can be written as [207]

$$B_{PMF} = \frac{\beta}{a} \left( \frac{\partial(-2\varepsilon_{xy})}{\partial x} - \frac{\partial(\varepsilon_{xx} - \varepsilon_{yy})}{\partial y} \right) \quad (2)$$

We note that a review on gauge fields in graphene was recently provided by Vozmediano et al. [208], while a very detailed discussion on the derivation of the pseudomagnetic fields in graphene as a function of strain starting from the classical tight-binding Hamiltonian was given by Ramezani Masir et al [209]. Other recent works have also suggested various modifications and corrections to the gauge fields of graphene, including geometric couplings [210], corrections for inhomogeneous strain [211], and corrections for atomic scale structural variations that are not captured by continuum elasticity [212].

The potential to control the motion of electrons in graphene via PMFs opens up opportunities to design new electronic device concepts, for which it is highly desirable to have uniform PMFs over a large area of planar graphene. However, existing experiments demonstrated PMFs in highly localized regions of graphene with a non-planar morphology [18, 19], which poses tremendous challenge for experimental control and characterization of the resulting fields. Further challenge originates from the dependence of the symmetry of the strain-induced PMF on the strain gradient in graphene. As a result, an axisymmetric strain field in graphene leads to a PMF of rotational threefold symmetry [19, 199, 202, 207, 213]. Theoretical insight to generate uniform PMFs in a planar graphene comes from Eq. (2), where the PMF magnitude is dependent on the gradient of the strain, rather than the strain itself. This insight has led to various attempts of creating uniform PMFs in graphene. One of the first and best-known works proposed to apply equal-triaxial strain to atomically thin graphene (Fig. 8B) [202], a technical challenge even prohibitive in bulk materials. Other studies have been performed to generate uniform PMFs by bending graphene into a circular arc [214, 215]. An interesting study showing the effects of geometry and mechanics on the electromechanical coupling was performed by Pereira et al. [216], who found that geometrical singularities such as cones can significantly modify the electronic properties in graphene. Nonetheless, bending graphene into a circular arc or forming conical shapes leads to non-planar contributions to the resulting PMFs.

Recently, an interesting idea possibly resolving the issue of obtaining tunable, uniform PMFs in planar graphene was put forth by Zhu et al [203]. In particular, they proposed using graphene ribbons with suitably designed non-uniform width to generate uniform strain gradients (Fig. 8C), and thus uniform PMFs in the graphene ribbons. Specifically, following Eq. (2), the magnitude of the PMF is directly proportional to the magnitude of the strain gradient as

$$B_{PMF} = \frac{3\beta}{a}(1+v)\frac{\partial \varepsilon_{yy}}{\partial y} \qquad (3)$$

where the gradient of the axial strain, $\frac{\partial \varepsilon_{yy}}{\partial y}$, is related to the variation of the ribbon width. Equation (3) further suggests that programmable PMFs of other natures (e.g., linear distribution of PMF) in graphene can be achievable by tailoring the ribbon width and thus the strain gradient in graphene. The essence of this idea is to program the strain gradient, and thus the PMF in graphene by tailoring the geometric shape of the planar graphene, a feasible approach given the recent advances in the ability to pattern and functionalize graphene [217-220].

Progress has also been made on developing multiphysics-based computational methodologies to analyze electromechanical coupling, and specifically PMFs in graphene. In such approaches, MD simulations are used to calculate the positions of atoms in the graphene lattice in response to external forces, at which point rigorous tight-binding calculations [221] are employed to calculate the electronic properties, including the local density of states (LDOS) and PMFs in graphene. This approach has been used to study PMFs in graphene bubbles of similar size and geometry as those studied experimentally by Levy et al. [18] and Lu et al. [198], where the potentially dominant role of substrate effects on the PMFs, as well as the potentially important role of bending curvature [222] on generating large PMFs in small graphene bubbles was revealed [199]. It has also been used to study and predict the existence of PMFs in graphene kirigami [175], which were recently demonstrated experimentally [9] and computationally [175] to exhibit stretchability that far exceeds that of pristine graphene. Such an approach enables an accurate representation of atomic scale physics, but results in high computational expense, with the largest systems studied on the order of a few tens of thousands of carbon atoms. To address this issue, Zhu et al. [207] have developed a bottom-up scalable coarse-grained modeling scheme that allows for high fidelity simulations of graphene structures at experimentally-accessible length scales, on the order of microns. This coarse-grained modeling scheme has been used to analyze the PMFs in the graphene drumhead experiments of Klimov et al. [19] and to investigate the dependence of the stretchabilty of graphene kirigami on its geometric parameters [223].

Another area for electromechanical coupling involves studying the transport properties of graphene due to strain. Such effects have been investigated experimentally [224] and theoretically [225]. The complex nature of this problem requires a computational approach, combining mechanical deformation, electronic structure and transport models. Specifically, by coupling MD simulations, tight binding calculations of electronic structures, and Landauer-Buttiker transport methods [226], electron transport in deformed graphene has been studied, including bubbles [227], triaxially stretched hexagons [228], and more recently graphene kirigami [176]. In the triaxially stretched graphene hexagon, it was found that the uniform PMF restricted transport to Landau level and edge state-assisted resonant tunneling. In the graphene kirigami case, it was found that the emergence of PMF at large strain gradients acts as an enabling mechanism to allow electron transport for large strains [176].

In summary, there have been remarkable advances in theoretical understanding of the electromechanics of graphene in recent years, which have shed light on the fertile opportunities to tailor the electronic properties of graphene via strain engineering. However, there remains a substantial need for experimentalists to demonstrate such potential while simultaneously exploring new nanoelectronic device concepts.

## 4.2 Phase transitions under mechanical constraints

The discovery of ultrathin 2D materials [1] has led to an extraordinary amount of interest for their fundamentally new physics [229, 230] and potential technological applications [3, 231, 232]. 2D materials are a diverse family of crystals ranging from graphene [233] and boron nitride [234] to single layers of TMDs [92, 231]. In particular, single-layered TMDs have received significant attention [92] because some of them are semiconductors, promising for potential applications in electronics. TMDs have the formula $MX_2$, where M is a transition metal atom and X is a chalcogen atom. Among these 2D TMDs, the Mo- and W-dichalcogenides (group VI TMDs) have attracted the most research interest because they are mostly semiconducting [231].

An intriguing feature of these group VI 2D TMDs is that they can exist in multiple crystal structures, each with distinct electrical properties [20, 21, 235, 236]. The two lowest-energy crystal structures are often referred to as H and T' [20, 21, 236]. Under ambient conditions, all Mo and W-based TMDs except $WTe_2$ are most stable in H phase, a semiconducting phase with photon adsorption band gap between 1 and 2 eV [231]. The semi-metallic T' phase has been found in $WTe_2$ under ambient conditions [235, 237], in $MoTe_2$ at high temperatures [237], and in lithium-intercalated $MoS_2$ [238].

Phase switching between a semiconducting phase and a semi-metallic phase is of technological importance in phase-change electronic devices, such as phase change memory (PCM). Structural phase switching in bulk TMDs has been reported via lithium-based chemical exfoliation [238-240], using an electron beam [241], and controlling chemical tellurization rate of a Mo film [242]. If such a phase control were to be realized in 2D TMDs, phase-change electronic devices may benefit from the mechanical flexibility and ultra-thinness of 2D materials, which could reduce the energy consumption required for phase switching. Recently, theoretical calculations have shown that phase switching in 2D TMDs could be introduced by a variety of stimuli, such as chemical doping [243], mechanical deformations [20], heating [21], alloying [21], and electrostatic gating [236]. In this section, we highlight some mechanisms to drive such a semiconductor-to-semimetal phase change in 2D TMDs, with an emphasis on the importance of mechanical constraints.

In 2D materials, there exist several different potential mechanical constraints under which a phase change could occur. The first one is constant stress, where the stress is fixed at a value (e.g. zero) when the phase change occurs. This is analogous to the constant pressure constraint for bulk materials. In this scenario, the lattice constants of the initial and transformed phases can differ and are the lattice constants at zero stress. The constant-stress (zero-stress) condition may apply when the 2D material is freely suspended such that the force applied to it is fixed. This constraint could also apply for a 2D material on a substrate if the substrate friction is sufficiently small that the 2D material is free to slide. A second type of mechanical constraint is fixed lattice constants, where both phases are constrained to the same lattice constants. This condition might be expected to hold when the 2D materials have strong interactions with the substrate that prevent the 2D materials from sliding. A third possible mechanical constraint applies when the total area of the 2D material is fixed, in analogy to the constant volume constraint for bulk materials. This might be applicable when a 2D material is freely suspended with rigidly clamped boundaries. One can also envision other types of mechanical constraints, e.g. when one of the in-plane principal tensions is zero and the other one is nonzero [20]. This constraint might apply for a film suspended over a trench. Which mechanical constraint should apply depends on how the 2D materials interact with the substrate or suspending mechanisms. The type of constraint can have an impact on the magnitude of the phase boundaries and also the potential to observe a stable mixed phase regime.

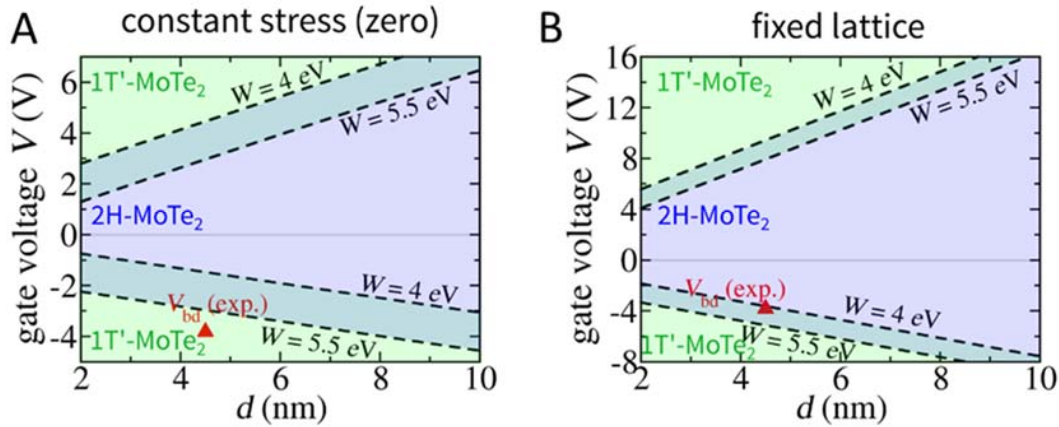

**Fig. 9**: The computed phase diagrams of monolayer MoTe$_2$ under electrostatic gating. These phase diagrams predict which phase is more stable given some gate voltage $V$ and dielectric thickness $d$. The phase boundaries of the H and T' phases vary with the the work function of the gate electrode $W$. The required transition gate voltage is smaller in constant-stress (zero-stress) case (A) than in fixed-lattice case (B). Figures adapted from [236].

The charge induced by electrostatic gating has been found to have the potential to drive structural phase transitions in some 2D TMDs by Li *et al.* [236]. By developing a density functional-based calculation approach, they discovered that the semiconductor-to-semimetal structural phase transition between H and T' phases in monolayer MoTe$_2$ can be driven by a gate voltage of several volts. Figure 9 shows the phase diagrams of MoTe$_2$ under electrostatic gating utilizing a capacitor structure. The phase stability was computed with respect to gate voltage $V$ and dielectric thickness $d$ for the cases of constant-stress (zero-stress) (Fig. 9A) and fixed-lattice (Fig. 9B) constraints. The dielectric medium was chosen to be $HfO_2$. The quantitative positions of the phase boundaries depend on the work function of the gate electrode $W$. The semiconducting H phase of $MoTe_2$ is stable between the two phase boundaries, and semi-metallic T' phase can be stabilized with application of sufficiently large positive or negative gate voltage. The phase boundaries are closer to ambient condition in the constant-stress case (Fig. 9A) than in the fixed-lattice case (Fig. 9B). Assuming a dielectric medium of 4.5-nm thickness, the magnitude of negative transition gate voltage can be reduced approximately from 4V to 2V if the mechanical constraint changes from fixed lattice to constant stress. The transition gate voltage is larger in the fixed-lattice case because the energy of the constrained T' phase is higher than that of the constant-stress T' phase, pushing the phase boundary further from ambient conditions.

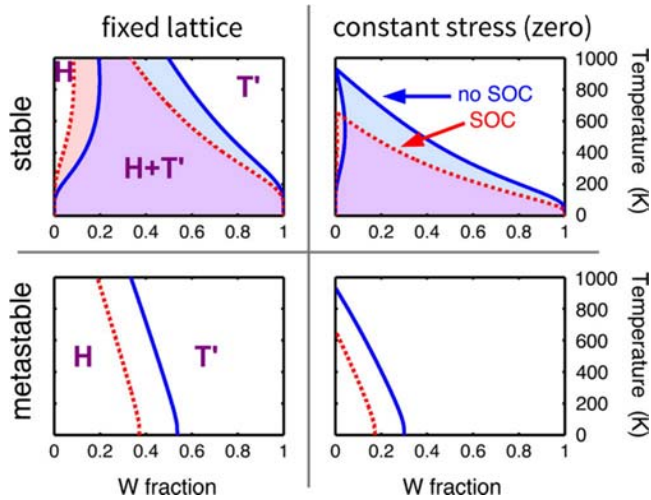

**Fig. 10**: Computed phase stabilities of monolayer $Mo_{1-x}W_xTe_2$ alloys with respect to temperature and W fraction. Stable (top) states include mixed phase regimes (shaded). Metastable (bottom) states exhibit random transition metal atom positions. When transition metal diffusion is quenched after high temperature growth, the metastable diagrams apply; otherwise, the stable diagrams apply. The transition temperatures vary by hundreds of degrees when the mechanical constraint changes from fixed lattice (left) to constant stress (right). Figure adapted from [21].

Duerloo and Reed [21] recently demonstrated that H-to-T' phase transition in 2D TMDs can be driven by heating and the transition temperature is tunable over a range of 0 to 933 K through alloying $Mo_{1-x}W_xTe_2$ monolayer. Figure 10 shows phase diagrams of the monolayer $Mo_{1-x}W_xTe_2$ alloy. The phase stability is computed with respect to temperature and W fraction. As shown by Fig. 10, the phase boundaries are much closer to ambient conditions in the constant-stress case (zero stress) than in the fixed-lattice case. For example, the transition temperature can vary by hundreds of degrees at a given W fraction depending on whether the material is at constant stress or fixed lattice.

The important role of mechanical constraints to phase changes in 2D TMDs has also been observed in other mechanisms, such as mechanical deformation [20] and chemical doping [243]. Duerloo *et al*. [20] computed the strains required to drive the H-to-T' phase transition in 2D TMDs for several mechanical constraints. A range from 0.3 to 3.0% tensile strain is identified to transform monolayer $MoTe_2$ under uniaxial conditions at room temperature, depending on the type of constraint. Mechanical deformation could be applied to 2D TMDs using flexible substrates, an AFM tip, and other standard experimental approaches. Zhou and Reed [243] studied the potential of using chemical doping to realize phase switching between semiconducting H-$MoTe_2$ and semi-metallic T'-$MoTe_2$. They discovered that the constant-stress (zero-stress) condition pushes the phase boundary toward the T' phase and has the potential to cause the T' phase favored for some types of chemical doping [243].

These recent studies demonstrated the potential of dynamic control of the structural phases and electrical properties in 2D materials, enabling potential applications in phase-change electronic devices with low energy consumption. These studies also showed that mechanical constraints play an important quantitative role in the magnitudes of the phase boundaries.

### 4.3 High pressure/strain effects

Applying hydrostatic pressure or uni-/biaxial strain to crystalline materials influences atomic bond lengths and angles, thus effectively altering phonon and electronic structures. 2D materials under pressure or strain exhibit especially unique behaviors thanks to its bond strength contrast between strong intra-layer covalent bonds and weak inter-layer van der Waals bonds. Considering its atom-thin nature and possible applications for flexible optoelectronic devices, it is crucial to understand the effect of pressure or strain on opto-electronic and vibrational properties of 2D materials. Hydrostatic pressure can be applied with a diamond anvil cell (DAC) setup (Fig. 11A), which offers precise determination and control of pressure, up to extremely high strain levels beyond what is typically seen in uniaxial test fixtures. DAC is also compatible with numerable in-situ measurements, including optical and electrical measurements [244].

Similar to many other crystalline materials, TMDs under pressure experience a blue-shift of the Raman modes, due to decreasing bond lengths. For instance, the two signature Raman modes of $MoS_2$, in-plane $E_{2g}$ and out-of-plane $A_{1g}$ vibrations, increase in wavelength with pressure-dependence of ~1.0 cm$^{-1}$/GPa and ~2.9 cm$^{-1}$/GPa, respectively [244]. The difference in the pressure-dependence originates from the unique van der Waals structure, where the c-axis is experiencing much more rapid compression, and therefore the out-of-plane $A_{1g}$ mode is under stronger hardening. At a pressure range of 10-19 GPa, $MoS_2$ undergoes a semiconductor-to-metal transition where the blue-shift of the $A_{1g}$ mode is drastically reduced. After the metallization pressure, both modes experience stronger blue-shift than under lower pressure (Fig. 11B). It is noteworthy that, while Raman modes of monolayer $MoS_2$ (A′ and E′ modes) under hydrostatic pressure exhibit a blue-shift similar to the case of bulk $MoS_2$ [22], uniaxial strain only has a marginal effect on the A′ mode, due to lack of strain in the direction of the c-axis [245, 246]. Doubly-degenerate E′ mode splitting was also reported in some uniaxial tensile strain setups [246].

In disordered 2D systems, such as $Mo_{1-x}W_xS_2$ alloy with randomly dispersed Mo and W at the metal atom cites of TMDs, disorder-activated Raman modes were observed in addition to the modes of pure end members [247, 248]. More interestingly, some additional disorder-activated Raman modes denoted $A^*$ and $A^†$ were observed only under high pressure, ~8 GPa and ~30 GPa, respectively. It is suggested that these modes are inter-layer disorder-related modes, activated when the layers are compressed close enough to develop and enhance inter-layer interactions (Fig. 11E) [249].

Electronic structures of 2D materials also evolve according to applied pressure and/or strain, allowing pressure/strain to be another degree of freedom to tune the electronic properties, sometimes referred to as *straintronics*. For example, bulk $MoS_2$ undergoes semiconductor-to-metal transition under hydrostatic pressure, with an abrupt drop of resistivity in an intermediate pressure range of 10-19 GPa (Fig. 11C). The band structure calculation of bulk $MoS_2$ with an indirect gap of ~1.2 eV was also predicted to monotonically decrease with increasing pressure until the gap closure at ~19 GPa [244]. Although the hydrostatic pressure was high enough to close the band gap, the metallization was irrelevant to structural phase transition, but an electronic transition [244]. The direct bandgap of monolayer $MoS_2$, on the other hand, increases under applied pressure with the compressive strain up to 12%, and undergoes direct-to-indirect transition at a pressure of ~16 GPa. Theoretical calculations predicted that the band gap decreases at higher pressure and eventually metallizes (~68 GPa, Fig. 11D) [22]. In contrast, uniaxial strain applied to monolayer MoS$_2$ exhibits a red-shift of photoluminescence (PL) and absorption spectrum peaks, and a decrease in PL intensity due to the transition toward indirect band gap [246, 250].

Large biaxial strain can be applied to 2D materials such as graphene and MoS$_2$ using a pressurized blister test [37, 251]. By applying a pressure difference across a suspended MoS$_2$ membrane (diameter ~ 5 – 10 µm) transferred onto etched microcavities in silicon oxide, Lloyd et al. [37] measured the optomechanical coupling directly using a combination of Raman spectroscopy, photoluminescence spectroscopy, and AFM. The pressure difference across the membrane deformed the membrane into a circular blister with a biaxial strain at its center, which resulted in a red-shift in the optical band gap of -99 meV/% strain, consistent with theoretical predictions [252, 253]. This technique enables one to apply biaxial strains as large as 5.6% and achieve a band gap shift of ~500 meV.

Hydrostatic pressure can also enable inter-layer charge transfer in heterostructures, which can further tune the electronic properties of 2D materials. In a heterostructure composed of monolayer graphene on top of monolayer MoS$_2$, the graphene was p-doped with high carrier concentration (~2.7 × 10$^{13}$ cm$^{-2}$ at a pressure of 30 GPa), an order of magnitude higher than the charge concentration of pristine graphene [23]. This concentration is also comparable to the doping concentration of graphene-MoS$_2$ heterostructures by gate bias.

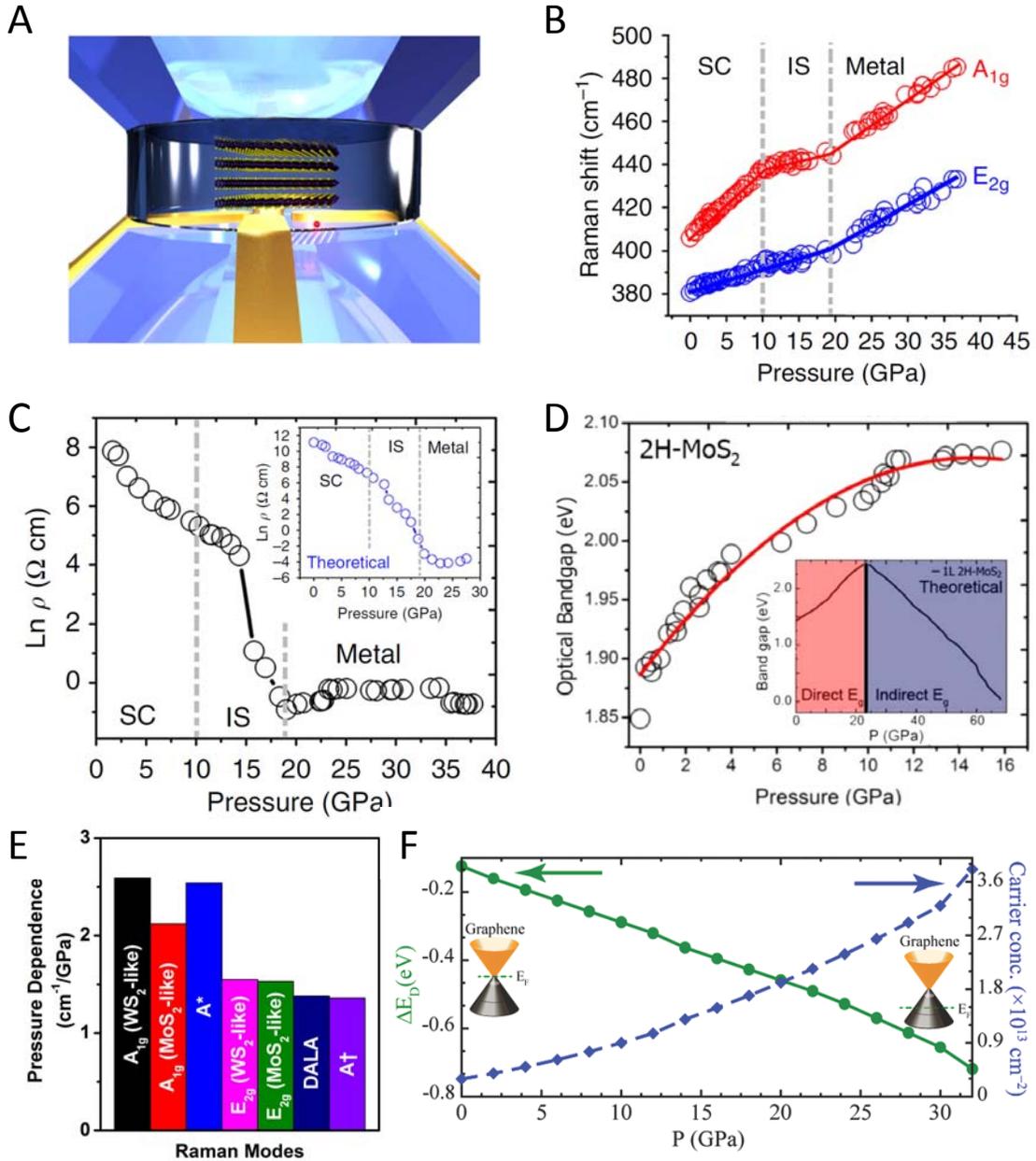

**Fig. 11**: (A) Schematic view of a DAC chamber with multi-layer TMD sample, red ruby pressure calibrate. (B) Evolution of bulk $MoS_2$ Raman modes with increasing pressure. (C) Abrupt drop of resistivity with increasing pressure clearly indicates metallization of $MoS_2$. (D) Optical band gap evolution of monolayer 2H-$MoS_2$ under pressure. (E) Pressure-dependence or Raman peaks of bulk $Mo_{1-x}W_xS_2$ alloy. (F) Relative shift of the Dirac point of graphene with respect to the Fermi level ($\Delta E_D$; left axis) and carrier concentration of graphene induced by charge transfer (right axis) in graphene-$MoS_2$ heterostructure as a function of hydrostatic pressure. Figures adapted from: (A-C) [244], (D) [22], (E) [249], and (F) [23].

## 4.4 Piezo- and Flexoelectricity

Flexoelectricity (FE) is a fairly recent discovery in physics [254-256]. It may be considered as an extension of the piezoelectric (PE) effect where PE relates uniform strain to polarization and FE relates strain gradients to polarization. These two electromechanical effects have applications in the fields of actuators, sensors, and energy harvesters [257-259] among others. Recently, PE and FE have been studied in 2D materials, adding to the long list of impressive characteristics of this family of materials [260, 261]. Studies of PE in 2D materials have already yielded substantial experimental data, but the study of FE in 2D materials is still in its infancy.

The relation between polarization and strain in materials can be summarized in the equation

$$P_i = e_{ijk}\varepsilon_{jk} + \mu_{ijkl}\frac{\partial \varepsilon_{jk}}{\partial x_l}, \qquad (4)$$

where $e_{ijk}$ is the PE coefficient, $\mu_{ijkl}$ is the FE coefficient, $\varepsilon_{jk}$ is the strain, and $\partial \varepsilon_{jk}/\partial x_l$ is the strain gradient [255]. A few studies have theoretically [262-264] and experimentally [265-268] explored PE in 2D materials. FE on the other hand is relatively understudied in 2D materials although its monatomic thickness affords a potentially greater platform for future studies. The main reason for the lack of study in FE is that it is largely unnoticeable in bulk, macro-scale materials because large strain gradient requires a large strain difference, which may fracture the material. As a result, the gradient of strain is very limited in bulk materials, silencing the FE term in Eq. (4). Once nanometer scales are reached, even small strain differences can lead to large strain gradients, allowing for an amplification of the FE term in Eq. (4).

On the theoretical side, there have been a handful of studies exploring the polarization arising from curvature of 2D materials. Most studies to date focus on carbon systems [269-271] and hexagonal boron nitride (h-BN) [25, 272]. Carbon systems, such as curved graphene [260] (Fig 12A) and graphitic nanocones [271] (Fig 12B), are predicted to have out-of-plane polarization that arises from the curvature of sp$^2$ bonds and redistribution of the electron gas in the normal direction. In contrast to carbon systems, h-BN tends to have polarization induced by curvature that arises in-plane. Bilayer h-BN was found to have enhanced electromechanical coupling compared to monolayer [272] (Fig 12C), but monolayer h-BN has still been predicted to have non-zero in-plane polarization when having a corrugated shape [25] (Fig 12D). FE can also be used to induce PE-like properties in non-piezoelectric 2D materials if non-symmetric holes are patterned into the material [273] (Fig 12E).

On the experimental side, studying FE can become challenging even for bulk materials. Isolating the FE effect completely from the PE effect is difficult and experimental methods for measuring FE coefficients often give results that are orders-of-magnitude different from theoretical predictions [255]. To date, the most promising experimental evidence for FE in 2D materials was observed using a method called piezoresponse force microscopy (PFM) where a conductive AFM tip was brought in contact with a sample to apply an alternating electric field [24]. The electric field causes a PE sample to expand/contract due to converse PE and can be measured by vertical displacement of the AFM tip. This technique is typically used to study PE materials, but can potentially be used to study FE if the electric field originating from the AFM tip is spatially varying, taking advantage of inverse FE. Alternatively, as observed for graphene nitride nanosheets, triangular holes in the 2D material cause a non-piezoelectric material to exhibit PE-like behavior as measured using PFM [24] (Fig 12E-H). The proposed origin of the PE-like behavior comes from the non-symmetric holes which create strain concentrations and strain gradients, thus resulting in a FE effect.

In summary, the theoretical and experimental study of FE in 2D materials has just begun and is full of potential for future discoveries. Very little experimental work has been reported on this topic, and any contribution in this area can significantly advance the entire field.

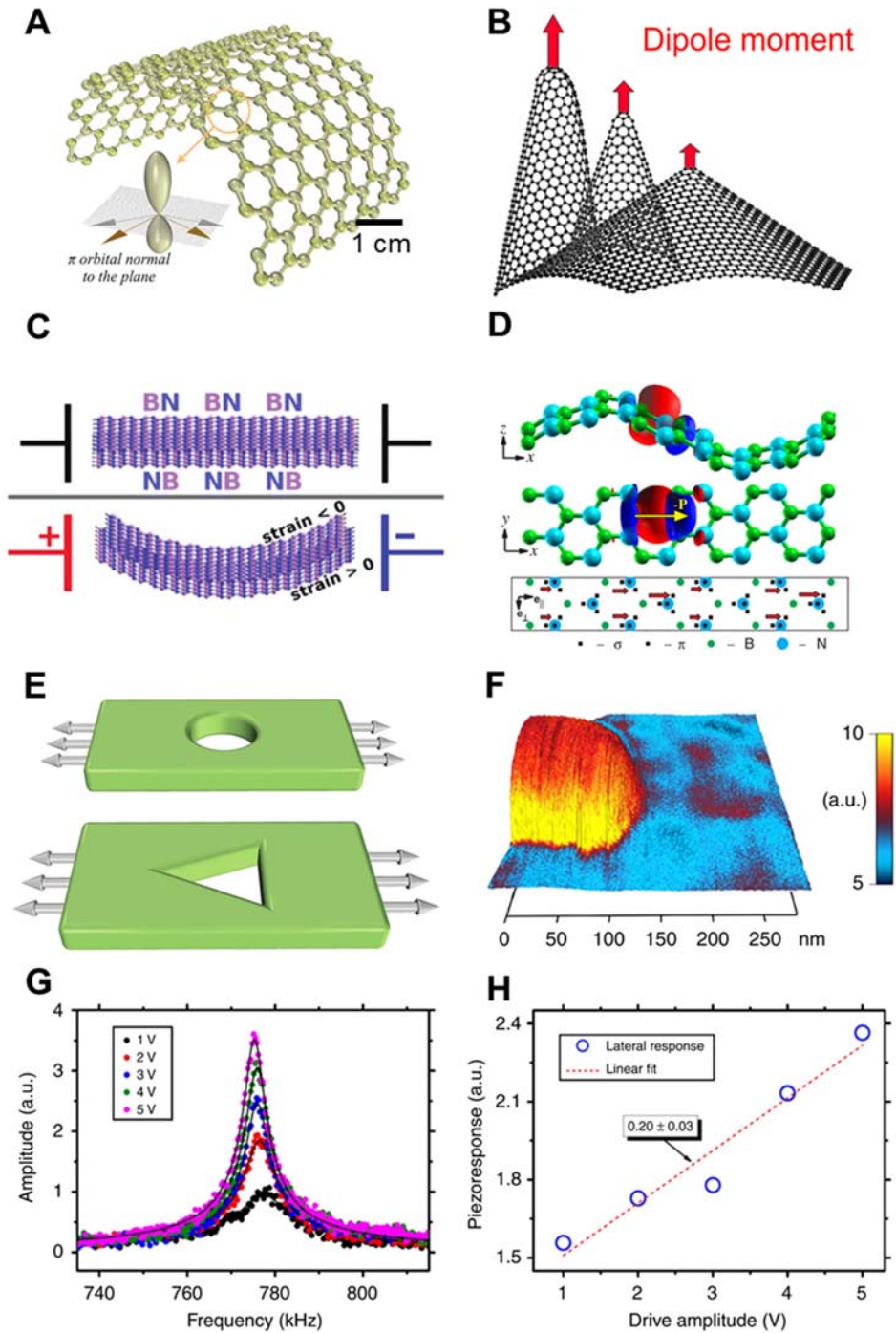

**Fig. 12**: Different ways that flexoelectricity can arise in 2D materials. (A) The electron distribution at each atomic site in curved graphene nanoribbons is non-symmetric and leads to a net out-of-plane polarization [260]. (B) Dipole moments arise at the tips of graphitic nanocones [271]. (C) Bilayer h-BN shows enhanced electromechanical coupling between film curvature and in-plane polarization [272]. (D) Polarization can occur in-plane in corrugated monolayer h-BN [25]. (E) Film with a symmetric circular hole under tension does not experience any polarization whereas film with a non-symmetric triangular hole under tension can exhibit a net polarization due to FE effects from strain concentration [260]. (F) A 3D PFM image of graphene nitride nanosheets showing PE-like response that arises from a FE effect with triangular holes [24] as illustrated in (E). (G) The amplitude of the PFM response of the graphene nitride nanosheets at the contact resonance point measured at various drive voltages [24]. (H) The piezoresponse vs. drive amplitude on the graphene nitride nanosheets. A linear relation indicates that there is PE-like behavior present [24]. Figures adapted from: (A and E) [260], (B) [271], (C) [272], (D) [25], and (F-H) [24].

## 5. Interfacial Properties: Adhesion and Friction

Graphene and other 2D materials have the highest surface to volume ratios of any class of materials. As a result, surface forces are expected to play a significant role as these materials are being integrated into microelectronics, MEMS and NEMS devices and composite materials. Surface forces are also important when the 2D materials have to be transferred from one substrate to another via selective delamination and adhesion. This section presents a brief review of recent developments in measuring interfacial properties of 2D materials including adhesion and friction, followed by a discussion on the nature of the interaction forces, such as van der Waals, capillary effects and the role of surface roughness.

### 5.1. Adhesion experiments

The majority of adhesion measurements between 2D materials have been made using blister and laminated beam fracture experiments. In some cases, the 2D material is supported by another layer in order to make sure that the monolayer does not break prior to delamination between the 2D material and its substrate. In others, the stress levels are low enough to allow freestanding membranes to be used. Examples of the latter approach have come mainly from Bunch's group [26, 251, 274] where a 2D material such as graphene is suspended over micro-cavities etched in an oxidized silicon wafer, thereby forming a sealed micro cavity (Fig. 13A). The 2D material is transferred using either the "scotch tape" method [1] or dry transfer utilizing a polymer stamp that incorporates films or flakes of the 2D material grown by chemical vapor deposition [275]. Utilizing the remarkable gas impermeability of 2D materials [251], one can apply a pressure difference across the membrane to deflect it upward or downward depending on the pressure difference.

Some of the earliest adhesion measurements of graphene to silicon oxide utilized a version of the blister test where the number of molecules was constant [26]. In this implementation, the sealed micro cavity is filled with a fixed number of gas molecules by diffusing gas through the silicon oxide. The external pressure is then lowered, which results in a membrane deflection and volume expansion of the gas in the micro cavity (Fig. 13B). At a critical pressure, the membrane begins to delaminate from the substrate and the blister diameter grows (Fig. 13C). Measuring the blister diameter as a function of internal pressure allows one to deduce the adhesion energy from a membrane mechanics model [26, 276]. Because the pressure in the sealed micro cavity decreases with increasing cavity volume, the blister growth is stable. This is in contrast to pressure-controlled blister tests which promote unstable crack growth [276]. The stable growth allows multiple measurements of the adhesion energy to be made on the same graphene membrane with varying blister diameters (Fig. 13C).

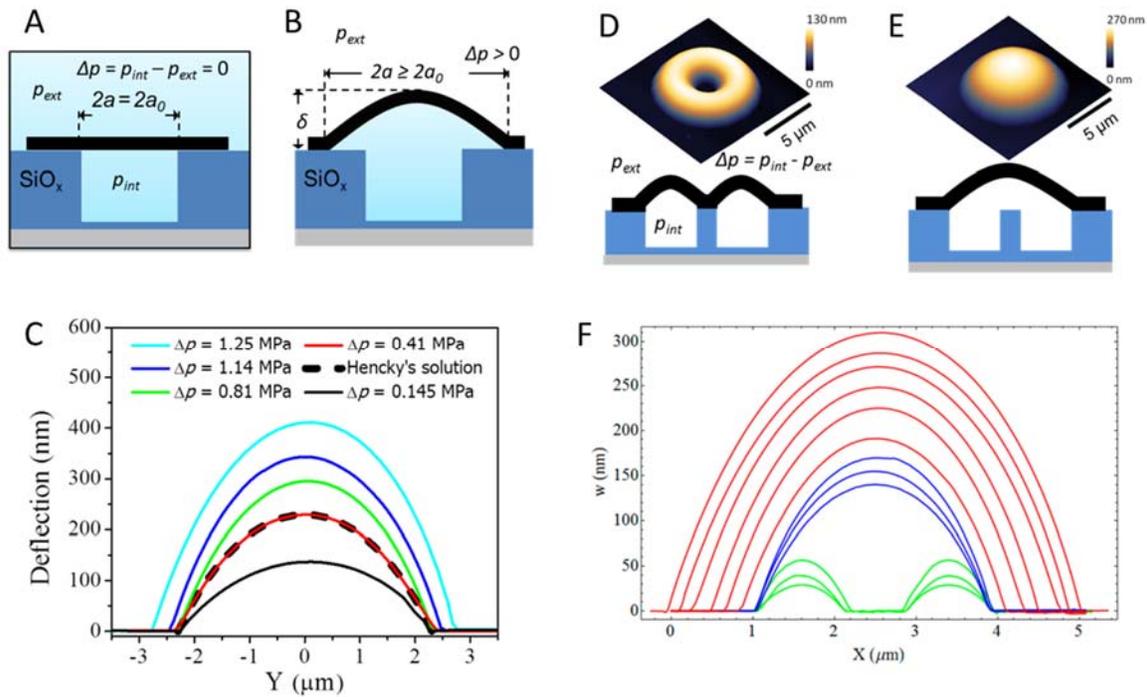

**Fig. 13:** (A) Schematic of the constant N blister test before pressurization, and (B) after pressurization and delamination. (C) Atomic force microscope line scans through the center of a pressurized graphene blister at varying pressure differences. (D) AFM image (upper) and schematic (lower) of a pressurized graphene membrane in the island blister test before and (E) after snap-off. (F) AFM line scan through the center of a pressurized graphene membrane in the island blister test. Figures adapted from: (A-C) [26], (D-E) [277], and (F) [278].

A variation of the standard blister test incorporates a micro fabricated inner post to the micro cavity [277, 278]. In this case, the graphene is bulged into an annular shape with the graphene remaining bonded to the inner post at small pressure differences (Fig. 13D). As the internal pressure increases, the graphene snaps off the inner post (Fig. 13E) and then at a higher pressure, begins delamination from the outer edge as in the standard blister test (Fig. 13F). The pressure at snap-off is used to measure the adhesion energy to the inner post while the growing blister diameter follows the classical blister test [278, 279]. The time reversal of this experiment is to keep the outside pressure fixed and let the internal pressure decrease slowly as gas diffuses out of the micro cavity. This results in a decreasing height of the blister and a pull-in instability (snap-back) at a critical height [277]. The pull-in instability was found to take place at separations of 10-20 nm and follow an inverse fourth power traction-separation relation [277, 279], which is consistent with van der Waals interactions.

A compilation of the adhesion energies determined by both the classical and island blister tests for single and multilayered graphene membranes is shown in Fig. 14 [26, 274, 278]. Adhesion energies of 1 to 5-layer graphene membranes to silicon oxide varied from 0.1 to 0.45 J/m$^2$. The measurements for a single flake showed remarkable reproducibility demonstrating that variations are not due to measurement error. However, there is considerable spread in values between devices [274]. This suggests that the adhesion between graphene and a surface may strongly depend on the surface conditions (e.g., roughness, moisture, chemical reactivity, etc.) and further work is needed to address this issue.

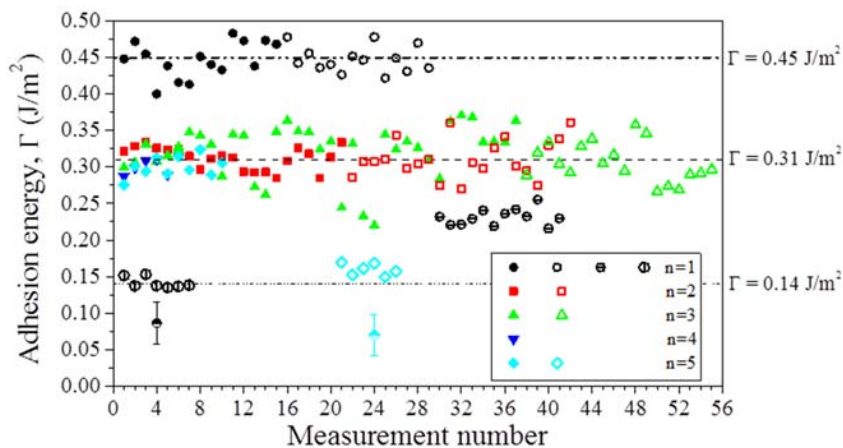

**Fig. 14**: Compilation of measured adhesion energy values for 1 to 5-layer graphene membranes ($n=1-5$) [26, 274, 278]. Each symbol represents a different flake. The points with the error bars represent adhesion energies from the center post of the island blister test [278].

Another variant of the blister test for determining the adhesion energy of exfoliated graphene to silicon oxide was to scatter nanoparticles on the silicon oxide and then drape graphene flakes over them. Measurements of the deformed shape of the graphene flakes allowed the adhesion energy (0.151 J/m$^2$) to be extracted from a simple analysis [280]. Strictly speaking, such measurements are truly attributed to adhesion rather than separation energy and can be expected to be lower than separation energies if adhesion hysteresis is active. In this context, differences in the energies associated with pull-in (adhesion) and snap-off (separation) phenomena can also be expected.

A related approach, which is particularly useful for extracting the adhesion energy between 2D materials and compliant substrates, is to make use of wrinkles and buckles that may form spontaneously due to in-plane compression or release of a pre-stretched substrate. AFM measurements of buckle delamination profiles were used in conjunction with a thin film mechanics model to obtain a rough estimate of the adhesion energy (~0.54 mJ/m$^2$) between graphene and polyethylene terephthalate (PET) [281]. A similar method was used to obtain an adhesion energy of ~18 mJ/m$^2$ between $MoS_2$ and polydimethylsiloxane (PDMS) [282]. Both values are much lower than the typical values for van der Waals interactions. The accuracy of this method may be improved from both experimental and theoretical sides. Experimentally, the resolution of the AFM measurements of the buckling profiles is rather limited due to convolution of the AFM tip and relatively sharp ridges of buckling. Theoretically, the mechanics model may be improved by considering the deformation of the compliant substrate and potentially large-scale bridging at the delamination front.

Larger scale blister tests on CVD graphene that had been wet-transferred to copper and silicon were conducted by Cao et al. [283]. Following transfer, the graphene was reinforced by an epoxy layer and separated from the substrates under volume-controlled pressurization. Measurements of the blister profile as a function of pressure allowed the separation energies to be extracted from mechanics models of a thin membrane or plate, depending on the thickness of the composite film specimen (graphene and epoxy) [41]. The separation energies were commensurate with the values obtained with exfoliated, single crystal graphene by Bunch's group [26, 274, 278].

While blister tests have yet to be used to determine the interfacial properties between CVD graphene and its seed copper, this has been achieved by making use of another common fracture test, the double cantilever beam (DCB). In this case, a graphene-coated copper foil is sandwiched between silicon strips with an epoxy [284] and then separated. At separation rates above 250 μm/s, it was found that delamination occurred between graphene and copper, effectively transferring the graphene to the epoxy. Below 25 μm/s, delamination occurred between graphene and the epoxy. The separation energies associated with delamination along the graphene/copper and graphene epoxy interfaces were 6 and 3.8 J/m$^2$, respectively. This selective delamination appears to have been facilitated by the rate dependence of the graphene/epoxy interface, which, if recent experiments on silicon/epoxy interfaces can be used as a proxy, suggests that the separation energy of the graphene/epoxy interface becomes higher than that of the graphene/copper interface above a characteristic separation rate. Such a rate dependence is most likely an interfacial effect, limited to the epoxy interphase, as the behavior of the bulk epoxy should be firmly in the glassy regime for the strain rates that are active at the separation front.

Selective delamination has also been observed for graphene-coated copper films on silicon wafers [285, 286]. In this case, the silicon substrate on which the copper and graphene are deposited forms one of the reinforcing layers and the second silicon strip is bonded to the copper foil with an epoxy. Yoon et al. [285] obtained a separation energy of 0.72 J/m$^2$. In the experiments by Na et al. [286], higher separation rates led to delamination along the silicon/copper interface at 1.7 J/m$^2$ compared to 1.5 J/m$^2$ for delamination along the graphene/copper interface at lower separation rates. The lower separation energy for graphene on copper film, as opposed to the copper foil, appears to be due to the lower surface roughness of the copper film following deposition of the graphene.

Another method to measure graphene adhesion is based on nanoindentation experiments. AFM has been widely used for adhesion measurements [287-289] due to the high-resolution imaging of surfaces and accurate measurement of interaction forces and displacements it provides. To convert the adhesion force measured by AFM to adhesion energy, a contact mechanics model is required. A number of such models have been developed including the well-known Johnson-Kendall-Roberts (JKR) [290], Derjaguin-Muller-Toporov (DMT) [291] and Maugis [292] models, which typically consider the interactions between an ideal sphere and an atomically flat surface. Due to the fact that the shape of a conventional AFM tip is typically not spherical and can be challenging to measure, a microsphere probe that can be attached to a tipless AFM cantilever has been used in adhesion studies [293, 294]. Recently, Jiang and Zhu [295] developed a similar method to measure adhesion between graphene and different materials (Fig. 15). The adhesion force between graphene and the spherical tip was measured by AFM in the force spectroscopy mode, and then the adhesion energy was calculated using the Maugis-Dugdale model. Their work addressed two challenges for measuring graphene adhesion using AFM: surface roughness of the AFM tip and pull-off instability that can occur during the experiment. To address the first challenge, a graphene flake was placed on top of an atomically flat mica substrate, which eliminates the effect of surface roughness due to the substrate, while the effect of surface roughness of the spherical tip was treated by the modified Rumpf model [296, 297]. To address the second challenge, their analysis showed that the adhesive force gradient is much larger than the cantilever stiffness but smaller than the contact stiffness. Hence, while the pull-off instability does occur, the pull-off force provides a reasonable estimate of the adhesion energy. Using the AFM-based method, adhesion energies of monolayer graphene to $SiO_2$ and Cu tips were obtained as 0.46 and 0.75 $J/m^2$, respectively.

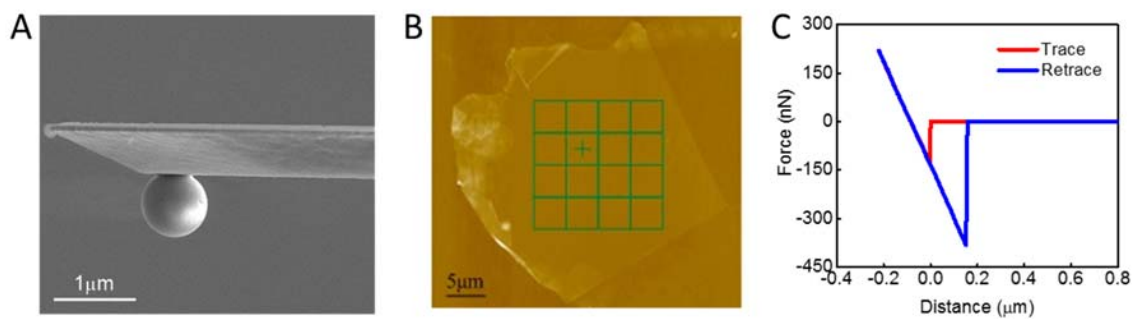

**Fig. 15**: (A) Side view of a microsphere tip on an AFM cantilever. (B) AFM measurements on a monolayer graphene flake. The grid shows 16 blocks and the force measurement was taken at the center of each block (denoted by the cross). (C) A typical AFM force profile with a silicon oxide sphere. Figures adapted from [295].

While classical contact mechanics analyses such as JKR and DMT can be used to extract adhesion and separation energies, the Maugis analysis assumes a traction-separation relation for a contacting pair in form of a hat function, rising sharply to the strength of the interface and then decaying to zero at a critical separation as the range of the interaction is exceeded. Additional information than just AFM force profiles, such as the contact radius [298], or assumptions [299] are required to extract the strength and the range of the hat-shaped interaction function. On the other hand, displacement-controlled nanoindentation experiments [300, 301] can provide much richer force profiles without the pull-off instability and allow the traction-separation relation to be extracted by comparing the measured force profiles with associated numerical analysis [302]. This method was recently used by Suk et al. [303] to compare the force profiles for a diamond tip indenting mono-, bi- and trilayer graphene membranes that had been transferred onto silicon oxide substrates. Bare silicon oxide and highly ordered pyrolytic graphite (HOPG) were also considered. As the number of graphene layers was increased, there was a transition in the adhesive interactions between the tip and the surfaces from that of bare silicon oxide to that of graphite. Examples of the force profiles are shown in Fig. 16 for silicon oxide, monolayer graphene and graphite during approaching and withdrawal in a dry nitrogen environment. On approaching, there was a snap which was associated with water bridge formation. The extent of this snap was greatest for bare silicon oxide (Fig. 16A), which is hydrophilic; it diminished as the number of graphene layers increased and was completely absent for graphite (Fig. 16C), which is hydrophobic. These observations suggest that graphene partially screened the force field between the diamond tip and the silicon oxide. The measured force profiles were compared with finite element method (FEM) based numerical simulations that accounted for the interactions between the probe and the target surfaces as well as between graphene and silicon oxide. The traction-separation relations that were required to bring the numerical and experimental force profiles into agreement suggested that both van der Waals and capillary forces were at play. These two effects can be seen clearly in Fig. 16 during withdrawal with the sharp drop from the peak tensile force being associated with van der Waals forces while the long tail that follows is likely due to thinning of capillary bridges.

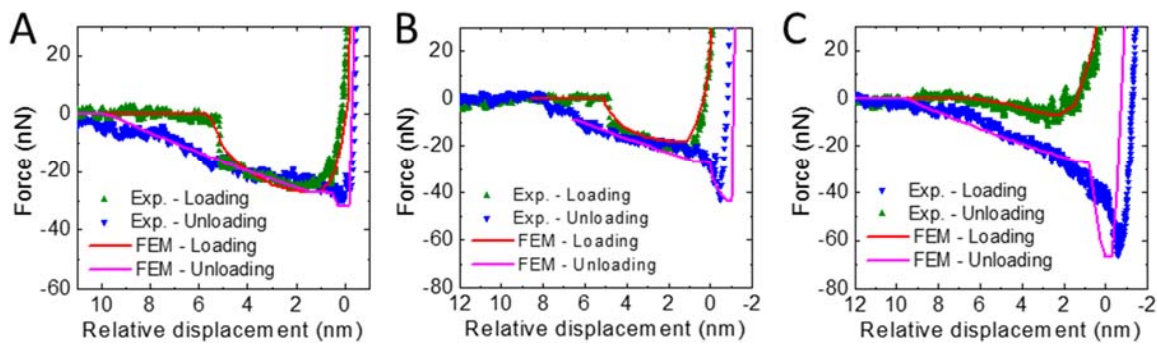

**Fig. 16:** Force profiles from displacement-controlled nanoindentation with a diamond tip indenting (A) bare silicon oxide, (B) monolayer graphene on silicon oxide and (C) HOPG in a dry nitrogen environment. The force profiles are compared with ones obtained from FEM simulations that accounted for interactions between the probe and target surfaces as well as between graphene and silicon oxide. Figures adapted from [303].

Adhesion or separation energies, by themselves, do not provide sufficient information as to the nature of the interactions between surfaces. The processes of adhesion and separation can be described by traction-separation relations (TSRs) as in cohesive zone modeling of nonlinear fracture mechanics [304, 305]. The TSR of an interface provides a functional form of the interaction during fracture, with which the interfacial strength and the range of interaction can be determined in addition to the adhesion energy or fracture toughness of the interface. The interfacial TSRs can be obtained directly or indirectly from experiments [306]. Recently, Na et al. [27] reported measurements of the TSRs between wet-transferred, CVD grown graphene and the native oxide surface of silicon substrates by combining the DCB experiments with interferometry measurements (Fig. 17A). The deduced TSRs (Fig. 17B) exhibited a much longer range (greater than 100 nm) than those normally associated with van der Waals forces. Similar to the displacement-controlled nanoindentation measurements [303], the TSRs suggest that interaction mechanisms other than van der Waals forces should be considered for adhesion of graphene and other 2D materials, as discussed further in Section 5.4.

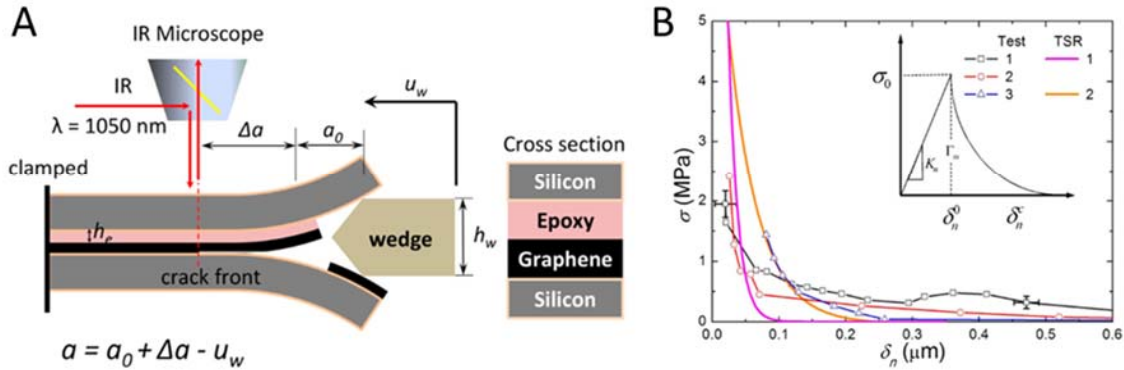

**Fig. 17**: (A) Schematic of a DCB experiment with Infrared (IR) crack opening interferometry. (B) The obtained traction-separation relations. TSR1 and TSR2 are the traction-separation relations used in the finite element analyses of tests 1-2 and test 3, respectively. Figures adapted from [27].

**5.2 Friction and wear**

The application of 2D materials to the field of lubrication [307-312], wear-prevention [313-315], and adhesion reduction [316] has been of recent interest in the field of nanotechnology. These materials have very high in-plane elastic moduli and high strengths [4, 13] as well as being chemically inert under most conditions [317-319]. Specifically, graphene has been used in tribological applications as it is derived from graphite, one of the most effective solid lubricants for engineering applications. Graphene, when applied to a surface, can substantially reduce friction at just one atomic layer, and approaches the low friction achieved using bulk graphite at four layers of graphene [309]. However, the extremely low bending stiffness of graphene has been thought to result in higher friction when the graphene lubricates a low adhesive material, or when the adhesive forces between the sliding asperity (typically an AFM tip) and the topmost graphene layer exceed the graphene interlayer cohesive forces [320]. Chemically modified graphene would increase the resistance to out-of-plane bending [321], and thus was thought to possibly reduce friction. However, in almost every case, friction was observed to increase on chemically modified graphene, when compared to that measured on pristine graphene [321-325]. Additionally, as shown in Fig. 18A chemical modification of graphene was detrimental to the excellent wear properties of pristine graphene, despite the fact that all chemical modifications did not increase the defect density or change the coverage of the graphene film on the surface of interest [324]. To understand this, three example studies on chemically modified graphene were examined, including friction of fluorinated graphene, friction and wear of graphene oxide using colloid probes, and stress-assisted wear of graphene oxide using thermal probes. The results showed that the change in mechanical properties of the graphene, altered through chemical modification, are likely less influential than the change in the local chemistry that results from this modification.

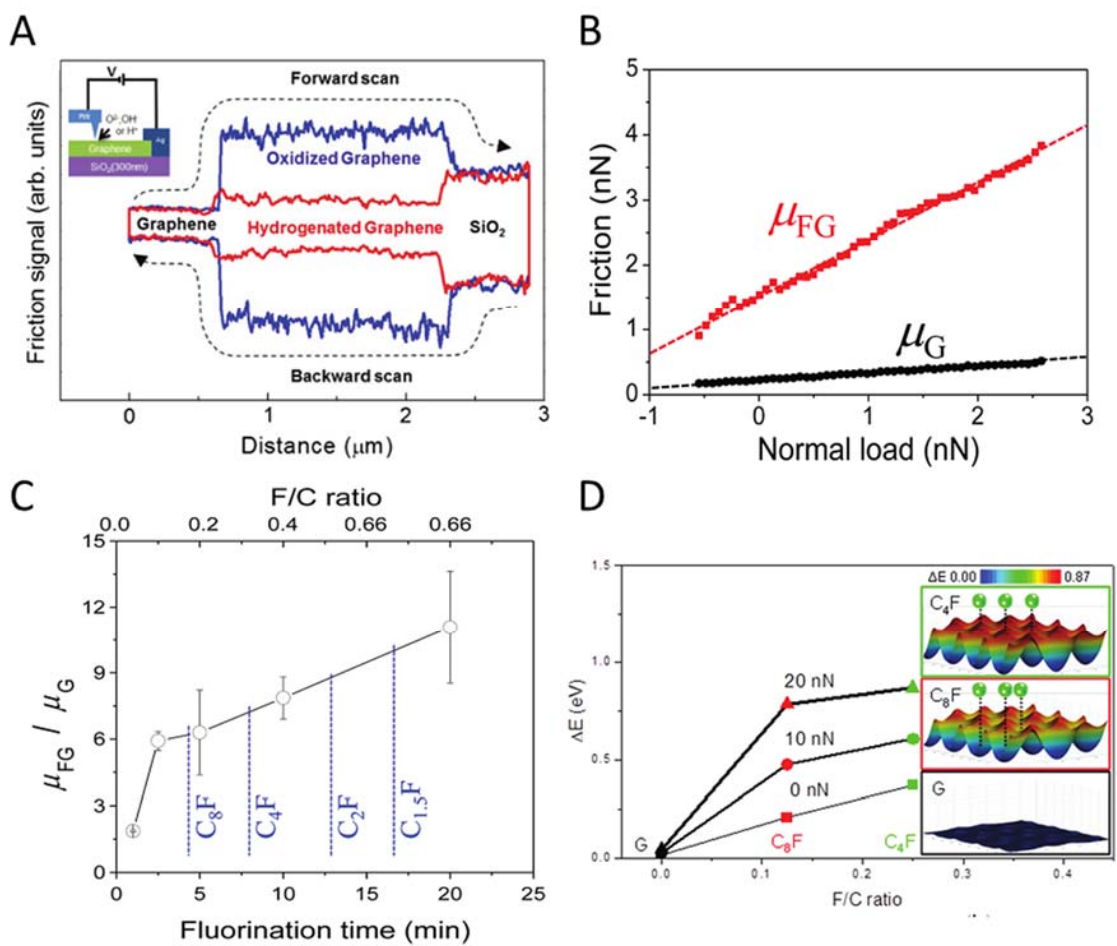

**Fig. 18:** (A) Variance of friction for pristine graphene, oxidized graphene, and hydrogenated graphene on a SiO$_2$ substrate. (B) Friction of fluorinated graphene and pristine graphene on a polycrystalline copper substrate. (C) Evolution of the friction coefficient for fluorinated graphene with increasing fluorination time. (D) Surface energy corrugation determined from MD simulations with increasing amount of fluorine on the surface. Figures adapted from: (A) [325] and (B – D) [322].

Fluorination of graphene, or single layers of polytetrafluorethylene (PTFE) with one of the lowest friction and adhesive surfaces, changes the bonding state of the carbon atoms from sp$^2$ in pristine graphene to sp$^3$ when the fluorine groups are attached to the graphene sheets [326]. The interlayer spacing for graphene fluoride is 1.24 nm [327] compared with 0.335 nm for pristine graphene. Thus the bending modulus of one layer of fluorinated graphene would be approximately the same as four layers of pristine graphene [321]. However, several studies have found that friction on fluorinated graphene increases substantially compared with pristine graphene [321, 322]. Figure 18B shows that the friction coefficient is increased for fluorinated graphene compared with pristine graphene, both measured on a polycrystalline copper substrate. Additionally, adhesion was found to decrease on fluorinated graphene compared with pristine graphene [321]. Thus the differing bending stiffness cannot be used to explain the increased friction of fluorinated graphene.

Two mechanisms for the friction observed on fluorinated graphene have been proposed, primarily supported by atomistic simulations. As depicted in Fig. 18D for the first model, the corrugation in the surface energy landscape in fluorinated graphene increases significantly resulting from the change in bonding state of the carbon atoms from sp$^2$ to sp$^3$ [328], as well as the introduction of the highly polarized fluorine atoms on the surface [322, 328]. Thus, the static coefficient of friction, as well as the average kinetic friction force increases substantially from the change in potential energy landscape. The second model, which is more globally applicable to chemically functionalized graphene, including fluorinated graphene and graphene oxide, suggests that friction increases because of an increase in atomic-scale roughness on the surface [329]. However, surface roughening of graphene through chemical modification was observed to saturate at low coverages, and perhaps cannot fully explain the increase in friction observed in Fig. 18 B-C.

Similar to fluorinated graphene, graphene oxide has emerged as an excellent lubricant that is significantly lower in cost than graphene to produce, as it can be produced through chemical exfoliation of graphite oxide [330]. Both the elastic modulus and strength of graphite oxide can be significantly improved by reducing the thickness to atomic dimensions, which effectively reduces the number of defects contained within the material [331]. However, graphene oxide still exhibits higher friction [323, 324] and lower resistance to wear than pristine graphene [324]. The reduction in the ability of graphene oxide to perform as well as pristine graphene was attributed to an increase in the interfacial sheer strength, or the stress required for two layers of graphene oxide to slide past each other [324]. This change in interfacial strength was attributed to a greater number of intercalated functional groups. Furthermore, at low oxidation treatments, the low number of functional groups resulted in a strongly reduced interfacial shear strength. Thus, although the increased bending stiffness and defect density can influence friction, the mechanochemical interaction between the sliding surfaces appears to have a stronger influence on the friction properties of graphene oxide films.

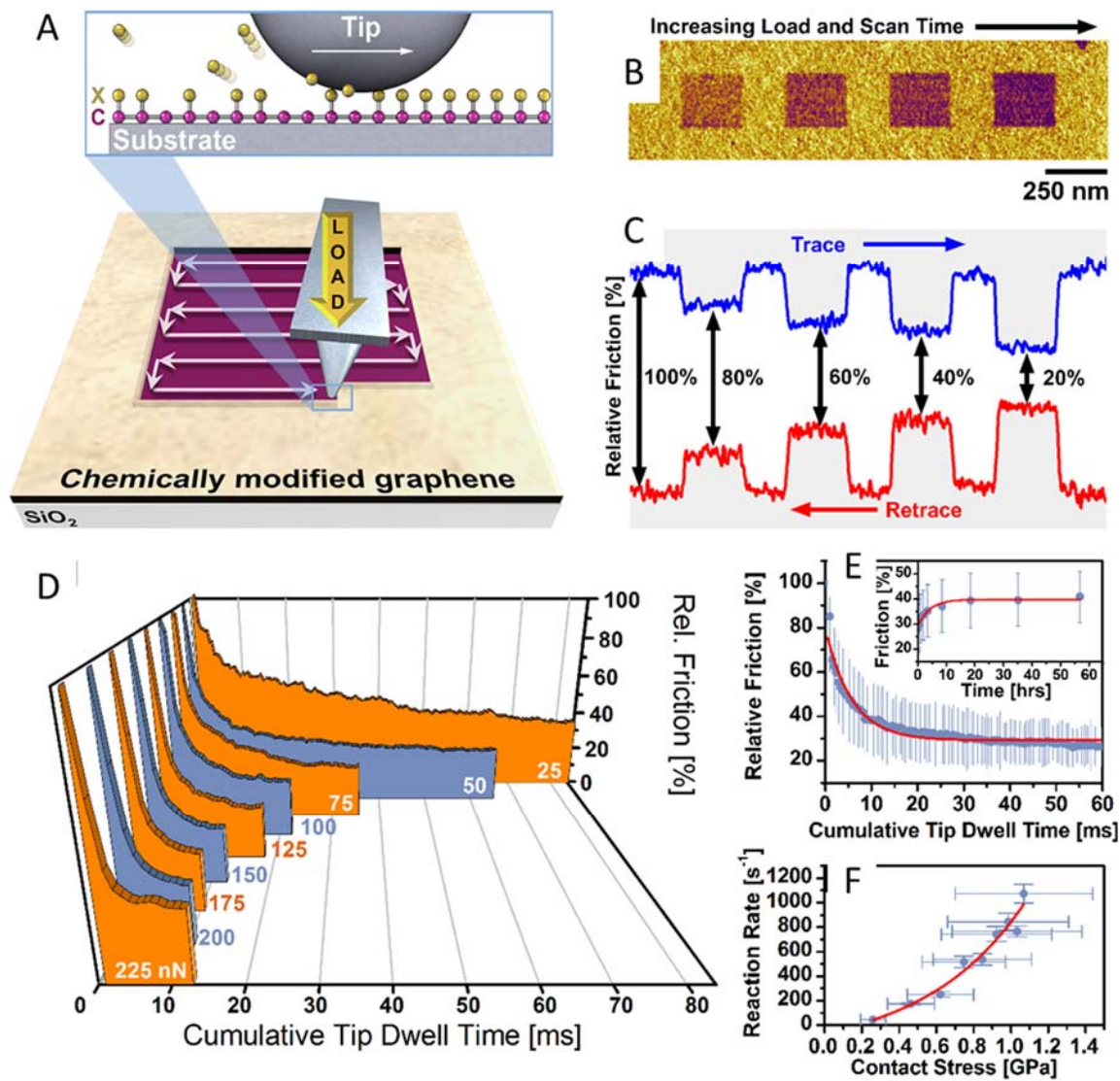

**Fig. 19:** (A) Reduction of chemically modified graphene using load applied with a sliding AFM tip. (B-C) Friction force measurements of reduced graphene oxide. (D) Friction evolution over time for various loads measuring the reduction of graphene oxide in situ. (E) Reduction evolves as a first-order reaction. (F) Reaction rate as a function of applied contact stress. Figures adapted from [332].

The mechanochemical interactions between the scanning probe tip and the functionalized graphene surface also significantly impact wear during the sliding process. A number of recent works on single asperity sliding in various silicon and carbon material systems suggests that tip wear occurs atom-by-atom following an Arrhenius process [333, 334]. Felts et al. [332] investigated how chemically functionalized graphene surfaces wear when in contact with a sliding nano-asperity. Figure 19A shows a schematic of the technique, where a tip scans on a region of chemically modified graphene at a defined normal load, and the composition of the graphene surface is interrogated *in situ via* lateral force friction measurements. Monitoring friction changes over time provides a measure of the dynamics of oxygen group cleavage from the graphene basal plane (Fig. 19 B-D), which develops as a first-order chemical reaction (Fig. 19E) due to the finite concentration of functional groups on the surface. Increasing the stress at the interface dramatically speeds up the removal process (Fig. 19F), and the exponential increase in the wear rate suggests that wear in this system is a stress-assisted chemical reaction.

Although evidence is growing for the stress-assisted chemical reaction model in atomic scale wear, many assumptions inherent to the model remain untested. A modified Arrhenius relationship has been used for the reaction rate in recent experimental studies:

$$k = k_0 \exp\left[\frac{E_a - f(F_b)}{k_B T}\right] \quad (5)$$

where $k_0$ is a pre-factor set by the attempt frequency of the bond, $E_a$ is the thermal activation energy of the bond at zero force, $k_B$ is Boltzmann's constant, $T$ is the temperature, and $f(E_b)$ captures how force applied to a bond modifies the thermal activation barrier. The recent studies of mechanochemical wear use the Zhurkov model to describe the effect of force, $f = \sigma V_a$, where $\sigma$ is the applied stress derived from the applied load using a contact mechanics model and $V_a$ is the activation volume loosely defined as a volume over which the stress acts. Bell's model describes similar bond scission behavior in molecule pulling experiments, which states $f(F_b) = F_b x$, where $F_b$ is the force exerted on the bond and $x$ is the reaction path. In both models, the applied load linearly reduces the thermal activation barrier, which can only be true assuming that the applied load does not significantly alter the physical structure of the molecules under investigation (i.e., the molecular system is infinitely stiff). Relaxing this assumption would lead to more complicated mechanochemical models, where the applied load shifts the observed energy barrier in a strongly non-linear fashion [335].

Unraveling precisely how the mechanochemical wear develops requires increasingly complex experimental measurements that can observe wear rates with atomic resolution, while controlling applied load, the sliding velocity, and tip-surface temperature. Further, fully describing wear as a chemical reaction requires knowledge of what the reactants and products are, and how one evolves towards the other, which would require integration of additional *in situ* spectroscopy techniques [336]. These observations would have a number of important implications for engineering surfaces to reduce friction and wear at the atomic level, where chemical interactions between sliding interfaces play a significant role relative to the mechanical properties of those surfaces that dominate friction and wear at larger scales. MD simulations of atomic-scale friction and wear would also provide critical insights into the temporal and spatial evolution of the precise interaction forces between tip and surface atoms, which could vastly improve atomic-scale contact mechanics models for wear at interfaces [337].

**5.3 van der Waals interactions**

It is well known that individual layers of graphene in bulk graphite are held together by van der Waals (vdW) forces with an equilibrium separation of ~0.335 nm. Similar interactions are expected between graphene and its supporting substrate. Indeed, DFT calculations have confirmed vdW interactions between graphene and silicon dioxide [338, 339], with an adhesion energy of ~0.3 J/m$^2$ and an equilibrium separation of ~0.3 nm. The interaction energy, calculated as a function of separation [339], is minimized at the equilibrium separation but has a relatively long tail (Fig. 20A), revealing the nature of dispersion interactions. Such interactions can be included in MD simulations using the empirical Lennard-Jones (LJ) potential for pairwise particle-particle interactions. In particular, for a monolayer graphene on a silicon oxide substrate, the interaction energy per unit area of graphene can be calculated as

$$U_{LJ}(\delta) = \sum_j \frac{2\pi \rho_j \varepsilon_{ij}}{A_0} \left( \frac{\sigma_{ij}^{12}}{45\delta^9} - \frac{\sigma_{ij}^6}{6\delta^3} \right), \tag{6}$$

where the summation takes both Si-C and O-C interactions into account with the subscript *i* representing C and *j* representing Si or O, $\sigma_{ij}$ and $\varepsilon_{ij}$ are the parameters for the pairwise interactions, $\rho_j$ is the number density of Si or O atoms in the substrate ($\rho_{Si} = 25.0$ nm$^{-3}$ and $\rho_O = 50.0 \text{ nm}^{-3}$ in SiO$_2$), and $A_0$ is the area of a unit cell of graphene. However, it was noted that the adhesion energy is underestimated by several empirical force fields (e.g., UFF, Charmm and Dreiding) as shown in Fig. 20A [339]. Eq. (6) may be further simplified to a two-parameter form as a continuum approximation of the vdW interactions [340]:

$$U_{vdW}(\delta) = \Gamma_0 \left( \frac{\delta_0^9}{2\delta^9} - \frac{3\delta_0^3}{2\delta^3} \right), \tag{7}$$

where $\Gamma_0$ is the adhesion energy and $\delta_0$ is the equilibrium separation.

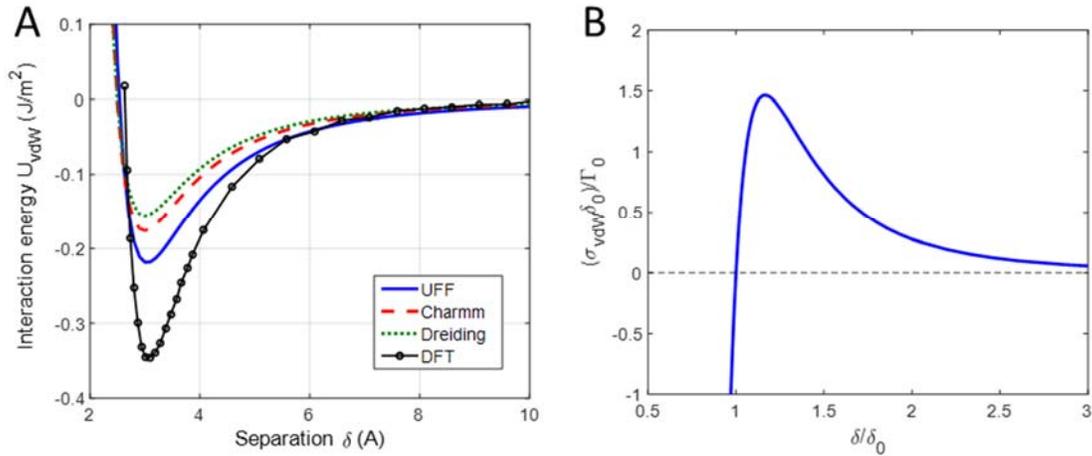

**Fig. 20**: (A) Interaction energy between graphene and silicon oxide, calculated by DFT and empirical force fields. (B) Normalized traction-separation relation with two parameters (adhesion energy $\Gamma_0$ and equilibrium separation $\delta_0$). Figures adapted from (A) [339] and (B) [340].

An immediate application of the continuum approximation in Eq. (7) is the associated traction-separation relation: $\sigma_{vdW}(\delta) = dU_{vdW}/d\delta$, as shown in Fig. 20B. Such a traction-separation relation has been used to model adhesion and delamination of graphene at much larger scales such as graphene bubbles and blisters [277-279, 341]. The peak traction, often called the interfacial (tensile) strength, is directly related to the two basic parameters as: $\sigma_{max} = 1.466\Gamma_0/\delta_0$. The range of vdW interactions typically extends to a few times of the equilibrium separation, although the pull-in instability observed in the island blister test [277] indicated much longer ranges (up to 10-20 nm). The continuum model was also used to predict the morphological corrugation of substrate-supported graphene [340], where the monolayer membrane could be fully conformal, partly conformal or non-conformal to the substrate surface depending on the surface roughness and the adhesive interactions. Moreover, the effective adhesion energy was found to depend on the surface roughness and corresponding graphene morphology [342], which led to apparently lower adhesion energy for multilayered graphene [26].

As discussed in Section 2.3, thermal rippling is inevitable for freestanding graphene. When placed on a solid substrate, the adhesive interactions between graphene and the substrate could considerably suppress thermal rippling. Meanwhile, the statistical nature of thermal rippling introduces an entropic repulsion to the graphene-substrate interactions. By statistical mechanics analysis and MD simulations [343], it was predicted that the equilibrium average separation increases and the effective adhesion energy decreases with increasing temperature, as a result of the entropic effect of thermal rippling. The temperature dependence of adhesion has yet to be confirmed experimentally for graphene and other 2D materials.

By smearing out the discrete structures of graphene and its substrate, the continuum approximation of the vdW interactions predicts zero resistance to graphene sliding along the interface. Experimentally, strain-dependent sliding friction was observed between graphene and silicon oxide [344], which was attributed to the close conformation of graphene to the surface roughness of the substrate. An intimate relationship between adhesion and friction may be established for graphene by considering the effects of surface roughness on the vdW interactions during sliding. On the other hand, relatively low sliding friction of graphene flakes on an atomically flat surface was predicted by atomistic simulations [345], which included other mechanisms (e.g., bond breaking/formation) in addition to vdW interactions.

It is expected that vdW interactions are equally important for interfacial properties of other 2D materials and for interactions between different atomic layers in heterostructures of 2D materials [346]. For example, vdW interactions during epitaxial growth of graphene on h-BN defined the preferential growth directions [347]. The vdW interactions between graphene and h-BN led to a commensurate-incommensurate transition and surface reconstruction with locally stretched graphene separated by incommensurate regions [348]. The so-called self-cleansing mechanism, which led to atomically flat interfaces between 2D materials (free of contamination) [349], was also attributed to relatively strong vdW interactions between certain pairs of 2D materials (e.g., graphene and $MoS_2$) [350]. A pick-and-lift technique was demonstrated to mechanically assemble different 2D materials into a heterostructure [351], which relied on the relative strength of vdW interactions to lift up individual flakes.

## 5.4 Other interaction mechanisms

Besides van der Waals interactions, other interaction mechanisms between graphene and its substrates may be active, as suggested by growing experimental evidence including: (1) A wide range of values have been reported for the adhesion energy of graphene: 0.1 to 0.45 J/m$^2$ on silicon oxide [26, 27, 274, 280], 0.7 to 1.7 J/m$^2$ on seed copper films [285, 286], up to 6 J/m$^2$ on seed copper foils [284], and ~3.4 J/m$^2$ for graphene cured on epoxy [284]. Values of adhesion energy greater than 1 J/m$^2$ cannot be attributed to van der Waals interactions alone. (2) The traction-separation relations extracted from experiments had much lower strength but longer range of interactions compared to those predicted for the van der Waals interactions [27, 284, 286]. (3) The adhesion/separation energy, strength and range of the interactions between graphene and its substrates were found to be dependent on the loading conditions in terms of the fracture mode-mix [283, 352, 353] and loading rate [284].

Theoretical understanding has been lacking on the interaction mechanisms beyond van der Waals. Gao et al. [354] considered the effect of water in their MD simulations and found that the presence of a thin layer of water at the interface could potentially extend the range of interaction by capillary bridging but the adhesion energy would remain low due to the relatively weak interactions between graphene and water. Higher adhesion energy of graphene under mixed-mode conditions was attributed to a toughening mechanism due to asperity locking of rough surfaces [352, 353]. More recently, Kumar et al. [341] suggested that discrete, short-range interactions originating from reactive defects on the surface of silicon oxide could be responsible for the ultrastrong adhesion as well as the high shear strength against sliding under mode II conditions. The rate effect, while not well understood at the moment, has facilitated selective transfer of graphene from seed copper foil [284]. It is most likely related to an interphase region that develops in the epoxy close to the graphene membrane, whose properties are quite different from those of the bulk epoxy [355].

While the interactions between graphene and target substrates may be complicated by capillary and contamination effects, the interaction between CVD graphene and its seed copper is expected to be due solely to vdW forces. However, the separation energy of 6 J/m$^2$, strength of 1 MPa and range of 24 μm obtained from the experiment [284] are far removed from the characteristics of vdW forces. It was noted that the CVD graphene conforms almost perfectly to the copper foil surface, forming regular ridges with a root-mean-square (RMS) roughness of about 10 nm within each copper grain. The surface roughness of the seed copper foil increases at larger spatial scales due to features produced during the fabrication of the foil itself. At millimeter scales, the RMS roughness of copper foil is about 0.8 μm. The surface roughness of graphene grown on copper films deposited on silicon wafers is much smaller and so was the separation energy [286]. Thus the effect of rough surfaces needs to be considered in further studies to close the gap between experimental data and the current understanding of interactions between 2D materials and their substrates (both seed and target).

Surface roughness appears to have played a role in the variation of interfacial toughness with fracture mode-mix between graphene and a target copper or silicon substrate [352]. Blister tests are inherently mixed mode in nature with both normal and shear tractions at the interface near the crack front [356]. The phase angle ($\psi$) of fracture mode-mix is generally defined by the ratio of the shear traction $\tau$ to the normal traction $\sigma$ (tension) at the interface, i.e., $\tan\psi = \tau/\sigma$. By varying the thickness of the backing layers in their blister tests, Cao et al. [352] showed that the interfacial toughness between graphene and both substrates (Cu and Si) had a strong dependence on the fracture mode-mix as shown in Fig. 21. Similar to the interfacial fracture between bulk materials [357], the interfacial toughness of graphene increases with increasing mode-mix for both positive and negative shear tractions. In the absence of plasticity effects, the most likely explanation of this effect is asperity locking [358] due to the surface roughness of the substrates.

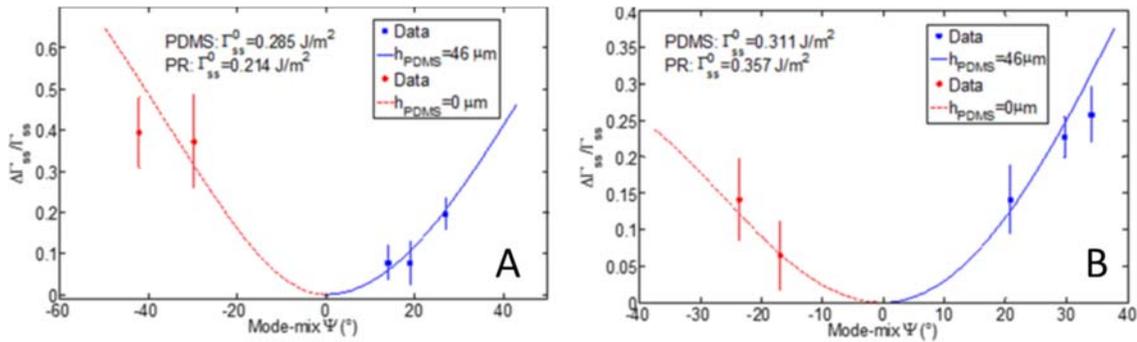

**Fig. 21**: The dependence of interfacial toughness on mode-mix for interactions between (A) graphene and polished copper, (B) graphene and silicon. The steady state toughness is a function of the mode-mix phase angle as: $\Gamma_{SS}(\psi) = \Gamma_{SS}^0 + \Delta\Gamma_{SS}(\psi)$, where $\Gamma_{SS}^0$ is the mode-I toughness for $\psi = 0$. Figures adapted from [352].

Recently, Cao et al. [353] devised a scheme to extract the mixed-mode traction-separation relations by blister tests. Graphene grown by CVD was backed by a photoresist film and transferred to a highly polished copper substrate from its seed copper foil. The graphene/photoresist composite film was then pressurized with deionized water through a hole in the substrate. The blister profiles and normal crack opening displacements (NCOD) were measured by two microscopes with synchronized cameras. Different mixed-mode conditions were again achieved by varying the thickness of the epoxy backing layer. Cohesive zone models associated with traction-separation relations were then developed to study the damage initiation and crack propagation under various mixed-mode conditions. The interactions between graphene and copper were found to be stronger in all respects than those associated with photoresist and copper. Because the monolayer graphene was sandwiched between photoresist and copper, this result suggested that graphene was not transparent to interactions between photoresist and copper, but opaque.

The mixed-mode interactions in general range from pure mode I ($\psi = 0$) to pure mode II ($\psi = \pm 90°$) cases. Traction-separation relations for nominally mode-I interactions between CVD graphene and the native oxide surface of silicon substrates were obtained by the DCB experiments (Fig. 17) [27]. In the pure mode II case, the normal traction at the interface is zero or negative (compression), while the shear traction is related to the relative sliding displacement, similar to friction. Like the normal adhesive interactions (mode I), the friction-like shear interactions (mode II) can also be described by traction-separation relations. For example, Jiang et al. [281] assumed a linear relation followed by a constant shear traction in their analysis of interfacial sliding of monolayer graphene flakes on a stretchable substrate (PET). Based on their measurements using *in-situ* Raman spectroscopy (Fig. 22), the interfacial shear strength was found to range between 0.46 and 0.69 MPa. Moreover, the maximum strain that can be transferred to graphene by stretching the substrate depends on the interfacial shear strength and the graphene memberane size. More recently, a similar traction-separation relation for shear interactions was used to understand cracking of polycrystalline graphene on copper foil under tension [359], where regularly spaced channel cracks were observed in the graphene during tension tests in a scanning electron microscope (SEM). The density of the channel cracks increased with increasing strain applied to the copper foil until a minimum spacing was reached. The increasing crack density was understood as a result of sequential channel cracking as in elastic thin films [360]. The minimum crack spacing was related to the interfacial shear strength by a classical shear lag analysis, yielding a shear strength of 0.49 MPa for the CVD graphene on copper foil. This value is similar to those obtained by Jiang et al. [281] for exfoliated graphene flakes on PET, although the underlying mechanisms may differ in the two cases. Further studies are needed to understand the mechanisms and to unify the normal (adhesive) and shear (frictional) interactions within a general framework of mixed-mode interactions for graphene and other 2D materials.

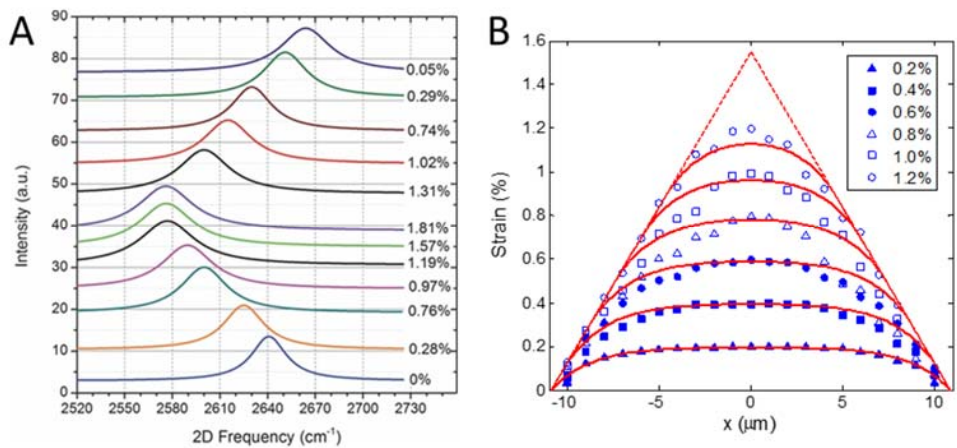

**Fig. 22**: (A) Evolution of the 2D Raman spectrum of a monolayer graphene as a function of the applied strain to the PET substrate. (B) Strain distributions in a monolayer graphene, comparing the Raman measurements (symbols) with a nonlinear shear lag analysis (lines). Figures adapted from [281].

## 6. Applications

This section briefly reviews a few applications of 2D materials where mechanics and mechanical properties play important roles, including synthesis and transfer, graphene origami and kirigami, flexible and biomedical applications.

### 6.1 Synthesis and transfer

When graphene was first isolated by stripping it from graphite with scotch tape, it was only available as small flakes with in-plane dimensions on the order of tens of microns [361] and there was little control over monolayer production. While such sizes were sufficient for establishing many of the unique properties of graphene, the extension to larger areas was clearly desirable in order to meet the expectations raised by such success. CVD turned out to be the most successful method of producing large area, monolayer graphene and was accomplished on thin (~35 μm) copper foils [362-364] and related metal foils such as commercially available Cu-Ni foils [318] to produce large area graphene, up to meters in the in-plane dimension [365]. Graphene produced in this way is compatible with roll-to-roll manufacture of graphene and its integration with flexible electronics applications. Graphene has also been grown on copper films (~1 μm thick) [366] that has been deposited on silicon wafers. In this case, the objective is to integrate graphene with very-large-scale integration (VLSI) electronics applications. Integration of graphene devices with silicon technology can benefit from both the maturity of silicon technology and the outstanding electronic, optical, thermal and mechanical properties of graphene. However, the high temperature growth of graphene is not compatible with silicon processing [121, 367, 368]. For these reasons, controlled transfer of graphene from its growth surface onto target substrates is crucial for enabling graphene to be integrated with silicon technology.

Transferring large-area graphene to its target substrate was first achieved by "wet transfer" where the seed copper foil is etched away [275]. Another option is an electrochemical process [369] that generates bubbles at the graphene/copper interface and separates the graphene from the foil. These processes have been incorporated in the production of large-scale graphene on polymer films as transparent electrodes for flexible applications [370-373]. However, etching is wasteful of copper, and both processes are relatively slow and may contaminate the graphene. Thus, a return to dry transfer as embodied in the original scotch tape method is attractive.

A few reports of dry transfer have appeared in the literature. The first one by Yoon et al. [285] demonstrated the possibility of transferring graphene from a copper film on silicon to epoxy using a DCB setup. Measurement of a separation energy of ~0.7 J/m$^2$ was also made in the process. By exploring a wide range of separation rates on a graphene coated copper foil sandwiched between silicon strips with epoxy, it was possible to control delamination paths to occur along the graphene/epoxy interface at relatively low separation rates but along the graphene/copper interface at higher rates [284]. It was postulated that the selective delamination was made possible by the rate dependence of the interactions between graphene and epoxy. As with other substrates [355], an interphase region is formed in the epoxy close to the graphene monolayer, whose mechanical and adhesive properties differ from the bulk. In fact, the interfacial toughness increases with rate, opposite to the behavior of bulk epoxy, to such an extent that the separation energy and strength of the graphene/epoxy interface becomes greater than that of the graphene/copper interface, thereby causing the latter interface to fail. At smaller spatial scales, the use of polymers and their rate dependent separation energy has been exploited for controlled dry transfer of flakes of 2D materials [374]. This parallels the development of pick and place strategies for dry transfer printing in nanomanufacturing that rely on so-called kinetic effects [375, 376]. Another fracture mechanics concept that is employed for selective delamination in transfer printing is to exploit the difference in separation energy under tension and shear [377, 378]. Some ground work for exploiting this for the transfer of 2D materials has been laid by Cao et al. [283, 352, 353], where the separation energies of graphene that had been transferred to silicon and polished copper, with RMS roughness values of 0.5 and 4.5 nm, respectively, increased with increasing shear component (Fig. 21). Such effects of tension and shear can readily be controlled in roll-to-roll transfer nanomanufacturing schemes by changing the angles of incoming and exiting feedstock.

Furthermore, for roll-to-roll transfer from seed copper, the limiting strain level (~0.5%) that polycrystalline graphene can tolerate before it starts to crack was established recently by applying tension to a graphene-coated copper foil and observing the successive formation of channel cracks in the graphene [359]. The mechanics of graphene cracking is related to the shear interactions between graphene and seed copper. Measurements of the crack spacing allowed the stiffness and strength of the shear interaction and the fracture toughness of graphene itself to be determined.

For integration of graphene with VLSI electronics, it has also been established that fracture mechanics concepts can be exploited for selective delamination [286]. Graphene was grown on a silicon wafer that was coated with a copper film prior to deposition. In order to effectively transfer the graphene to a target silicon wafer, the graphene was bonded to the target wafer using an epoxy and then separated in a double cantilever beam (DCB) setup. Graphene/copper or copper/silicon oxide delamination paths could be selected by slow and fast separation rates, respectively. Thus graphene can be transferred to a target wafer, either exposed or protected by the seed copper film, which can later be removed by etching. Delamination paths were identified by SEM and Raman spectroscopy. The sheet resistance of the graphene produced by the two approaches (exposed or protected) was slightly higher than graphene transferred by the wet-transfer process, indicating reduced impurity doping, and the variation in the sheet resistance values was much lower. Copper contamination levels, quantitatively established by Time-of-Flight Secondary Ion Mass Spectrometry (ToF-SIMS), were several orders of magnitude lower than the values for wet transfer. In addition, it was demonstrated that top-gated transistor devices from delamination-transferred graphene exhibited superior transistor behavior to PMMA-assisted wet transfer graphene. The separation energy, strength and range of the interactions were quantitatively determined by nonlinear fracture mechanics analyses, which again suggest that the roughness of the interface between graphene and copper plays an important role with implications for further improvements in the manufacturing processes.

## 6.2 Graphene origami and kirigami

Extra-large surface-to-volume ratio renders graphene a highly flexible morphology, giving rise to intriguing observations such as ripples, wrinkles and folds [85, 379-382]. Such a malleable nature of graphene makes it a potential candidate material for nanoscale origami and kirigami, a promising bottom-up nanomanufacturing approach to fabricating nano-building blocks of desirable shapes beyond conventional material preparation techniques [379, 382-389]. The success of graphene origami and kirigami hinges upon precise and facile control of graphene morphology, which remains as a significant challenge. Nonetheless, recent progress in patterning graphene with atomic-scale precision has further paved the way toward achieving graphene origami and kirigami in a programmable fashion [9, 172, 175, 223, 390-394].

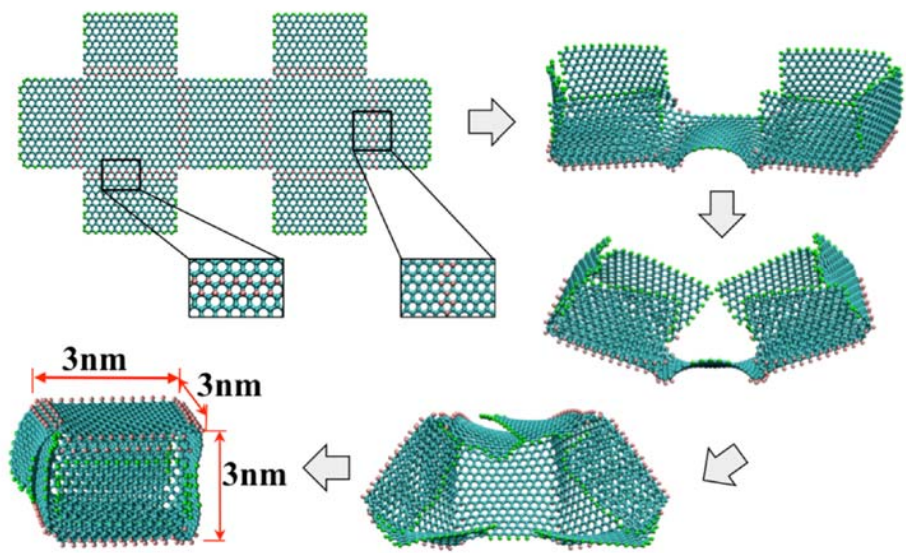

**Fig. 23**: A monolayer graphene patterned into a double-cross shape and suitably hydrogenated (insets) can spontaneously fold and form a graphene nanocage. Figure adapted from [394].

The 2D nature of graphene makes the chemical functionalization of graphene a promising approach to modulating the graphene properties. For example, hydrogenation of graphene involves bonding atomic hydrogen to the carbon atoms in graphene [395]. Such a reaction changes the hybridization of the C-C bonds in graphene from $sp^2$ into $sp^3$. As a result, the 2D atomic structure of pristine graphene is distorted [389]. It has been shown that suitable single-sided hydrogenation of graphene can lead to folding of graphene in a programmable fashion. This feature can be leveraged to achieve the hydrogenation assisted graphene origami (HAGO), in which initially planar, suitably patterned graphene can self-assemble into three dimensional nanoscale objects of desirable shapes (Fig. 23) [394]. A unique feature of the HAGO process is that the resulting nanostructure can be modulated by an external electric field, enabling programmable opening and closing of the nano-objects, a desirable feature to achieve molecular mass manipulation, storage and delivery. For example, MD simulations have demonstrated HAGO-enabled nanocages for controllable uptake and release of nanoparticles as well as ultra-high density of hydrogen storage.

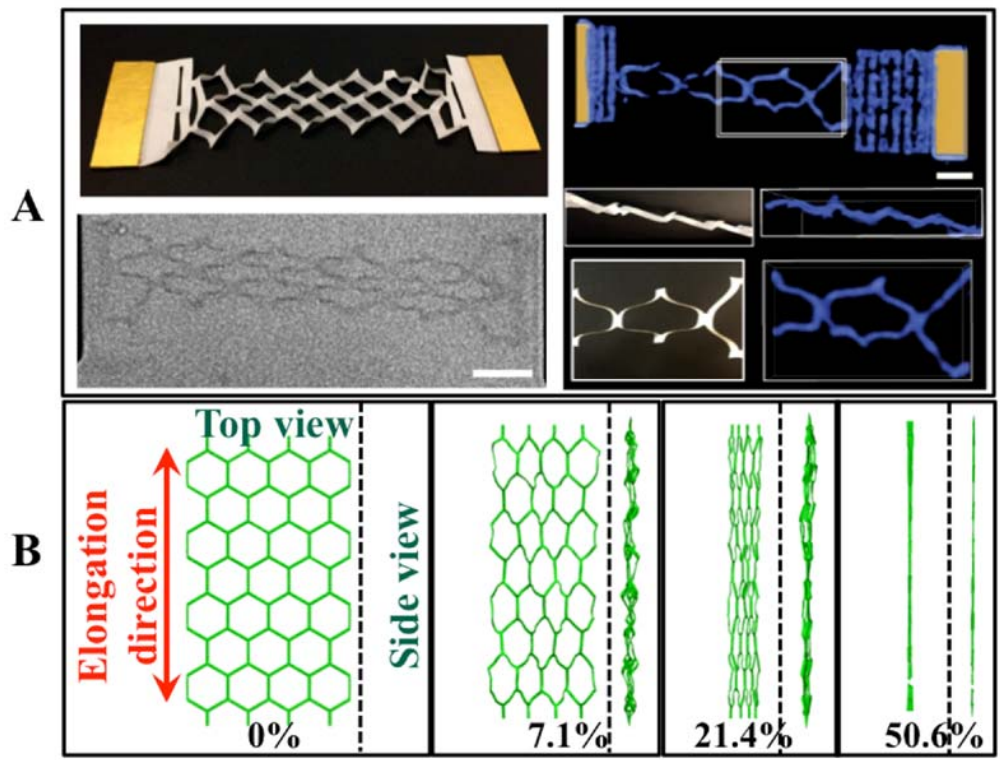

**Fig. 24**: A. (Left) Paper and graphene in-plane kirigami springs, respectively. (Right) Three-dimensional reconstruction of the deformed graphene spring under large elongation. B. Deformation sequence of a graphene nanomesh subject to an elongation beyond 50%. Figures adapted from: (A) [9] and (B) [223].

Recent experiments showed that monolayer graphene can be patterned via optical lithography into desirable shapes, such as a spiral spring, a kirigami pyramid, a cantilever or a nanomesh [9, 223]. Such a patterned graphene structure is planar as fabricated, but can deflect and twist out of the plane when stretched. The out-of-plane deformation can accommodate huge in-plane elongation without substantial strain in a suitably patterned structure [396]. As a result, the patterned graphene achieves significant elastic stretchability [9, 175, 223], way beyond the elastic limit of pristine graphene (Fig. 24). For example, systematic coarse-grained simulations revealed that a suitably patterned graphene nanomesh can be made extremely compliant with nearly zero stiffness up to about 20% elongation and then remain highly compliant up to about 50% elongation (Fig. 24B). These features of graphene kirigami are desirable for graphene-based functional devices such as epidermal electronics and sensing prosthesis [223] as well as stretchable and tunable quantum dot arrays [176].

**6.3 Biomedical applications**

Unlike conventional wafer-based electronics, which is rigid, planar, and brittle, biological systems are soft, curvilinear, and dynamic. The emergence of 2D materials can effectively bridge this gap. On the one hand, 2D materials have superior electronic performance that are on par or even better than silicon and metal. On the other hand, their mechanical properties such as flexibility and stretchability allow them to intimately integrate with the soft and curvilinear bio-systems without causing significant disruptions such as rupture and detachment or imposing any mechanical constraint. As a result, 2D materials have been increasingly sought after for biomedical applications over the past few years.

The bio-compatibility, electrical conductivity, flexibility, and transparency afforded by graphene have enabled its applications in bio-integrated soft electronics. For example, serpentine-shaped stretchable graphene interconnects, resistance temperature detectors and strain gauges have been integrated on transparent epidermal electronic systems for noninvasive physiology monitoring [397]. As for *in vivo* sensing such as neural interfacing, flexible and transparent micro-electrode array enabled by graphene could allow for simultaneous electrophysiology and optical imaging, as well as optogenetic modulation of the underlying brain tissue [398, 399]. In tissue engineering, graphene nanoribbons supported by ultrasoft PDMS have been demonstrated as a soft cell-culture platform which is capable of aligning plated cells as well as *in situ* monitoring of cellular physiological characteristics during proliferation and differentiation [400].

As a semiconductor, monolayer $MoS_2$ exhibits superior piezoresistive properties, meaning that its resistivity can change significantly with mechanical deformation due to its strain-dependent bandgap [246], as mentioned in Section 4.2. This enabled a recent development of an ultra-sensitive and ultra-conformable, transparent tactile sensor for human-mimetic electronic skins [401]. The $MoS_2$-based tactile sensor remained intact and functional at a strain level of 2% for 10,000 loading cycles.

Beyond physical and physiological sensors, electrochemical sensors and biosensors based on graphene and other 2D materials, such as boron nitride (BN), graphite-carbon nitride (g-$C_3N_4$), transition metal dichalcogenides (TMDs), transition metal oxides, and graphane have been applied for the detection of important biomarkers such as glucose, hydrogen peroxide, and other biomarkers [402]. Examples include graphene-based enzymatic [397] and non-enzymatic [403] glucose sensors, graphene-based enzymatic [404] and non-enzymatic [405] hydrogen peroxide sensors, $MoS_2$ based enzymatic and non-enzymatic biosensors for glucose detection [406], as well as graphene oxide (GO) or reduced GO (rGO) [407] and $MoS_2$ [406] based immunosensors for detection of dopamine.

In addition to bio-sensing, 2D materials have also been demonstrated as promising nano-platforms for therapy and diagnostic imaging attributing to their large surface-area-to-mass ratio and unique physicochemical properties [408]. For example, biocompatible graphene derivatives, such as GO and rGO, have been widely applied for anticancer drug delivery [409], gene transportation [410], photothermal therapy (PTT) [411], as well as photodynamic therapy (PDT) [412]. These GO-based therapeutics showed superior performances compared with other conventional nanostructures such as particles, tubes, wires, and cages. Graphene-based hybrid nanomaterials by integrating graphene with other functional nanomaterials have been developed for diagnostic imaging such as fluorescent imaging [413], magnetic resonance imaging (MRI) [414], computed tomography (CT) [415], and radionuclide imaging [416].

In parallel to the intensive studies of promising biomedical applications of 2D materials, a significant effort has also been devoted to the understanding of biological and environmental impacts of 2D materials [417, 418], with the aim to safely harness their application potentials as well as to control their cytotoxicity to living creatures. It has been shown that 2D materials exhibit unique interaction mechanisms with cell membranes due to their high aspect ratio, sheet-like nano-structures and atomically sharp edges [419-421]. For example, experimental and simulation studies have shown that sharp corners or edge asperities can enable graphene to spontaneously penetrate a cell membrane [419, 420]. Lipid extraction by graphene and graphene oxide was recently identified as an important damage mechanism to cell membranes [421] and intracellular vesicles [422]. On the other hand, the toxicity of graphene and other 2D materials, if well controlled, may also provide potential therapeutics by utilizing their cytotoxicity against bacteria cells [418, 421].

## 6.4 Flexible applications

2D materials are naturally suited for flexible, stretchable, foldable, and wearable devices owing to their atomic thickness that offers many sought after attributes including maximum optical transparency, optimum electrostatic control, high strain limit, and large surface to volume ratio [423, 424]. These attributes benefits a wide range of applications such as electronics, photonics, sensors, and mechanical and energy devices [2]. Furthermore, 2D materials can be dispersed in a host of solvents to make functional inks for printed electronics [231]. The individual inks can serve the function of semiconductors, dielectrics, and electrodes; components needed to make advanced printed electronic systems.

In recent years, advances in large-area manufacturing (synthesis and transfer) of 2D materials, particularly graphene, has led to the introduction of the first consumer products, where graphene is used as a transparent conductive film, replacing indium tin oxide (ITO) in the touch panels of smartphones. The first graphene touch-panel smartphone was released in 2014 by 2D Carbon [425]. Subsequently, a collaboration between Moxi and Galapad resulted in a graphene touch-panel smartphone that also featured graphene in other smartphone components for cooling and energy benefits [426]. In the near-term, flexible/bendable touch panels are expected as graphene technology matures. In the long-term, active devices based on highly integrated 2D circuits (beyond graphene) are likely to emerge and usher in complex high-performance flexible nanosystems.

Towards this end, basic research on lab-scale flexible devices and circuits have been investigated and developed over the past 10 years [423]. This sustained research has led to many notable achievements such as: (i) ~100 GHz graphene transistors on flexible glass [427], (ii) 20 GHz BP transistors that also have sufficient bandgap to be employed for both analog and high-speed digital circuits [428], and (iii) low-power radio-frequency $MoS_2$ transistors and circuits operating in the GHz regime suitable for the internet of things applications on low-cost plastic substrates [429]. It must be emphasized that the flexible applications of 2D materials appears to be compelling for high-density large-area nanosystems. For flexible applications where a few discrete active devices are needed, contemporary thin-film transistor solutions based on metal oxides or bulk semiconductors are more suitable. With this in mind, research to advance large-area manufacturing and integration technology of the portfolio of 2D materials is essential and a matter that requires sustained effort for the foreseeable future.

## 7. Summary and Outlook

The family of 2D materials has grown beyond graphene, and together they hold great promise for a wide range of applications. The mechanics and mechanical properties of 2D materials play important roles in many applications including large-scale manufacturing and integration. Fundamental research on the mechanics and mechanical properties of 2D materials has made significant progress over the last decade, both theoretically and experimentally. As an outlook for future studies, a few topics of interest are summarized as follows.

Although the elastic properties of 2D materials have been thought to be well predicted by first-principles calculations, recent experiments have suggested significant effects of temperature and rippling, for both in-plane and bending stiffnesses. Further experiments are needed to confirm such effects. Theoretically, statistical mechanics approaches may be developed along with MD simulations to predict the elastic behavior of 2D materials at finite temperatures.

The presence of various defects has profound influence on the mechanical properties of 2D materials. Defect engineering may be explored to tailor the strength and toughness of 2D materials, but a few fundamental questions have yet to be addressed regarding fracture mechanics of 2D materials from both continuum and atomistic points of view.

Strain engineering has a great potential for 2D materials, where the mechanics and mechanical properties are inherently coupled with other physical properties. Scaling up from the first-principles calculations, a multiphysics-based theoretical framework may be developed to couple the mechanics of 2D materials with pseudomagnetic fields, phase transitions, phonon and electronic structures. Experimental methods may be further developed to uncover the coupling phenomena and characterize the coupling properties (such as the piezo- and flexo-electric coefficients).

For the interfacial properties of 2D materials, theoretical understanding beyond van der Waal interactions is lacking. Among others, the effects of surface roughness and capillary bridging need to be better understood. A generally mixed-mode traction-separation relation may be developed to unify adhesion and friction at the interface. Both theoretical and experimental studies are needed to unravel the mechanochemical interactions that lead to wear of the 2D materials.

Finally, novel applications that take advantage the superior mechanical properties of 2D materials will continue to grow out of the fundamental research. Of particular interest is the structural design utilizing the concept of origami and kirigami, which may find unprecedented applications for flexible and biomedical devices.

## Acknowledgements

This review results from discussions at the AmeriMech Symposium on Mechanical Behavior of 2D Materials – Graphene and Beyond, which was held at the University of Texas at Austin on April 4-6, 2016. We gratefully acknowledge financial support of the symposium by the National Science Foundation through Grant No. 1625862 and the National Academy of Sciences through the US National Committee of Theoretical and Applied Mechanics (USNCTAM) with Grant No. 2000006243.

**References**

1. Novoselov, K.S., et al., *Two-dimensional atomic crystals.* Proceedings of the National Academy of Sciences, 2005. **102**: p. 10451-10453.
2. Bhimanapati, G.R., et al., *Recent advances in two-dimensional materials beyond graphene.* ACS Nano, 2015. **9**: p. 11509–11539.
3. Butler, S.Z., et al., *Progress, challenges, and opportunities in two-dimensional materials beyond graphene.* ACS Nano, 2013. **7**: p. 2898-2926.
4. Lee, C., et al., *Measurement of the elastic properties and intrinsic strength of monolayer graphene.* Science, 2008. **321**: p. 385-388.
5. Cooper, R.C., et al., *Nonlinear elastic behavior of two-dimensional molybdenum disulfide.* Physical Review B, 2013. **87**: p. 035423.
6. Bertolazzi, S., J. Brivio, and A. Kis, *Stretching and breaking of ultrathin MoS2.* ACS Nano, 2011. **5**: p. 9703-9709.
7. López-Polín, G., et al., *Increasing the elastic modulus of graphene by controlled defect creation.* Nature Physics, 2015. **11**: p. 26-31.
8. Lindahl, N., et al., *Determination of the bending rigidity of graphene via electrostatic actuation of buckled membranes.* Nano Lett, 2012. **12**: p. 3526–3531.
9. Blees, M.K., et al., *Graphene kirigami.* Nature, 2015. **524**: p. 204-207.
10. Kosmrlj, A. and D.R. Nelson, *Response of thermalized ribbons to pulling and bending.* Physical Review B, 2016. **93**: p. 125431.
11. Gao, W. and R. Huang, *Thermomechanics of monolayer graphene: Rippling, thermal expansion and elasticity.* Journal of the Mechanics and Physics of Solids, 2014. **66**: p. 42-58.
12. Banhart, F., J. Kotakoski, and A.V. Krasheninnikov, *Structural defects in graphene.* ACS Nano, 2011. **5**: p. 26–41.
13. Lee, G.-H., et al., *High-strength chemical-vapor–deposited graphene and grain boundaries.* Science, 2013. **340**: p. 1073-1076.
14. Rasool, H.I., et al., *Measurement of the intrinsic strength of crystalline and polycrystalline graphene.* Nature Commun, 2013. **4**: p. 2811.
15. Zhang, P., et al., *Fracture toughness of graphene.* Nature Commun, 2014. **5**: p. 3782.
16. Zhang, T., X. Li, and H. Gao, *Fracture of graphene: a review.* Int. J. Fracture, 2015. **196**: p. 1-31.
17. Amorim, B., et al., *Novel effects of strains in graphene and other two dimensional materials.* Physics Reports, 2016. **617**: p. 1-54.
18. Levy, N., et al., *Strain-induced pseudo-magnetic fields greater than 300 Tesla in graphene nanobubbles.* Science, 2010. **329**: p. 544-547.
19. Klimov, N.N., et al., *Electromechanical properties of graphene drumheads.* Science, 2012. **336**: p. 1557-1561.
20. Duerloo, K.-A.N., Y. Li, and E.J. Reed, *Structural phase transitions in two-dimensional Mo- and W-dichalcogenide monolayers.* Nature Commun, 2014. **5**: p. 4214.
21. Duerloo, K.-A.N. and E.J. Reed, *Structural phase transitions by design in monolayer alloys.* ACS Nano, 2016. **10**: p. 289-297.
22. Nayak, A.P., et al., *Pressure-dependent optical and vibrational properties of monolayer molybdenum disulfide.* Nano Lett, 2015. **15**: p. 346-353.
23. Pandey, T., et al., *Pressure-induced charge transfer doping of monolayer graphene/MoS2 heterostructure.* Small, 2016. **12**: p. 4063-4069.
24. Zelisko, M., et al., *Anomalous piezoelectricity in two-dimensional graphene nitride nanosheets.* Nature Commun, 2014. **5**: p. 4284.

25. Naumov, I., A.M. Bratkovsky, and V. Ranjan, *Unusual flexoelectric effect in two-dimensional noncentrosymmetric sp2-bonded crystals.* Phys Rev Lett, 2009. **102**: p. 217601.
26. Koenig, S.P., et al., *Ultrastrong adhesion of graphene membranes.* Nature Nanotechnology, 2011. **6**: p. 543-546.
27. Na, S.R., et al., *Ultra long-range interactions between large area graphene and silicon.* ACS Nano, 2014. **8**: p. 11234-11242.
28. Lu, Q. and R. Huang, *Nonlinear mechanics of single-atomic-layer graphene sheets.* International Journal of Applied Mechanics, 2009. **1**: p. 443-467.
29. Lu, Q., M. Arroyo, and R. Huang, *Elastic bending modulus of monolayer graphene.* Journal of Physics D: Applied Physics, 2009. **42**: p. 102002.
30. Wei, X. and J.W. Kysar, *Experimental validation of multiscale modeling of indentation of suspended circular graphene membranes.* Int. J. Solids Struct. , 2012. **49**: p. 3201–3209.
31. Ruiz-Vargas, C.S., et al., *Softened elastic response and unzipping in chemical vapor deposition graphene membranes.* Nano Lett, 2011. **11**: p. 2259-2263.
32. Song, L., et al., *Large scale growth and characterization of atomic hexagonal boron nitride layers.* Nano Lett, 2010. **10**: p. 3209-3215.
33. López-Polín, G., et al., *Strain dependent elastic modulus of graphene.* arXiv:1504.05521, 2015.
34. Los, J.H., A. Fasolino, and M.I. Katsnelson, *Scaling behavior and strain dependence of in-plane elastic properties of graphene.* Phys Rev Lett, 2016. **116**: p. 015901.
35. Song, Z. and Z. Xu, *Geometrical effect 'stiffens' graphene membrane at finite vacancy concentrations.* Extreme Mechanics Letters, 2016. **6**: p. 82-87.
36. Freund, L.B. and S. Suresh, *Thin Film Materials: Stress, Defect Formation and Surface Evolution*. 2003: Cambridge University Press.
37. Lloyd, D., et al., *Band gap engineering with ultralarge biaxial strains in suspended monolayer MoS2.* Nano Lett, 2016. **16**: p. 5836–5841.
38. Nicholl, R.J.T., et al., *The effect of intrinsic crumpling on the mechanics of free-standing graphene.* Nature Commun, 2015. **6**: p. 8789.
39. Castellanos-Gomez, A., et al., *Elastic properties of freely suspended MoS2 nanosheets.* Advanced Materials, 2012. **24**: p. 772–775.
40. Castellanos-Gomez, A., et al., *Mechanical properties of freely suspended atomically thin dielectric layers of mica.* Nano Research, 2012. **5**: p. 550-557.
41. Wang, P., et al., *Numerical analysis of circular graphene bubbles.* Journal of Applied Mechanics, 2013. **80**: p. 040905.
42. Castellanos-Gomez, A., et al., *Mechanics of freely-suspended ultrathin layered materials.* Ann. Phys. (Berlin), 2015. **527**: p. 27-44.
43. Nicklow, R., N. Wakabayashi, and H.G. Smith, *Lattice dynamics of pyrolytic graphite.* Physical Review B, 1972. **5**: p. 4951-4962.
44. Koskinen, P. and O.O. Kit, *Approximate modeling of spherical membranes.* Physical Review B, 2010. **82**: p. 235420.
45. Zhang, D.B., E. Akatyeva, and T. Dumitrica, *Bending ultrathin graphene at the margins of continuum mechanics.* Phys Rev Lett, 2011. **106**: p. 255503.
46. Kudin, K.N., G.E. Scuseria, and B.I. Yakobson, *C2F, BN, and C nanoshell elasticity from ab initio computations.* Physical Review B, 2001. **64**: p. 235406.
47. Liu, F., P. Ming, and J. Li, *Ab initio calculation of ideal strength and phonon instability of graphene under tension.* Physical Review B, 2007. **76**: p. 064120.
48. Wei, X., et al., *Nonlinear elastic behavior of graphene: Ab initio calculations to continuum description.* Physical Review B, 2009. **80**: p. 205407.

49. Xu, M., et al., *A constitutive equation for graphene based on density functional theory.* Int. J. Solids Struct., 2012. **49**: p. 2582–2589.
50. Kumar, S. and D.M. Parks, *On the hyperelastic softening and elastic instabilities in graphene.* Proc. R. Soc. A 2015. **471**: p. 20140567.
51. Brenner, D.W., *Empirical potential for hydrocarbons for use in simulating the chemical vapor deposition of diamond films.* Physical Review B 1990. **42**: p. 9458-9471
52. Brenner, D.W., et al., *A second-generation reactive empirical bond order (REBO) potential energy expression for hydrocarbons.* Journal of Physics: Condensed Matter 2002. **14**: p. 783-802.
53. Stuart, S.J., A.B. Tutein, and J.A. Harrison, *A reactive potential for hydrocarbons with intermolecular interactions.* J. Chem. Phys., 2000. **112**: p. 6472–6486.
54. Arroyo, M. and T. Belytschko, *Finite crystal elasticity of carbon nanotubes based on the exponential Cauchy-Born rule.* Physical Review B, 2004. **69**: p. 115415.
55. Huang, Y., J. Wu, and K.C. Hwang, *Thickness of graphene and single-wall carbon nanotubes.* Physical Review B, 2006. **74**: p. 245413.
56. Lindsay, L. and D.A. Briodo, *Optimized Tersoff and Brenner empirical potential parameters for lattice dynamics and phonon thermal transport in carbon nanotubes and graphene.* Physical Review B 2010. **81**: p. 205441.
57. Lu, Q., W. Gao, and R. Huang, *Atomistic simulation and continuum modeling of graphene nanoribbons under uniaxial tension.* Modelling and Simulation in Materials Science and Engineering, 2011. **19**: p. 054006.
58. Wei, Y., et al., *Bending rigidity and gaussian bending stiffness of single-layered graphene.* Nano Lett, 2013. **13**: p. 26-30.
59. Los, J.H., et al., *Improved long-range reactive bond-order potential for carbon. I. Construction.* Physical Review B 2005. **72**: p. 214102.
60. Peng, Q., W. Ji, and S. De, *Mechanical properties of the hexagonal boron nitride monolayer: Ab initio study.* Computational Materials Science, 2012. **56**: p. 11-17.
61. Li, T., *Ideal strength and phonon instability in single-layer MoS2.* Physical Review B, 2012. **85**: p. 235407.
62. Zhao, H., *Strain and chirality effects on the mechanical and electronic properties of silicene and silicane under uniaxial tension.* Physics Letters A, 2012. **376**: p. 3546-3550.
63. Peng, Q., X. Wen, and S. De, *Mechanical stabilities of silicene.* RSC Advances, 2013. **3**: p. 13772-13781.
64. Wei, Q. and X. Peng, *Superior mechanical flexibility of phosphorene and few-layer black phosphorus.* Applied Physics Letters, 2014. **104**: p. 251915.
65. Jiang, J.-W. and H.S. Park, *Mechanical properties of single-layer black phosphorus.* Journal of Physics D: Applied Physics, 2014. **47**(38): p. 385304.
66. Jiang, J.-W., *Parametrization of Stillinger–Weber potential based on valence force field model: application to single-layer MoS2 and black phosphorus.* Nanotechnology, 2015. **26**: p. 315706.
67. Jiang, J.-W., H.S. Park, and T. Rabczuk, *Molecular dynamics simulations of single-layer molybdenum disulphide (MoS2): Stillinger-Weber parametrization, mechanical properties, and thermal conductivity.* Journal of Applied Physics, 2013. **114**: p. 064307.
68. Jiang, J.-W., T. Rabczuk, and H.S. Park, *A Stillinger–Weber potential for single-layered black phosphorus, and the importance of cross-pucker interactions for a negative Poisson's ratio and edge stress-induced bending.* Nanoscale, 2015. **7**: p. 6059-6068.
69. Stewart, J.A. and D.E. Spearot, *Atomistic simulations of nanoindentation on the basal plane of crystalline molybdenum disulfide (MoS2).* Modelling and Simulation in Materials Science and Engineering, 2013. **21**: p. 045003.

70. Liang, T., S.R. Philpott, and S.B. Sinnott, *Parametrization of a reactive many-body potential for Mo-S systems.* Physical Review B, 2009. **79**: p. 245110.
71. Lorenz, T., et al., *Theoretical study of the mechanical behavior of individual TiS2 and MoS2 nanotubes.* J. Phys. Chem. C, 2012. **116**: p. 11714–11721.
72. Wang, L., et al., *Electro-mechanical anisotropy of phosphorene.* Nanoscale, 2015. **7**: p. 9746-9751.
73. Evans, K.E., et al., *Molecular network design.* Nature, 1991. **353**: p. 124.
74. Jiang, J.W. and H.S. Park, *Negative Poisson's ratio in single-layer black phosphorus.* Nature Commun, 2014. **5**: p. 4727.
75. Lakes, R.S., *Foam structures with a negative Poisson's ratio.* Science, 1987. **235**: p. 1038-1040.
76. Han, J., et al., *Negative Poisson's ratios in few-layer orthorhombic arsenic: First-principles calculations.* Applied Physics Letters, 2015. **8**: p. 041801.
77. Jiang, J.W., et al., *Intrinsic negative Poisson's ratio for single-layer graphene.* Nano Lett, 2016. **16**: p. 5286-5290.
78. Ho, V.H., et al., *Negative Poisson's ratio in periodic porous graphene structures.* Phys. Status Solidi B, 2016. **253**: p. 1303-1309.
79. Wang, H., et al., *Strain effects on borophene: Ideal strength, negative Poisson's ratio and phonon instability.* New Journal of Physics, 2016. **18**: p. 073016.
80. Grima, J.N., et al., *Tailoring graphene to achieve negative Poisson's ratio.* Advanced Materials, 2015. **27**: p. 1455-1459.
81. Jiang, J.W. and H.S. Park, *Negative Poisson's ratio in single-layer graphene ribbons.* Nano Lett, 2016. **16**: p. 2657-2662.
82. Meyer, J., et al., *The structure of suspended graphene sheets.* Nature, 2007. **446**: p. 60-63.
83. Bangert, U., et al., *Manifestation of ripples in free-standing graphene in lattice images obtained in an aberration-corrected scanning transmission electron microscope.* Phys. Status Solidi A 2009. **206**: p. 1117-1122.
84. Xu, P., et al., *Unusual ultra-low-frequency fluctuations in freestanding graphene.* Nature Commun, 2014. **5**: p. 3720.
85. Fasolino, A., J. Los, and M. Katsnelson, *Intrinsic ripples in graphene.* Nature Materials, 2007. **6**: p. 858-861.
86. Zhao, H. and N. Aluru, *Temperature and strain-rate dependent fracture strength of graphene.* Journal of Applied Physics, 2010. **108**: p. 064321.
87. Zakharchenko, K.V., M. Katsnelson, and A. Fasolino, *Finite temperature lattice properties of graphene beyond the quasiharmonic approximation.* Phys Rev Lett, 2009. **102**: p. 046808.
88. Chen, S. and D. Chrzan, *Monte carlo simulation of temperature-dependent elastic properties of graphene.* Physical Review B, 2011. **84**: p. 195409.
89. Mounet, N. and N. Marzari, *First-principles determination of the structural, vibrational and thermodynamic properties of diamond, graphite, and derivatives.* Physical Review B, 2005. **71**: p. 205214.
90. Jiang, J.W., J. Wang, and B. Li, *Thermal expansion in single-walled carbon nanotubes and graphene: Nonequilibrium green's function approach.* Physical Review B, 2009. **80**: p. 205429.
91. Pozzo, M., et al., *Thermal expansion of supported and freestanding graphene: Lattice constant versus interatomic distance.* Phys Rev Lett, 2011. **106**: p. 135501.
92. Chhowalla, M., et al., *The chemistry of two-dimensional layered transition metal dichalcogenide nanosheets.* Nature Chemistry, 2013. **5**: p. 263-275.
93. Sorkin, V., et al., *Nanoscale transition metal dichalcogenides: structures, properties, and applications.* Crit. Rev. Solid State Mater. Sci, 2014. **39**: p. 319-367.
94. Sorkin, V., et al., *Recent advances in the study of phosphorene and its nanostructures.* Crit. Rev. Solid State Mater. Sci, 2016: p. in press.

95. Cai, Y., et al., *Highly itinerant atomic vacancies in phosphorene.* J. Am. Chem. Soc., 2016. **138**: p. 10199–10206.
96. Zou, X., Y. Liu, and B.I. Yakobson, *Predicting dislocations and grain boundaries in two-dimensional metal-disulfides from the first principles.* Nano Lett, 2013. **13**: p. 253–258.
97. Zhou, W., et al., *Intrinsic structural defects in monolayer molybdenum disulphide.* Nano Lett, 2013. **13**: p. 2615–2622.
98. Xu, F., et al., *Riemann surfaces of carbon as graphene nanosolenoids.* Nano Lett, 2016. **16**: p. 34-39.
99. Ly, T.H., et al., *Vertically conductive MoS2 spiral pyramid.* Advanced Materials, 2016. **28**: p. 7723-7728.
100. Yu, Z.G., Y.-W. Zhang, and B.I. Yakobson, *An anomalous formation pathway for dislocation-sulfur vacancy complexes in polycrystalline monolayer MoS2.* Nano Lett, 2015. **15**: p. 6855–6861.
101. Grantab, R., V.B. Shenoy, and R.S. Ruoff, *Anomalous strength characteristics of tilt grain boundaries in graphene.* Science, 2010. **330**: p. 946-948.
102. Wei, Y., et al., *The nature of strength enhancement and weakening by pentagon–heptagon defects in graphene.* Nature Materials, 2012. **11**: p. 759-763.
103. Yakobson, B.I. and F. Ding, *Observational geology of graphene, at the nanoscale.* ACS Nano, 2011. **5**: p. 1569–1574.
104. Song, Z., et al., *Pseudo Hall–Petch strength reduction in polycrystalline graphene.* Nano Lett, 2013. **13**: p. 1829-1833.
105. Sha, Z., et al., *Inverse pseudo Hall-Petch relation in polycrystalline graphene.* Scientific Reports, 2014. **4**: p. 5991.
106. Liu, Y., et al., *Two-dimensional mono-elemental semiconductor with electronically inactive defects: The case of phosphorus.* Nano Lett, 2014. **14**: p. 6782–6786.
107. Guo, Y., et al., *Atomic structures and electronic properties of phosphorene grain boundaries.* 2D Materials, 2016. **3**: p. 025008.
108. Berman, D., A. Erdemir, and A.V. Sumant, *Few layer graphene to reduce wear and friction on sliding steel surfaces.* Carbon, 2013. **54**: p. 454-459.
109. Lee, J.-H., et al., *Dynamic mechanical behavior of multilayer graphene via supersonic projectile penetration.* Science, 2014. **346**: p. 1092-1096.
110. Wetzel, E.D., R. Balu, and T.D. Beaudet, *A theoretical consideration of the ballistic response of continuous graphene membranes.* Journal of the Mechanics and Physics of Solids, 2015. **82**: p. 23-31.
111. Stankovich, S., et al., *Graphene-based composite materials.* Nature, 2006. **442**: p. 282-286.
112. Xu, Y., et al., *Strong and ductile poly (vinyl alcohol)/graphene oxide composite films with a layered structure.* Carbon, 2009. **47**: p. 3538-3543.
113. Wu, J., et al., *Mechanics and mechanically tunable band gap in single-layer hexagonal boron-nitride.* Materials Research Letters, 2013. **1**: p. 200-206.
114. Peng, Q. and S. De, *Outstanding mechanical properties of monolayer MoS2 and its application in elastic energy storage.* Physical Chemistry Chemical Physics, 2013. **15**: p. 19427-19437.
115. Gao, E., B. Xie, and Z. Xu, *Two-dimensional silica: Structural, mechanical properties, and strain-induced band gap tuning.* Journal of Applied Physics, 2016. **119**: p. 014301.
116. Wang, H., et al., *The ideal tensile strength and phonon instability of borophene.* arXiv:1602.00456, 2016.
117. Marianetti, C.A. and H.G. Yevick, *Failure mechanisms of graphene under tension.* Phys Rev Lett, 2010. **105**: p. 245502.
118. Xu, Z., *Graphene nano-ribbons under tension.* Journal of Computational and Theoretical Nanoscience, 2009. **6**: p. 625-628.

119. Zhao, H., K. Min, and N. Aluru, *Size and chirality dependent elastic properties of graphene nanoribbons under uniaxial tension.* Nano Lett, 2009. **9**: p. 3012-3015.
120. Huang, P.Y., et al., *Grains and grain boundaries in single-layer graphene atomic patchwork quilts.* Nature, 2011. **469**: p. 389-392.
121. Li, X., et al., *Large-area synthesis of high-quality and uniform graphene films on copper foils.* Science, 2009. **324**: p. 1312-1314.
122. Yu, Q., et al., *Control and characterization of individual grains and grain boundaries in graphene grown by chemical vapour deposition.* Nature Materials, 2011. **10**: p. 443-449.
123. Liu, T.-H., C.-W. Pao, and C.-C. Chang, *Effects of dislocation densities and distributions on graphene grain boundary failure strengths from atomistic simulations.* Carbon, 2012. **50**: p. 3465-3472.
124. Zhang, J., J. Zhao, and J. Lu, *Intrinsic strength and failure behaviors of graphene grain boundaries.* ACS Nano, 2012. **6**: p. 2704-2711.
125. Cao, A. and J. Qu, *Atomistic simulation study of brittle failure in nanocrystalline graphene under uniaxial tension.* Applied Physics Letters, 2013. **102**: p. 071902.
126. Cao, A. and Y. Yuan, *Atomistic study on the strength of symmetric tilt grain boundaries in graphene.* Applied Physics Letters, 2012. **100**: p. 211912.
127. Yi, L., et al., *A theoretical evaluation of the temperature and strain-rate dependent fracture strength of tilt grain boundaries in graphene.* Carbon, 2013. **51**: p. 373-380.
128. Zhang, Z., et al., *Unraveling the sinuous grain boundaries in graphene.* Advanced Functional Materials, 2015. **25**: p. 367-373.
129. Kotakoski, J. and J.C. Meyer, *Mechanical properties of polycrystalline graphene based on a realistic atomistic model.* Physical Review B, 2012. **85**: p. 195447.
130. Shekhawat, A. and R.O. Ritchie, *Toughness and strength of nanocrystalline graphene.* Nature Commun, 2016. **7**: p. 10546.
131. Tabarraei, A. and X. Wang, *A molecular dynamics study of nanofracture in monolayer boron nitride.* Materials Science and Engineering A, 2015. **641**: p. 225-230.
132. Mortazavi, B. and G. Cuniberti, *Mechanical properties of polycrystalline boron-nitride nanosheets.* RSC Advances, 2014. **4**: p. 19137-19143.
133. Becton, M. and X. Wang, *Grain-size dependence of mechanical properties in polycrystalline boron-nitride: a computational study.* Physical Chemistry Chemical Physics, 2015. **17**: p. 21894-21901.
134. Liu, Y., X. Zou, and B.I. Yakobson, *Dislocations and grain boundaries in two-dimensional boron nitride.* ACS Nano, 2012. **6**: p. 7053-7058.
135. Huang, P.Y., et al., *Imaging atomic rearrangements in two-dimensional silica glass: Watching silica's dance.* Science, 2013. **342**: p. 224-227.
136. Min, K. and N.R. Aluru, *Mechanical properties of graphene under shear deformation.* Applied Physics Letters, 2011. **98**: p. 013113.
137. Liu, Y. and B.I. Yakobson, *Cones, pringles, and grain boundary landscapes in graphene topology.* Nano Lett, 2010. **10**: p. 2178-2183.
138. Chen, S. and D.C. Chrzan, *Continuum theory of dislocations and buckling in graphene.* Physical Review B, 2011. **84**: p. 214103
139. Liu, T.-H., et al., *Structure, energy, and structural transformations of graphene grain boundaries from atomistic simulations.* Carbon, 2011. **49**: p. 2306-2317.
140. Zhang, T., X. Li, and H. Gao, *Defects controlled wrinkling and topological design in graphene.* Journal of the Mechanics and Physics of Solids, 2014. **67**: p. 2-13.
141. Zhang, T., X. Li, and H. Gao, *Designing graphene structures with controlled distributions of topological defects: A case study of toughness enhancement in graphene ruga.* Extreme Mechanics Letters, 2014. **1**: p. 3-8.

142. Song, Z., et al., *Defect-detriment to graphene strength is concealed by local probe: the topological and geometrical effects.* ACS Nano, 2014. **9**: p. 401-408.
143. Jiang, J.-W., et al., *Elastic bending modulus of single-layer molybdenum disulfide (MoS2): finite thickness effect.* Nanotechnology, 2013. **24**: p. 435705.
144. Cohen-Tanugi, D. and J.C. Grossman, *Water desalination across nanoporous graphene.* Nano Lett, 2012. **12**: p. 3602-3608.
145. O'Hern, S.C., et al., *Selective ionic transport through tunable subnanometer pores in single-layer graphene membranes.* Nano Lett, 2014. **14**: p. 1234-1241.
146. Koenig, S.P., et al., *Selective molecular sieving through porous graphene.* Nature Nanotechnology, 2012. **7**: p. 728-732.
147. Wang, L., et al., *Molecular valves for controlling gas phase transport made from discrete ångström-sized pores in graphene.* Nature Nanotechnology, 2015. **10**: p. 785-790.
148. Merchant, C.A., et al., *DNA translocation through graphene nanopores.* Nano Lett, 2010. **10**: p. 2915-2921.
149. Schneider, G.F., et al., *DNA translocation through graphene nanopores.* Nano Lett, 2010. **10**: p. 3163-3167.
150. Khare, R., et al., *Coupled quantum mechanical/molecular mechanical modeling of the fracture of defective carbon nanotubes and graphene sheets.* Physical Review B, 2007. **75**: p. 075412.
151. Xu, M., et al., *A coupled quantum/continuum mechanics study of graphene fracture.* Int. J. Fracture, 2012. **173**: p. 163-173.
152. Terdalkar, S.S., et al., *Nanoscale fracture in graphene.* Chemical Physics Letters, 2010. **494**: p. 218-222.
153. Zhang, T., et al., *Flaw insensitive fracture in nanocrystalline graphene.* Nano Lett, 2012. **12**: p. 4605-4610.
154. Jung, G., Z. Qin, and M.J. Buehler, *Molecular mechanics of polycrystalline graphene with enhanced fracture toughness.* Extreme Mechanics Letters, 2015. **2**: p. 52-59.
155. Liu, N., et al., *Fracture patterns and the energy release rate of phosphorene.* Nanoscale, 2016. **8**: p. 5728-5736.
156. Wang, X., A. Tabarraei, and D.E. Spearot, *Fracture mechanics of monolayer molybdenum disulfide.* Nanotechnology, 2015. **26**: p. 175703.
157. Sen, D., et al., *Tearing graphene sheets from adhesive substrates produces tapered nanoribbons.* Small, 2010. **6**: p. 1108-1116.
158. Moura, M.J.B. and M. Marder, *Tearing of free-standing graphene.* Physical Review E, 2013. **88**: p. 032405.
159. Curtin, W.A., *On lattice trapping of cracks.* Journal of Materials Research, 1990. **5**: p. 1549-1560.
160. Zhang, S., T. Zhu, and T. Belytschko, *Atomistic and multiscale analyses of brittle fracture in crystal lattices.* Physical Review B, 2007. **76**: p. 094114.
161. Lu, Q. and R. Huang, *Excess energy and deformation along free edges of graphene nanoribbons.* Physical Review B, 2010. **81**: p. 155410.
162. Zhang, Z., A. Kutana, and B.I. Yakobson, *Edge reconstruction-mediated graphene fracture.* Nanoscale, 2015. **7**: p. 2716-2722.
163. Yin, H., et al., *Griffith criterion for brittle fracture in graphene.* Nano Lett, 2015. **15**: p. 1918-1924.
164. Brochard, L., I.G. Tejada, and K. Sab, *From yield to fracture, failure initiation captured by molecular simulation.* Journal of the Mechanics and Physics of Solids, 2016. **95**: p. 632-646.
165. Kumar, K.S., H. Van Swygenhoven, and S. Suresh, *Mechanical behavior of nanocrystalline metals and alloys.* Acta Materialia, 2003. **51**: p. 5743-5774.
166. Lu, K., L. Lu, and S. Suresh, *Strengthening materials by engineering coherent internal boundaries at the nanoscale.* Science, 2009. **324**: p. 349-352.

167. Tian, Y., et al., *Ultrahard nanotwinned cubic boron nitride.* Nature, 2013. **493**: p. 385-388.
168. Huang, Q., et al., *Nanotwinned diamond with unprecedented hardness and stability.* Nature, 2014. **510**: p. 250-253.
169. Zhang, T. and H. Gao, *Toughening graphene with topological defects: a perspective.* Journal of Applied Mechanics, 2015. **82**: p. 051001.
170. Meng, F., C. Chen, and J. Song, *Dislocation shielding of a nanocrack in graphene: Atomistic simulations and continuum modeling.* Journal of Physical Chemistry Letters, 2015. **6**: p. 4038-4042.
171. Mitchell, N.P., et al., *Fracture in sheets draped on curved surfaces.* Nature Materials, 2016. **In press**.
172. Ci, L., et al., *Controlled nanocutting of graphene.* Nano Research, 2008. **1**: p. 116-122.
173. Campos, L.C., et al., *Anisotropic etching and nanoribbon formation in single-layer graphene.* Nano Lett, 2009. **9**: p. 2600-2604.
174. Feng, J., et al., *Patterning of graphene.* Nanoscale, 2012. **4**: p. 4883-4899.
175. Qi, Z., D.K. Campbell, and H.S. Park, *Atomistic simulations of tension-induced large deformation and stretchability in graphene kirigami.* Physical Review B, 2014. **90**: p. 245437.
176. Bahamon, D., et al., *Graphene kirigami as a platform for stretchable and tunable quantum dot arrays.* Physical Review B, 2016. **93**: p. 235408.
177. Hanakata, P.Z., et al., *Highly stretchable MoS2 kirigami.* Nanoscale, 2016. **8**: p. 458-463.
178. Hutchinson, J.W., *Crack tip shielding by micro-cracking in brittle solids.* Acta Metallurgica, 1987. **35**: p. 1605-1619.
179. Evans, A.G., *Perspective on the development of high-toughness ceramics.* Journal of the American Ceramic Society, 1990. **73**: p. 187-206.
180. Evans, A.G. and K.T. Faber, *Crack-growth resistance of microcracking brittle materials.* Journal of the American Ceramic Society, 1984. **67**: p. 255-260.
181. Evans, A.G. and Y. Fu, *Some effects of microcracks on the mechanical properties of brittle solids— II. Microcrack toughening.* Acta Metallurgica, 1985. **33**: p. 1525-1531.
182. Shin, Y.A., et al., *Nanotwin-governed toughening mechanism in hierarchically structured biological materials.* Nature Commun, 2016. **7**: p. 10772.
183. López-Polín, G., J. Gómez-Herrero, and C. Gómez-Navarro, *Confining crack propagation in defective graphene.* Nano Lett, 2015. **15**: p. 2050-2054.
184. Gao, H., et al., *Materials become insensitive to flaws at nanoscale: lessons from nature.* Proceedings of the National Academy of Sciences, 2003. **100**: p. 5597-5600.
185. Gao, H. and S. Chen, *Flaw tolerance in a thin strip under tension.* Journal of Applied Mechanics, 2005. **72**: p. 732-737.
186. Pereira, V.M. and A.H. Castro Neto, *Strain engineering of graphene's electronic structure.* Phys Rev Lett, 2009. **103**: p. 046801.
187. Gui, G., J. Li, and J. Zhong, *Band structure engineering of graphene by strain: First-principles calculations.* Physical Review B, 2008. **78**: p. 075435.
188. Ni, Z.H., et al., *Uniaxial strain on graphene: Raman spectroscopy study and band-gap opening.* ACS Nano, 2008. **2**: p. 2301-2305.
189. Pereira, V.M., A.H. Castro Neto, and N.M.R. Peres, *Tight-binding approach to uniaxial strain in graphene.* Physical Review B, 2009. **80**: p. 045401.
190. Cocco, G., E. Cadelano, and L. Colombo, *Gap opening in graphene by shear strain.* Physical Review B, 2010. **81**: p. 241412(R).
191. Naumov, I.I. and A.M. Bratkovsky, *Gap opening in graphene by simple periodic inhomogeneous strain.* Physical Review B, 2011. **84**: p. 245444.
192. Xu, K., P. Cao, and J.R. Heath, *Scanning tunneling microscopy characterization of the electrical properties of wrinkles in exfoliated graphene monolayers.* Nano Lett, 2009. **9**: p. 4446-4451.

193. Zhu, W., et al., *Structure and electronic transport in graphene wrinkles.* Nano Lett, 2012. **12**: p. 3431-3436.
194. Castro Neto, A.H., et al., *The electronic properties of graphene.* Reviews of Modern Physics, 2009. **81**: p. 109-162.
195. Das Sarma, S., et al., *Electronic transport in two-dimensional graphene.* Reviews of Modern Physics, 2011. **83**: p. 407-470.
196. Kotov, V.N., et al., *Electron-electron interactions in graphene: Current status and perspectives.* Reviews of Modern Physics, 2012. **84**: p. 1067-1125.
197. *Strongest non-destructive magnetic field: world record set at 100-tesla level.* 2014; Available from: https://www.lanl.gov/science-innovation/features/science-digests/world-record-set-magnetic-field.php.
198. Lu, J., A.H. Castro Neto, and K.P. Loh, *Transforming moire blisters into geometric graphene nano-bubbles.* Nature Commun, 2012. **3**: p. 823.
199. Qi, Z., et al., *Pseudomagnetic fields in graphene nanobubbles of constrained geometry: A molecular dynamics study.* Physical Review B, 2014. **90**: p. 125419.
200. Neek-Amal, M. and F.M. Peeters, *Strain-engineered graphene through a nanostructured substrate. II. Pseudomagnetic fields.* Physical Review B, 2012. **85**: p. 195446.
201. Gill, S.T., et al., *Mechanical control of graphene on engineered pyramidal strain arrays.* ACS Nano, 2015. **9**: p. 5799-5806.
202. Guinea, F., M.I. Matsnelson, and A.K. Geim, *Energy gaps and a zero-field quantum hall effect in graphene by strain engineering.* Nature Physics, 2010. **6**: p. 30-33.
203. Zhu, S., J.A. Stroscio, and T. Li, *Programmable extreme pseudomagnetic fields in graphene by a uniaxial stretch.* Phys Rev Lett, 2015. **115**: p. 245501.
204. Kane, C.L. and E.J. Mele, *Size, shape, and low energy electronic structure of carbon nanotubes.* Phys Rev Lett, 1997. **78**: p. 1932-1935.
205. Suzuura, H. and T. Ando, *Phonons and electron-phonon scattering in carbon nanotubes.* Physical Review B, 2002. **65**: p. 235412.
206. Guinea, F., B. Horovitz, and P. Le Doussal, *Gauge field induced by ripples in graphene.* Physical Review B, 2008. **77**: p. 205421.
207. Zhu, S., et al., *Pseudomagnetic fields in a locally strained graphene drumhead.* Physical Review B, 2014. **90**: p. 075426.
208. Vozmediano, M.A.H., M.I. Katsnelson, and F. Guinea, *Gauge fields in graphene.* Physics Reports, 2010. **496**: p. 109-148.
209. Ramezani Masir, M., D. Moldovan, and F.M. Peeters, *Pseudo magnetic field in strained graphene: Revisited.* Solid State Communications, 2013. **175-176**: p. 76-82.
210. de Juan, F., J.L. Mañes, and M.A.H. Vozmediano, *Gauge fields from strain in graphene.* Physical Review B, 2013. **87**: p. 165131.
211. Kitt, A.L., et al., *Lattice-corrected strain-induced vector potentials in graphene.* Physical Review B, 2012. **85**: p. 115432.
212. Sloan, J.V., et al., *Strain gauge fields for rippled graphene membranes under central mechanical load: an approach beyond first-order continuum elasticity.* Physical Review B, 2013. **87**: p. 155436.
213. Kim, K., Y. Blanter, and K. Ahn, *Interplay between real and pseudomagnetic field in graphene with strain.* Physical Review B, 2011. **84**: p. 081401(R).
214. Guinea, F., et al., *Generating quantizing pseudomagnetic fields by bending graphene ribbons.* Physical Review B, 2010. **81**: p. 035408.
215. Low, T. and F. Guinea, *Strain-induced pseudomagnetic field for novel graphene electronics.* Nano Lett, 2010. **10**: p. 3551-3554.

216. Pereira, V.M., et al., *Geometry, mechanics, and electronics of singular structures and wrinkles in graphene.* Phys Rev Lett, 2010. **105**: p. 156603.
217. Bai, J., et al., *Graphene nanomesh.* Nature Nanotechnology, 2010. **5**: p. 190-194.
218. Baringhaus, J., et al., *Exceptional ballistic transport in epitaxial graphene nanoribbons.* Nature, 2014. **506**: p. 349-354.
219. Sun, Z., et al., *Towards hybrid superlattices in graphene.* Nature Commun, 2011. **2**: p. 559.
220. Balog, R., et al., *Bandgap opening in graphene induced by patterned hydrogen adsorption.* Nature Materials, 2010. **9**: p. 315-319.
221. Ribeiro, R.M., et al., *Strained graphene: tight-binding and density functional calculations.* New Journal of Physics, 2009. **11**: p. 115002.
222. Kim, E.-A. and A.H. Castro Neto, *Graphene as an electronic membrane.* EPL, 2008. **84**: p. 57007.
223. Zhu, S., Y. Huang, and T. Li, *Extremely compliant and highly stretchable patterned graphene.* Applied Physics Letters, 2014. **104**: p. 173103.
224. Teague, M.L., et al., *Evidence for strain-induced local conductance modulations in single-layer graphene on SiO2.* Nano Lett, 2009. **9**: p. 2542-2546.
225. Gradinar, D.A., et al., *Transport signatures of pseudomagnetic Landau levels in strained graphene ribbons.* Phys Rev Lett, 2013. **110**: p. 266801.
226. Caroli, C., et al., *Direct calculation of the tunneling current.* Journal of Physics C: Solid State Physics, 1971. **4**: p. 916-929.
227. Bahamon, D.A., et al., *Conductance signatures of electron confinement induced by strained nanobubbles in graphene.* Nanoscale, 2015. **7**: p. 15300-15309.
228. Qi, Z., et al., *Resonant tunneling in graphene pseudomagnetic quantum dots.* Nano Lett, 2013. **13**: p. 2692-2697.
229. Zhang, Y., et al., *Experimental observation of the quantum Hall effect and Berry's phase in graphene.* Nature, 2005. **438**: p. 201-204.
230. Rao, C.N.R., et al., *Graphene: the new 2wo-dimensional nanomaterial.* Angewandte Chemie, 2009. **48**: p. 7752-7777.
231. Wang, Q.H., et al., *Electronics and optoelectronics of two-dimensional transition metal dichalcogenides.* Nature Nanotechnology, 2012. **7**: p. 699-712.
232. Jariwala, D., et al., *Emerging device applications for semiconducting two-dimensional transition metal dichalcogenides.* ACS Nano, 2014. **8**: p. 1102-1120.
233. Novoselov, K.S., et al., *Two-dimensional gas of massless Dirac fermions in graphene.* Nature, 2005. **438**: p. 197-200.
234. Pacilé, D., et al., *The two-dimensional phase of boron nitride: Few-atomic-layer sheets and suspended membranes.* Applied Physics Letters, 2008. **92**: p. 133107.
235. Wilson, J.A. and A.D. Yoffe, *The transition metal dichalcogenides discussion and interpretation of the observed optical, electrical and structural properties.* Advances in Physics, 1969. **18**: p. 193-335.
236. Li, Y., et al., *Structural semiconductor-to-semimetal phase transition in two-dimensional materials induced by electrostatic gating.* Nature Commun, 2016. **7**: p. 10671.
237. Brown, B.E., *The crystal structures of WTe2 and high-temperature MoTe2.* Acta Crystallographica, 1966. **20**: p. 268-274.
238. Jiménez Sandoval, S., et al., *Raman study and lattice dynamics of single molecular layers of MoS2.* Physical Review B, 1991. **44**: p. 3955-3962.
239. Eda, G., et al., *Coherent atomic and electronic heterostructures of single-layer MoS2.* ACS Nano, 2012. **6**: p. 7311-7317.
240. Gordon, R.A., et al., *Structures of exfoliated single layers of WS2, MoS2 and MoSe2 in aqueous suspension.* Physical Review B, 2002. **65**: p. 125407.

241. Lin, Y.-C., et al., *Atomic mechanism of the semiconducting-to-metallic phase transition in single-layered MoS2.* Nature Nanotechnology, 2014. **9**: p. 391-396.
242. Park, J.C., et al., *Phase-engineered synthesis of centimeter-scale 1T'- and 2H-molybdenum ditelluride thin films.* ACS Nano, 2015. **9**: p. 6548-6554.
243. Zhou, Y. and E.J. Reed, *Structural phase stability control of monolayer MoTe2 with adsorbed atoms and molecules.* Journal of Physical Chemistry C, 2015. **119**: p. 21674-21680.
244. Nayak, A.P., et al., *Pressure-induced semiconducting to metallic transition in multilayered molybdenum disulphide.* Nature Commun, 2014. **5**: p. 3731.
245. Rice, C., et al., *Raman-scattering measurements and first-principles calculations of strain-induced phonon shifts in monolayer MoS2.* Physical Review B, 2013. **87**: p. 081307(R).
246. Conley, H.J., et al., *Bandgap engineering of strained monolayer and bilayer MoS2.* Nano Lett, 2013. **13**: p. 3626-3630.
247. Dumcenco, D.O., et al., *Raman study of 2H-Mo1–xWxS2 layered mixed crystals.* Journal of Alloys and Compounds, 2010. **506**: p. 940-943.
248. Nayak, A.P., et al., *Pressure-modulated conductivity, carrier density, and mobility of multilayered tungsten disulfide.* ACS Nano, 2015. **9**: p. 9117-9123.
249. Kim, J.-S., et al., *High pressure Raman study of layered Mo0.5W0.5S2 ternary compound.* 2D Materials, 2016. **3**: p. 025003.
250. He, K., et al., *Experimental demonstration of continuous electronic structure tuning via strain in atomically thin MoS2.* Nano Lett, 2013. **13**: p. 2931-2936.
251. Bunch, J.S., et al., *Impermeable atomic membranes from graphene sheets.* Nano Lett, 2008. **8**: p. 2458-2462.
252. Yun, W.S., et al., *Thickness and strain effects on electronic structures of transition metal dichalcogenides: 2H-MX2 semiconductors (M = Mo, W; X = S, Se, Te).* Physical Review B, 2012. **85**: p. 033305.
253. Feng, J., et al., *Strain-engineered artificial atom as a broad-spectrum solar energy funnel.* Nature Photonics, 2012. **6**: p. 866–872.
254. Nguyen, T.D., et al., *Nanoscale flexoelectricity.* Advanced Materials, 2013. **25**: p. 946-974.
255. Yudin, P.V. and A.K. Tagantsev, *Fundamentals of flexoelectricity in solids.* Nanotechnology, 2013. **24**: p. 432001.
256. Zubko, P., G. Catalan, and A.K. Tagantsev, *Flexoelectric effect in solids.* Annual Review of Materials Research, 2013. **43**: p. 387-421.
257. Jiang, X., W. Huang, and S. Zhang, *Flexoelectric nano-generator: Materials, structures and devices.* Nano Energy, 2013. **2**: p. 1079-1092.
258. Majdoub, M., P. Sharma, and T. Çağin, *Dramatic enhancement in energy harvesting for a narrow range of dimensions in piezoelectric nanostructures.* Physical Review B, 2008. **78**: p. 121407.
259. Majdoub, M.S. and P. Sharma, *Erratum: Dramatic enhancement in energy harvesting for a narrow range of dimensions in piezoelectric nanostructures.* Physical Review B, 2009. **79**: p. 159901.
260. Ahmadpoor, F. and P. Sharma, *Flexoelectricity in two-dimensional crystalline and biological membranes.* Nanoscale, 2015. **7**: p. 16555-16570.
261. Krichen, S. and P. Sharma, *Flexoelectricity: A perspective on an unusual electromechanical coupling.* Journal of Applied Mechanics, 2016. **83**: p. 030801.
262. Blonsky, M.N., et al., *Ab initio prediction of piezoelectricity in two-dimensional materials.* ACS Nano, 2015. **9**: p. 9885-9891.
263. Ong, M.T. and E.J. Reed, *Engineered piezoelectricity in graphene.* ACS Nano, 2011. **6**: p. 1387-1394.
264. Duerloo, K.A.N., M.T. Ong, and E.J. Reed, *Intrinsic piezoelectricity in two-dimensional materials.* Journal of Physical Chemistry Letters, 2012. **3**: p. 2871-2876.

265. Kim, S.K., et al., *Directional dependent piezoelectric effect in CVD grown monolayer MoS2 for flexible piezoelectric nanogenerators.* Nano Energy, 2016. **22**: p. 483-489.
266. Zhu, H., et al., *Observation of piezoelectricity in free-standing monolayer MoS2.* Nature Nanotechnology, 2014. **10**: p. 151-155.
267. Qi, J., et al., *Piezoelectric effect in chemical vapour deposition-grown atomic-monolayer triangular molybdenum disulfide piezotronics.* Nature Commun, 2015. **6**: p. 7430.
268. Wu, W., et al., *Piezoelectricity of single-atomic-layer MoS2 for energy conversion and piezotronics.* Nature, 2014. **514**: p. 470-474.
269. Dumitrică, T., C.M. Landis, and B.I. Yakobson, *Curvature-induced polarization in carbon nanoshells.* Chemical Physics Letters, 2002. **360**: p. 182-188.
270. Kalinin, S.V. and V. Meunier, *Electronic flexoelectricity in low-dimensional systems.* Physical Review B, 2008. **77**: p. 033403.
271. Kvashnin, A.G., P.B. Sorokin, and B.I. Yakobson, *Flexoelectricity in carbon nanostructures: nanotubes, fullerenes, and nanocones.* Journal of Physical Chemistry Letters, 2015: p. 2740-2744.
272. Duerloo, K.-A.N. and E.J. Reed, *Flexural electromechanical coupling: a nanoscale emergent property of boron nitride bilayers.* Nano Lett, 2013. **13**: p. 1681-1686.
273. Chandratre, S. and P. Sharma, *Coaxing graphene to be piezoelectric.* Applied Physics Letters, 2012. **100**: p. 2014-2017.
274. Boddeti, N.G., et al., *Mechanics of adhered, pressurized graphene blisters.* Journal of Applied Mechanics, 2013. **80**: p. 040909.
275. Suk, J.W., et al., *Transfer of CVD-grown monolayer graphene onto arbitrary substrates.* ACS Nano, 2011. **5**: p. 6916-6924.
276. Wan, K.-T. and Y.-W. Mai, *Fracture mechanics of a new blister test with stable crack growth.* Acta Metallurgica et Materialia, 1995. **43**: p. 4109-4115.
277. Liu, X., et al., *Observation of pull-in instability in graphene membranes under interfacial forces.* Nano Lett, 2013. **13**: p. 2309-2313.
278. Boddeti, N.G., et al., *Graphene blisters with switchable shapes controlled by pressure and adhesion.* Nano Lett, 2013. **13**: p. 6216-6221.
279. Wang, P., K.M. Liechti, and R. Huang, *Snap transitions of pressurized graphene blisters.* Journal of Applied Mechanics, 2016. **83**: p. 071002.
280. Zong, Z., et al., *Direct measurement of graphene adhesion on silicon surface by intercalation of nanoparticles.* Journal of Applied Physics, 2010. **107**(2): p. 026104.
281. Jiang, T., R. Huang, and Y. Zhu, *Interfacial sliding and buckling of monolayer graphene on a stretchable substrate.* Advanced Functional Materials, 2014. **24**: p. 396-402.
282. Brennan, C.J., et al., *Interface adhesion between 2D materials and elastomers measured by buckle delaminations.* Advanced Materials Interfaces, 2015. **2**: p. 1500176.
283. Cao, Z., et al., *A blister test for interfacial adhesion of large-scale transferred graphene.* Carbon, 2014. **69**: p. 390-400.
284. Na, S.R., et al., *Selective mechanical transfer of graphene from seed copper foil using rate effects.* ACS Nano, 2015. **9**: p. 1325-1335.
285. Yoon, T., et al., *Direct measurement of adhesion energy of monolayer graphene as-grown on copper and its application to renewable transfer process.* Nano Lett, 2012. **12**: p. 1448-1452.
286. Na, S.R., et al., *Clean graphene interfaces by selective dry transfer for large area silicon integration.* Nanoscale, 2016. **8**: p. 7523-7533.
287. Carpick, R.W. and J.D. Batteas, *Scanning Probe Studies of Nanoscale Adhesion Between Solids in the Presence of Liquids and Monolayer Films*, in *Springer Handbook of Nanotechnology*. 2004, Springer. p. 605-629.

288. Butt, H.-J., B. Cappella, and M. Kappl, *Force measurements with the atomic force microscope: Technique, interpretation and applications.* Surface Science Reports, 2005. **59**: p. 1-152.
289. Burnham, N.A., et al., *Probing the surface forces of monolayer films with an atomic-force microscope.* Phys Rev Lett, 1990. **64**: p. 1931-1934.
290. Johnson, K.L., Kendall, K., and Roberts, A. D., *Surface energy and the contact of elastic solids.* Proc. R. Soc. A, 1971. **324**: p. 301-313.
291. Derjaguin, R.V., V.M. Muller, and Y.P. Toporov, *Effect of contact deformations on the adhesion of particles.* Journal of Colloid and Interface Science, 1975. **53** p. 314-326.
292. Maugis, D., *Adhesion of spheres: The JKR-DMT transition using a Dugdale model.* Journal of Colloid and Interface Science, 1992. **150**: p. 243-269.
293. Heim, L.-O., et al., *Adhesion and friction forces between spherical micrometer-sized particles.* Phys Rev Lett, 1999. **83**: p. 3328-3331.
294. Ducker, W.A., T.J. Senden, and R.M. Pashley, *Direct measurement of colloidal forces using an atomic force microscope.* Nature, 1991. **353**: p. 239-241.
295. Jiang, T. and Y. Zhu, *Measuring graphene adhesion using atomic force microscopy with a microsphere tip.* Nanoscale, 2015. **7**: p. 10760-10766.
296. Jacobs, T.B., et al., *The effect of atomic-scale roughness on the adhesion of nanoscale asperities: A combined simulation and experimental investigation.* Tribology Letters, 2013. **50**: p. 81-93.
297. Rabinovich, Y.I., et al., *Adhesion between nanoscale rough surfaces: I. Role of asperity geometry.* Journal of Colloid and Interface Science, 2000. **232**: p. 10-16.
298. Liechti, K.M., Schnapp, S.T., and Swadener, J.G., *Contact angle and contact mechanics of a glass/epoxy interface.* Int. J. Fracture, 1998. **86**: p. 361-374.
299. Grierson, D.S., E.E. Flater, and R.W. Carpick, *Accounting for the JKR-DMT transition in adhesion and friction measurements with atomic force microscopy.* Journal of Adhesion Science and Technology, 2005. **19**: p. 291-311.
300. Goertz, M.P. and N.W. Moore, *Mechanics of soft interfaces studied with displacement-controlled scanning force microscopy.* Progress in Surface Science, 2010. **85**: p. 347-397.
301. Joyce, S.A. and J.E. Houston, *A new force sensor incorporating force-feedback control for interfacial force microscopy.* Review of Scientific Instruments, 1991. **62**: p. 710-715.
302. Wang, M., et al., *A hybrid molecular-continuum analysis of IFM experiments of a self-assembled monolayer.* Journal of Applied Mechanics, 2006. **73**: p. 769-777.
303. Suk, J.W., et al., *Probing the adhesion interactions of graphene on silicon oxide by nanoindentation.* Carbon, 2016. **103**: p. 63-72.
304. Hutchinson, J.W. and A.G. Evans, *Mechanics of materials: top-down approaches to fracture.* Acta Materialia, 2000. **48**: p. 125–135.
305. Bao, G. and Z. Suo, *Remarks on crack-bridging concepts.* Appl. Mech. Rev., 1992. **45**: p. 355-366.
306. Wu, C., et al., *On determining mixed-mode traction-separation relations for interfaces.* Int. J. Fracture, 2016. **202**: p. 1-19.
307. Egberts, P., et al., *Frictional behavior of atomically thin sheets: hexagonal-shaped graphene islands grown on copper by chemical vapor deposition.* ACS Nano, 2014. **8**: p. 5010-5021.
308. Filleter, T., et al., *Friction and dissipation in epitaxial graphene films.* Phys Rev Lett, 2009. **102**: p. 086102.
309. Lee, C., et al., *Frictional characteristics of atomically thin sheets.* Science, 2010. **328**: p. 76-80.
310. Shin, Y.J., et al., *Frictional characteristics of exfoliated and epitaxial graphene.* Carbon, 2011. **49**: p. 4070-4073.
311. Marsden, A.J., M. Phillips, and N.R. Wilson, *Friction force microscopy: a simple technique for identifying graphene on rough substrates and mapping the orientation of graphene grains on copper.* Nanotechnology, 2013. **24**: p. 255704.

312. Kim, K.-S., et al., *Chemical vapor deposition-grown graphene: the thinnest solid lubricant.* ACS Nano, 2011. **5**: p. 5107-5114.
313. Wählisch, F., et al., *Friction and atomic-layer-scale wear of graphitic lubricants on SiC (0001) in dry sliding.* Wear, 2013. **300**: p. 78-81.
314. Klemenz, A., et al., *Atomic scale mechanisms of friction reduction and wear protection by graphene.* Nano Lett, 2014. **14**: p. 7145-7152.
315. Marchetto, D., et al., *Friction and wear on single-layer epitaxial graphene in multi-asperity contacts.* Tribology Letters, 2012. **48**: p. 77-82.
316. Deng, Z., et al., *Nanoscale interfacial friction and adhesion on supported versus suspended monolayer and multilayer graphene.* Langmuir, 2012. **29**: p. 235-243.
317. Hui, L., et al., *Effect of post-annealing on the plasma etching of graphene-coated-copper.* Faraday Discussions, 2014. **173**: p. 79-93.
318. Chen, S., et al., *Oxidation resistance of graphene-coated Cu and Cu/Ni alloy.* ACS Nano, 2011. **5**: p. 1321-1327.
319. Nie, S., et al., *Growth from below: bilayer graphene on copper by chemical vapor deposition.* New Journal of Physics, 2012. **14**: p. 093028.
320. Deng, Z., et al., *Adhesion-dependent negative friction coefficient on chemically modified graphite at the nanoscale.* Nature Materials, 2012. **11**: p. 1032-1037.
321. Kwon, S., et al., *Enhanced nanoscale friction on fluorinated graphene.* Nano Lett, 2012. **12**: p. 6043-6048.
322. Li, Q., et al., *Fluorination of graphene enhances friction due to increased corrugation.* Nano Lett, 2014. **14**: p. 5212-5217.
323. Daly, M., et al., *Interfacial shear strength of multilayer graphene oxide films.* ACS Nano, 2016. **10**: p. 1939-1947.
324. Chen, H. and T. Filleter, *Effect of structure on the tribology of ultrathin graphene and graphene oxide films.* Nanotechnology, 2015. **26**: p. 135702.
325. Ko, J.H., et al., *Nanotribological properties of fluorinated, hydrogenated, and oxidized graphenes.* Tribology Letters, 2013. **50**: p. 137-144.
326. Nair, R.R., et al., *Fluorographene: A two-dimensional counterpart of Teflon.* Small, 2010. **6**: p. 2877-2884.
327. Zboril, R., et al., *Graphene fluoride: A stable stoichiometric graphene derivative and its chemical conversion to graphene.* Small, 2010. **6**: p. 2885-2891.
328. Wang, L.F., et al., *Ab initio study of the friction mechanism of fluorographene and graphane.* Journal of Physical Chemistry C, 2013. **117**: p. 12520-12525.
329. Dong, Y., X. Wu, and A. Martini, *Atomic roughness enhanced friction on hydrogenated graphene.* Nanotechnology, 2013. **24**: p. 375701.
330. Dreyer, D.R., et al., *The chemistry of graphene oxide.* Chemical Society Reviews, 2010. **39**: p. 228-240.
331. Gao, Y., et al., *The effect of interlayer adhesion on the mechanical behaviors of macroscopic graphene oxide papers.* ACS Nano, 2011. **5**: p. 2134-2141.
332. Felts, J.R., et al., *Direct mechanochemical cleavage of functional groups from graphene.* Nature Commun, 2015. **6**: p. 6467.
333. Jacobs, T.D.B. and R.W. Carpick, *Nanoscale wear as a stress-assisted chemical reaction.* Nature Nanotechnology, 2013. **8**: p. 108-112.
334. Vahdat, V., et al., *Mechanics of interaction and atomic-scale wear of amplitude modulation atomic force microscopy probes.* ACS Nano, 2013. **7**: p. 3221-3235.
335. Spikes, H. and W. Tysoe, *On the commonality between theoretical models for fluid and solid friction, wear and tribochemistry.* Tribology Letters, 2015. **59**: p. 1-14.

336. Ovchinnikova, O.S., et al., *Co-registered topographical, band excitation nanomechanical, and mass spectral imaging using a combined atomic force microscopy/mass spectrometry platform.* ACS Nano, 2015. **9**: p. 4260-4269.
337. Aghababaei, R., D.H. Warner, and J.-F. Molinari, *Critical length scale controls adhesive wear mechanisms.* Nature Commun, 2016. **7**: p. 11816.
338. Fan, X., et al., *Interaction between graphene and the surface of SiO2.* Journal of Physics: Condensed Matter, 2012. **24**: p. 305004.
339. Gao, W., et al., *Interfacial adhesion between graphene and silicon dioxide by density functional theory with van der Waals corrections.* Journal of Physics D: Applied Physics, 2014. **47**: p. 255301.
340. Aitken, Z.H. and R. Huang, *Effects of mismatch strain and substrate surface corrugation on morphology of supported monolayer graphene.* Journal of Applied Physics, 2010. **107**: p. 123531.
341. Kumar, S., D. Parks, and K. Kamrin, *Mechanistic origin of the ultrastrong adhesion between graphene and a-SiO2: beyond van der Waals.* ACS Nano, 2016. **10**: p. 6552-6562.
342. Gao, W. and R. Huang, *Effect of surface roughness on adhesion of graphene membranes.* Journal of Physics D: Applied Physics, 2011. **44**: p. 452001.
343. Wang, P., W. Gao, and R. Huang, *Entropic effects of thermal rippling on van der Waals interactions between monolayer graphene and a rigid substrate.* Journal of Applied Physics, 2016. **119**: p. 074305.
344. Kitt, A.L., et al., *How graphene slides: measurement and theory of strain-dependent frictional forces between graphene and SiO2.* Nano Lett, 2013. **13**: p. 2605-2610.
345. Bonelli, F., et al., *Atomistic simulations of the sliding friction of graphene flakes.* European Physical Journal B, 2009. **70**: p. 449-459.
346. Novoselov, K., et al., *2D materials and van der Waals heterostructures.* Science, 2016. **353**: p. aac9439.
347. Yang, W., et al., *Epitaxial growth of single-domain graphene on hexagonal boron nitride.* Nature Materials, 2013. **12**: p. 792-797.
348. Woods, C., et al., *Commensurate-incommensurate transition in graphene on hexagonal boron nitride.* Nature Physics, 2014. **10**: p. 451-456.
349. Haigh, S., et al., *Cross-sectional imaging of individual layers and buried interfaces of graphene-based heterostructures and superlattices.* Nature Materials, 2012. **11**: p. 764-767.
350. Kretinin, A., et al., *Electronic properties of graphene encapsulated with different two-dimensional atomic crystals.* Nano Lett, 2014. **14**: p. 3270-3276.
351. Wang, L., et al., *One-dimensional electrical contact to a two-dimensional material.* Science, 2013. **342**: p. 614-617.
352. Cao, Z., et al., *Mixed-mode interactions between graphene and substrates by blister tests.* Journal of Applied Mechanics, 2015. **82**: p. 081008.
353. Cao, Z., et al., *Mixed-mode traction-separation relations between graphene and copper by blister tests.* Int. J. Solids Struct., 2016. **84**: p. 147-159.
354. Gao, W., K.M. Liechti, and R. Huang, *Wet adhesion of graphene.* Extreme Mechanics Letters, 2015. **3**: p. 130-140.
355. Rakestraw, M., et al., *Time dependent crack growth and loading rate effects on interfacial and cohesive fracture of adhesive joints.* The Journal of Adhesion, 1995. **55**: p. 123-149.
356. Jensen, H.M., *Analysis of mode mixity in blister tests.* Int. J. Fracture, 1998. **94**: p. 79-88.
357. Hutchinson, J.W. and Z. Suo, *Mixed mode cracking in layered materials.* Adv. Appl. Mech., 1991. **29**: p. 63-191.
358. Evans, A.G. and J.W. Hutchinson, *Effects of non-planarity on the mixed-mode fracture-resistance of bimaterial interfaces.* Acta Metallurgica, 1989. **37**: p. 909-916.

359. Na, S.R., et al., *Cracking of polycrystalline graphene on copper under tension.* ACS Nano, 2016. **10**: p. 9616–9625.
360. Xia, Z.C. and J.W. Hutchinson, *Crack patterns in thin films.* Journal of the Mechanics and Physics of Solids, 2000. **48**: p. 1107-1131.
361. Novoselov, K.S., et al., *Electric field effect in atomically thin carbon films.* Science, 2004. **306**: p. 666-669.
362. Li, X.S., et al., *Large-area synthesis of high-quality and uniform graphene films on copper foils.* Science, 2009. **324**: p. 1312-1314.
363. Li, X., et al., *Large-area graphene single crystals grown by low-pressure chemical vapor deposition of methane on copper.* Journal of the American Chemical Society, 2011. **133**: p. 2816-2819.
364. Li, X., et al., *Graphene films with large domain size by a two-step chemical vapor deposition process.* Nano Lett, 2010. **10**: p. 4328-4334.
365. *Commercial scale large-area graphene now available from Bluestone Global Tech in 24"x300" films*. 2013; Available from: http://www.azonano.com/news.aspx?newsID=26553.
366. Tao, L., et al., *Synthesis of high quality monolayer graphene at reduced temperature on hydrogen-enriched evaporated copper (111) films.* ACS Nano, 2012. **6**: p. 2319-2325.
367. Rahimi, S., et al., *Toward 300 mm wafer-scalable high-performance polycrystalline chemical vapor deposited graphene transistors.* ACS Nano, 2014. **8**: p. 10471-10479.
368. Emtsev, K.V., et al., *Towards wafer-size graphene layers by atmospheric pressure graphitization of silicon carbide.* Nature Materials, 2009. **8**: p. 203-207.
369. Wang, Y., et al., *Electrochemical delamination of CVD-grown graphene film: toward the recyclable use of copper catalyst.* ACS Nano, 2011. **5**: p. 9927-9933.
370. Bae, S., et al., *Roll-to-roll production of 30-inch graphene films for transparent electrodes.* Nature Nanotechnology, 2010. **5**: p. 574-578.
371. Kobayashi, T., et al., *Production of a 100-m-long high-quality graphene transparent conductive film by roll-to-roll chemical vapor deposition and transfer process.* Applied Physics Letters, 2013. **102**: p. 023112.
372. Chandrashekar, B.N., et al., *Roll-to-roll green transfer of CVD graphene onto plastic for a transparent and flexible triboelectric nanogenerator.* Advanced Materials, 2015. **27**: p. 5210-5216.
373. Deng, B., et al., *Roll-to-roll encapsulation of metal nanowires between graphene and plastic substrate for high-performance flexible transparent electrodes.* Nano Lett, 2015. **15**: p. 4206-4213.
374. Castellanos-Gomez, A., et al., *Deterministic transfer of two-dimensional materials by all-dry viscoelastic stamping.* 2D Materials, 2014. **1**: p. 011002.
375. Feng, X., et al., *Competing fracture in kinetically controlled transfer printing.* Langmuir, 2007. **23**: p. 12555-12560.
376. Meitl, M.A., et al., *Transfer printing by kinetic control of adhesion to an elastomeric stamp.* Nature Materials, 2006. **5**: p. 33-38.
377. Carlson, A., et al., *Shear-enhanced adhesiveless transfer printing for use in deterministic materials assembly.* Applied Physics Letters, 2011. **98**: p. 264104.
378. Cheng, H., et al., *An analytical model for shear-enhanced adhesiveless transfer printing.* Mechanics Research Communications, 2012. **43**: p. 46-49.
379. Patra, N., B. Wang, and P. Kral, *Nanodroplet activated and guided folding of graphene nanostructures.* Nano Lett, 2009. **9**: p. 3766-3771.
380. Li, T., *Extrinsic morphology of graphene.* Modelling and Simulation in Materials Science and Engineering, 2011. **19**: p. 054005.
381. Kim, K., et al., *Multiply folded graphene.* Physical Review B, 2011. **83**: p. 245433.
382. Zhu, S., J. Galginaitis, and T. Li, *Critical dispersion distance of silicon nanoparticles intercalated between graphene layers.* Journal of Nanomaterials, 2012: p. 375289.

383. Braga, S., et al., *Structure and dynamics of carbon nanoscrolls.* Nano Lett, 2004. **4**: p. 881-884.
384. Zhang, Z. and T. Li, *Carbon nanotube initiated formation of carbon nanoscrolls.* Applied Physics Letters, 2010. **97**: p. 081909.
385. Yu, D. and F. Liu, *Synthesis of carbon nanotubes by rolling up patterned graphene nanoribbons using selective atomic adsorption.* Nano Lett, 2007. **7**: p. 3046-3050.
386. Zhang, Z. and T. Li, *Ultrafast nano-oscillators based on interlayer-bridged carbon nanoscrolls.* Nanoscale Research Letters, 2011. **6**: p. 470.
387. Shi, X., N. Pugno, and H. Gao, *Tunable core size of carbon nanoscrolls.* Journal of Computational and Theoretical Nanoscience, 2010. **7**: p. 517-521.
388. Xie, X., et al., *Controlled fabrication of high-quality carbon nanoscrolls from monolayer graphene.* Nano Lett, 2009. **9**: p. 2565-2570.
389. Zhu, S. and T. Li, *Hydrogenation enabled scrolling of graphene.* Journal of Physics D-Applied Physics, 2013. **46**: p. 075301.
390. Qi, J., et al., *The possibility of chemically inert, graphene-based all-Carbon electronic devices with 0.8 eV gap.* ACS Nano, 2011. **5**: p. 3475-3482.
391. Bell, D., et al., *Precision cutting and patterning of graphene with helium ions.* Nanotechnology, 2009. **20**: p. 455301.
392. Ci, L., et al., *Graphene shape control by multistage cutting and transfer.* Advanced Materials, 2009. **21**: p. 4487-4491.
393. Fischbein, M. and M. Drndic, *Electron beam nanosculpting of suspended graphene sheets.* Applied Physics Letters, 2008. **93**: p. 113107.
394. Zhu, S. and T. Li, *Hydrogenation-assisted graphene origami and its application in programmable molecular mass uptake, storage, and release.* ACS Nano, 2014. **8**: p. 2864–2872.
395. Elias, D., et al., *Control of graphene's properties by reversible hydrogenation: Evidence for graphane.* Science, 2009. **323**: p. 610-613.
396. Li, T., et al., *Compliant thin film patterns of stiff materials as platforms for stretchable electronics.* Journal of Materials Research, 2005. **20**: p. 3274-3277.
397. Lee, H., et al., *A graphene-based electrochemical device with thermoresponsive microneedles for diabetes monitoring and therapy.* Nature Nanotechnology, 2016. **11**: p. 566–572.
398. Kuzum, D., et al., *Transparent and flexible low noise graphene electrodes for simultaneous electrophysiology and neuroimaging.* Nature Commun, 2014. **5**: p. 5259.
399. Park, D.W., et al., *Graphene-based carbon-layered electrode array technology for neural imaging and optogenetic applications.* Nature Commun, 2014. **5**: p. 5258.
400. Kim, S.J., et al., *Multifunctional cell-culture platform for aligned cell sheet monitoring, transfer printing, and therapy.* ACS Nano, 2015. **9**: p. 2677-2688.
401. Park, M., et al., *MoS2-based tactile sensor for electronic skin applications.* Advanced Materials, 2016. **28**: p. 2556-2562.
402. Bollella, P., et al., *Beyond graphene: Electrochemical sensors and biosensors for biomarkers detection.* Biosensors & Bioelectronics, 2016. **In press**.
403. Liu, M.M., R. Liu, and W. Chen, *Graphene wrapped Cu2O nanocubes: Non-enzymatic electrochemical sensors for the detection of glucose and hydrogen peroxide with enhanced stability.* Biosensors & Bioelectronics, 2013. **45**: p. 206-212.
404. Zhou, K.F., et al., *A novel hydrogen peroxide biosensor based on Au-graphene-HRP-chitosan biocomposites.* Electrochimica Acta, 2010. **55**: p. 3055-3060.
405. Song, H.Y., Y.N. Ni, and S. Kokot, *A novel electrochemical biosensor based on the hemin-graphene nano-sheets and gold nano-particles hybrid film for the analysis of hydrogen peroxide.* Analytica Chimica Acta, 2013. **788**: p. 24-31.

406. Wu, S.X., et al., *Electrochemically reduced single-layer MoS2 nanosheets: Characterization, properties, and sensing applications.* Small, 2012. **8**: p. 2264-2270.
407. Palanisamy, S., S.H. Ku, and S.M. Chen, *Dopamine sensor based on a glassy carbon electrode modified with a reduced graphene oxide and palladium nanoparticles composite.* Microchimica Acta, 2013. **180**: p. 1037-1042.
408. Chen, Y., et al., *Two-dimensional graphene analogues for biomedical applications.* Chemical Society Reviews, 2015. **44**: p. 2681-2701.
409. Liu, Z., et al., *PEGylated nanographene oxide for delivery of water-insoluble cancer drugs.* Journal of the American Chemical Society, 2008. **130**: p. 10876-10877.
410. Zhang, L.M., et al., *Enhanced chemotherapy efficacy by sequential delivery of siRNA and anticancer drugs using PEI-grafted graphene oxide.* Small, 2011. **7**: p. 460-464.
411. Yang, K., et al., *Graphene in mice: Ultrahigh in vivo tumor uptake and efficient photothermal therapy.* Nano Lett, 2010. **10**: p. 3318-3323.
412. Tian, B., et al., *Photothermally enhanced photodynamic therapy delivered by nano-graphene oxide.* ACS Nano, 2011. **5**: p. 7000-7009.
413. Yang, K., et al., *Nano-graphene in biomedicine: theranostic applications.* Chemical Society Reviews, 2013. **42**: p. 530-547.
414. Narayanan, T.N., et al., *Hybrid 2D nanomaterials as dual-mode contrast agents in cellular imaging.* Advanced Materials, 2012. **24**: p. 2992-2998.
415. Jin, Y.S., et al., *Graphene oxide modified PLA microcapsules containing gold nanoparticles for ultrasonic/CT bimodal imaging guided photothermal tumor therapy.* Biomaterials, 2013. **34**: p. 4794-4802.
416. Shi, S.X., et al., *Tumor vasculature targeting and imaging in living mice with reduced graphene oxide.* Biomaterials, 2013. **34**: p. 3002-3009.
417. Wang, Z., et al., *Biological and environmental interactions of emerging two-dimensional nanomaterials.* Chemical Society Reviews, 2016. **45**: p. 1750-1780.
418. Zhou, R. and H. Gao, *Cytotoxicity of graphene: recent advances and future perspective.* WIREs Nanomed Nanobiotechnol, 2014. **6**: p. 452–474.
419. Li, Y., et al., *Graphene microsheets enter cells through spontaneous membrane penetration at edge asperities and corner sites.* Proceedings of the National Academy of Sciences, 2013. **110**: p. 12295–12300.
420. Wang, J., et al., *Cellular entry of graphene nanosheets: the role of thickness, oxidation and surface adsorption.* RSC Advances, 2013. **3**: p. 15776-15782.
421. Tu, Y., et al., *Destructive extraction of phospholipids from Escherichia coli membranes by graphene nanosheets.* Nature Nanotechnology, 2013. **8**: p. 594–601.
422. Zhu, W., et al., *Nanomechanical mechanism for lipid bilayer damage induced by carbon nanotubes confined in intracellular vesicles.* Proceedings of the National Academy of Sciences, 2016: p. in press.
423. Akinwande, D., N. Petrone, and J. Hone, *Two-dimensional flexible nanoelectronics.* Nature Commun, 2014. **5**: p. 5678.
424. Yao, S. and Y. Zhu, *Nanomaterial-enabled stretchable conductors: strategies, materials and devices.* Advanced Materials, 2015. **27**: p. 1480-1511.
425. Akinwande, D., et al., *Large-area graphene electrodes: Using CVD to facilitate applications in commercial touchscreens, flexible nanoelectronics, and neural interfaces.* IEEE Nanotechnology Magazine, 2015. **9**: p. 6-14.
426. Xiao, X., Y. Li, and Z. Liu, *Graphene commercialization.* Nature Materials, 2016. **15**: p. 697-698.
427. Park, S., et al., *Extremely high frequency flexible graphene thin film transistors.* IEEE Electron Device Letters, 2016. **37**: p. 512 - 515.

428. Zhu, W., et al., *Black phosphorus flexible thin film transistors at GHz frequencies.* Nano Lett, 2016. **16**: p. 2301–2306.
429. Chang, H.-Y., et al., *Large-area monolayer MoS2 for flexible low-power RF nanoelectronics in the GHz regime.* Advanced Materials, 2016. **28**: p. 1818–1823.